\documentclass[11pt,a4paper]{article} 

\usepackage{amsmath,amsfonts,epsfig,graphicx}
\usepackage{jheppub}
\usepackage{multicol}
\usepackage{float}
\numberwithin{equation}{section}
\def\nn{\nonumber}

\def\beq{\begin{equation}}
\def\eeq{\end{equation}}
\def\bea{\begin{eqnarray}}
\def\eea{\end{eqnarray}}

\newcommand{\ie}{{\it i.e.}}
\newcommand{\eg}{{\it e.g.}}

\newcommand{\pythia}{{\sc Pythia}}
\newcommand{\herwig}{{\sc Herwig}}
\newcommand{\mgme}{{\sc MadGraph/MadEvent}}

\newcommand{\madgraph}{{\sc MadGraph}}
\newcommand{\madevent}{{\sc MadEvent}}

\proceeding{\begin{flushright} PSI-PR-10-12 \\ CP3-10-41 \\ KEK-TH-1432 \end{flushright}}

\title{Simulating graviton production at hadron colliders}

\author[a,b]{Priscila de Aquino}
\author[c]{Kaoru Hagiwara}
\author[d]{Qiang Li}
\author[a]{Fabio Maltoni}

\affiliation[a]{Center for Cosmology, Particle Physics and Phenomenology (CP3),\\
Universit\'e Catholique de Louvain, B-1348 Louvain-la-Neuve, Belgium}
\affiliation[b]{Instituut voor Theoretische Fysica, Katholieke Universiteit Leuven,\\ B-3001 Leuven, Belgium}
\affiliation[c]{KEK Theory Center and Sokendai,\\ Tsukuba 305-0801, Japan}
\affiliation[d]{Paul Scherrer Institut,\\ CH--5232 Villigen PSI, Switzerland}

\emailAdd{priscila@itf.fys.kuleuven.be}
\emailAdd{kaoru.hagiwara@kek.jp}
\emailAdd{qiang.li@psi.ch}
\emailAdd{fabio.maltoni@uclouvain.be}

\abstract {Spin-2 particles and in particular gravitons are predicted in
many new physics scenarios at the TeV scale. Depending on the details of
models such new states might show up as a continuum, massless particles,
or TeV scale resonances. Correspondingly, very different discovery
signatures should be exploited, from the search of excesses in events with
multi jets and large missing transverse energy, to resonances in weak boson
or jet pair productions. We present a very general and flexible implementation in \mgme~of spin-2 particles interacting with the standard model particles via
the energy momentum tensor, which encompasses all of the most popular TeV scale 
models featuring gravitons. By merging matrix elements with parton shower,
we can generate inclusive samples of graviton+jets  at the hadron colliders in several scenarios (ADD, zero-mass graviton and RS). We compare and validate our results against the corresponding next-to-leading order QCD calculations. }

\date{\Date}


\arxivnumber{1101.5499}

\begin{document}
\maketitle

% --------------------------------------------------------------------

\section{Introduction}
\label{intr}

Currently, the Standard Model (SM) of particle physics matches to a great accuracy a plethora of the experimental data. However, there are many reasons to expect new physics beyond the SM appearing at the LHC (TeV) energy scale, such as the demand for a dark matter candidate or  the problem on the large hierarchy between the electroweak and Planck scales. Many models attempt to provide a solution to these and other issues by extending the SM to larger symmetries (such as in Supersymmetric or Little Higgs models) or larger space-time (extra space dimensional models), with the condition that they reproduce the SM at low energies, while new physics effects might only become apparent at high energies.
The mechanism of how these new degrees of freedom appear at the TeV scales is particularly intriguing for
extra dimensional theories.

Take the flat extra dimensional models (ADD)~\cite{Antoniadis:1998ss,Arkani-Hamed:1998fc,Arkani-Hamed:1999qc} for example. In this $D=4+\delta$ dimensional model, only gravitons ($G$) can propagate in the $D$ dimensional space-time, while all the SM particles are confined onto the usual 3+1-dimensional space-time. Assuming for simplicity that the additional $\delta$-dimensional space to be compactified on the torus with a common radius $R$, the Planck scale ($M_{\rm Pl}$)
can then be related to the fundamental scale $M_S$ as follows:
\begin{eqnarray}\label{scale}
M^2_{\rm Pl}=8\pi R^\delta M^{\delta+2}_{\rm S}\,.
\end{eqnarray}
If one chooses a large compactification radius ($R\gg 1/M_{\rm Pl}$), the hierarchy between the Planck and the electroweak scale can be understood by the fact that the fundamental scale is now allowed to be 
closer to the weak scale, \ie, $M_{\rm S} \sim {\rm TeV}$, where the effects of extra dimensions will start manifest
themselves through an apparent non-conservation of energy and momentum in the usual 3+1 dimensions.

In addition to the ADD model, other TeV scale theories on gravity have been proposed that attempt to explain the hierarchy problem in similar or different ways. For instance, if a huge number of particles are in a hidden sector that interacts only gravitationally with the SM particles, the Newton's constant might be affected and the fundamental gravity's scale could then be in the TeV region (TeV scale Massless Graviton Model or MGM for short) \cite{Dvali:2002mw,Dvali:2007hz,Dvali:2007ud, Calmet:2007uq, Calmet:2008kx, Calmet:2008vn}. Alternatively, in warped extra dimensional models  proposed by Randall and Sundrum (RS) in 1999 \cite{Randall:1999fk, Randall:1999vf}, the large hierarchy is addressed by the introduction of a geometrical exponential factor in the metric, a solution that does not need the extra dimension to be significantly  larger than the Planck length.

The search for this kind of new physics at the TeV scale is today among the major tasks of the LHC. The most
natural signatures involve  missing energy in association with jet(s) (\eg, in the ADD and MGM models), or spin-2 resonances in lepton or jet pair productions (\eg, in the RS model). The experimental analyses of such signatures 
demand accurate simulations within Monte Carlo (MC) event generators not only to design efficient signal selection strategies  but also to extract information on the new physics parameters (masses, coupling strengths, compositeness, structures and spin). Several tools exist nowadays that are able to perform such simulations for a wide class of models~\cite{Butterworth:2010ym}. In this work, we will focus on \madgraph~\cite{Stelzer:1994ta, Maltoni:2002qb, Alwall:2007ys}, where  a spin-2 particle implementation at the matrix element level has been presented recently~\cite{Hagiwara:2008zr}. In this work we have extended it to full-fledged automatic event generator and embedded it in the \mgme~(MG/ME) package. 
This now includes  a wide class of models involving spin-2 particles, such as:
\begin{itemize}
{\item weakly coupled (Planck scale suppressed) massive graviton models, with infinite number of nearly continuously distributed Kaluza-Klein (KK) modes of the graviton and the overall integrated effects are significant at the TeV scale, \eg, the ADD model;} 
{\item TeV scale massless graviton models, in which the graviton effects are negligible below the fundamental scale of quantum gravity, \ie, the TeV scale, due to the running of the gravitational constant, \eg, in theories with many particles that interacts with the SM particles only gravitationally;}
{\item strongly coupled (TeV scale) massive graviton models, in which the graviton KK modes are widely separated and the lowest one gets mass at the TeV scale and can decay into the SM particles within detectors, \eg, the RS model.}
\end{itemize}

To show and validate the capabilities of the new implementation we have chosen to study the most important signatures in representative models of the three classes above.
For the ADD and MGM models, the gravitons couple weakly to the SM particles and thus can appear as missing energy signals at colliders. The main searching channel is $pp(\bar{p}) \rightarrow G + {\rm jet(s)}$, in which at least one accompanying hard jet is needed. For the RS model, resonance production  $pp(\bar{p}) \rightarrow G$ is expected, with $G$ decaying into $e^+e^-$ or $\mu^+\mu^-$ for example. In fact, for completeness we further consider the $G$ plus mono-jet channel  to compare  the RS model with the ADD and MGM results. The corresponding NLO QCD calculations to those channels have been performed in Refs.~\cite{Karg:2009dq,Mathews:2005cr,Li:2006nx}.

In this paper we have extended the NLO QCD calculations in \cite{Karg:2009dq} to the cases of the RS and MGM models. While current NLO fixed-order results are useful to reliably predict cross sections and observables  involving at most one jet, multi-jet based observables and more exclusive quantities in general can have an important role in the analyses. For example, NLO QCD calculations for $G$ + mono-jet production can only provide LO distributions for the 2nd jet $G\,+\,{\rm jet}$. Merging matrix elements with a parton shower can, on the other hand, reliably predict multi-jet final states and generate samples that can be directly used in experimental simulations. Several methods of \emph{merging} or \emph{matching schemes}  are now available~\cite{Catani:2001ve,Krauss:2002ly, Hoeche:qf} which have been shown to be in good agreement~\cite{Alwall:2007bh}. 
 
Here, we present the first results from the parton shower merging with matrix elements, for spin-2 graviton productions at hadron colliders, where the new implementation of graviton in \mgme~and $k_{\perp}-$MLM merging scheme~\cite{Alwall:2008qv} are used. We further present a detailed comparison between the matching and the relevant NLO/LO results \cite{Karg:2009dq,Mathews:2005cr,Li:2006nx}. We also include the results for an irreducible and representative background process from hadro-productions of $Z$ obtained with {\sc MCFM}~\cite{Campbell:2002oq}.

The article is organized as follows. In section~\ref{model} we review briefly the three models mentioned above. The complete implementation of spin-2 particle in \mgme~is detailed in Section~\ref{implem}. In Section~\ref{matching}, we give an introduction for parton shower matching in \mgme. The numerical results of the matching and the NLO/LO calculations for $G$ hadro-productions are presented in Section~\ref{numres}. We draw our conclusions in Section~\ref{sec:end}.

% --------------------------------------------------------------------
\section{Description of the models}
\label{model}

\subsection{Extra space Dimensions: ADD and RS models}
\label{exdm}
Various extra dimension models which have been proposed so far can be divided into two major classes according to the geometry of the background space-time manifold. The first one includes the ADD and its variants, which extend the dimension of the total space-time to $D=4+\delta$, with a factorisable metric and large size of the compact extra dimensions ($\gg 1/M_{\rm Pl}$). The second one includes the 5-dimensional RS model~\cite{Randall:1999fk,Randall:1999vf} and its variants, in which a warped metric is introduced along the 5-th dimension  the size of the extra dimension can be at the order of the Planck length.

In both classes of extra dimension models  Kaluza-Klein (KK) towers of massive spin-2 gravitons appear, that can interact with the Standard Model (SM) fields. The effective interaction Lagrangian is given by~\cite{Han:1999kl, Giudice:1999tg}:
\begin{align}
 \label{IL}
 {\cal L}_{\rm int} = - \frac{1}{\Lambda}
  \sum_{\vec{n}} T^{(\vec{n})\mu \nu} {\cal T}_{\mu \nu},
\end{align}
where $T^{(\vec{n})\mu \nu}$ is the $\vec{n}$-th graviton KK modes,
and $\Lambda$ is the relevant coupling scale. In the ADD model
\begin{align}
 \Lambda = {\overline M}_{\rm Pl}\equiv M_{\rm Pl}/\sqrt{8\pi}
           \sim 2.4\times 10^{18}\ {\rm GeV},
\end{align}
where ${\overline M}_{\rm Pl}$ is the 4-dimensional reduced Planck mass, while in the RS model 
\begin{align}
 \Lambda=e^{-kr_c\pi}\overline{M}_{\rm Pl}\,,
\end{align}
where $k$ is a scale of order of the Planck scale and $r_c$ is the compactification radius. By tuning $k r_c$ one can place $\Lambda$  at the electroweak scale.

In Eq.~(\ref{IL}), ${\cal T}_{\mu \nu}$ is the energy-momentum tensor of the SM fields,
\begin{align}
\label{tensor}
 {\cal T}_{\mu\nu}
 &= \Big(-\eta_{\mu\nu} {\cal L}_{\rm SM}
   +2\frac{\delta{\cal L}_{\rm SM}}{\delta g^{\mu\nu}}\Big)
    {\Big |}_{g^{\mu\nu}=\eta^{\mu\nu}},
\end{align}
where $g^{\mu\nu}$ is the metric and $\eta^{\mu\nu}={\rm
diag}(1,-1,-1,-1)$ is the Minkowski value. Note the difference on the effective strength of the gravity/SM coupling for each class of models. 

In the ADD model the individual KK resonances have masses equal to $m_{(\vec{n})} = |\vec{n}|/R$. The mass gap between neighboring modes $\Delta m=R^{-1}$ is thus small for $\delta$ not too large. Quantitatively one finds $\Delta m \approx$ 20~keV, 7~MeV and 0.1~GeV for $M_{\rm S} =1$~TeV and $\delta=$4, 6 and 8, respectively~\cite{Giudice:1999tg}. The discrete mass spectrum can be approximated by a continuum with a density of states ${\rm
d}N = \rho(m) {\rm d}m$~\cite{Giudice:1999tg,Mirabelli:1999hc}, where
\begin{eqnarray}
\label{rho}
\rho(m)= S_{\delta-1}\frac{{\overline M}_{\rm Pl}^2}{M_{\rm
S}^{2+\delta}}m^{\delta-1},\, \,\, \, \rm{and}\, \,
S_{\delta-1}=\frac{2\pi^{\delta/2}}{\Gamma(\delta/2)}.
\end{eqnarray}
In other words, the 4-dimensional graviton appears as an infinite sum of excited states of the graviton. Current terrestrial test of gravity set a limit on $M_{\rm S} \ge 3.6$~TeV for $\delta = 2$~\cite{Kapner:2007ij}. Further constraints have been derived from astrophysics and cosmology, in particular for $\delta < 4$. However, they can be evaded in specific models~\cite{Kaloper:2000bs,Dienes:2002fv,Giudice:2005dz} and do not lessen the importance of collider searches for extra dimensions. 
At high energy colliders, both virtual graviton exchange between SM particles and real graviton emission provide viable signatures of large extra dimensions. Since the coupling of gravitons with matter is suppressed $\propto 1/{\overline M}_{\rm Pl}$, direct graviton production gives rise to missing energy signals. Searches for the ADD graviton production have been performed in the processes $e^+e^- \to \gamma(Z) + E^{\rm miss}$ at LEP and $p\bar{p}\to \gamma({\rm jet}) + p_T^{\rm miss}$ at the Tevatron. 
The combined LEP limits~\cite{LEP} read $M_{\rm S}>$ 1.60, 1.20, 0.94, 0.77, 0.66~TeV, for $\delta=$ 2,$\cdots$,6 respectively, while Tevatron searches exclude $M_{\rm S} >$ 1.40, 1.15, 1.04, 0.98, 0.94~TeV, for $\delta=$ 2,$\cdots$,6 respectively~\cite{Collaboration:2008fu, Collaboration:2008kl, Collaboration:2006qa}.

In the RS model, the mass of the $n$th graviton KK excitation
mode  at the electroweak scale is given by 
\begin{eqnarray}\label{RSMass} m_n=kx_ne^{-kr_c\pi}=m_1\frac{x_n}{x_1},
\end{eqnarray}
where the $x_n$'s are the $n$th roots of the first order Bessel function. The graviton sector of the RS model is completely determined by the two parameters $m_1$ and $\Lambda$. Current constraints \cite{Davoudiasl:2009mi, Aaltonen:2008pi,
Collaboration:2007ff} for the parameters of the RS model are from the theoretical requirement, the low energy precise measurement and also the data from Tevatron, from which $0.01\leq k/\overline{M}_{\rm Pl}<0.1$ and $\Lambda\leq 10$\,TeV. The relation between $k/\overline{M}_{\rm Pl}$ and $\Lambda$ is given by: 
\begin{equation}
\Lambda=\frac{m_1\overline{M}_{\rm Pl}}{x_1\,k}.
\end{equation}
At variance with ADD, the lightest RS massive graviton can have a mass of TeV scale, and may be produced copiously at the LHC. More importantly, it has much larger couplings to the SM particles than the ones in the ADD model, which allows the graviton to decay into observable particles and hence be detected as a resonance.

% --------------------------------------------------------------------
\subsection{The Massless Graviton Model}
\label{MGMsection}

In the previous section, we have seen that if Nature entails more than 4 space-time dimensions, gravity could become strong at much lower scale than the Planck scale. However, it has recently been shown that even for a 4-dimensional model, TeV scale gravity effects could appear \cite{Dvali:2007hz,Calmet:2008kx}. 

For instance, Ref.~\cite{Dvali:2007hz} suggests that strong gravity should emerge at the TeV scale if there are huge number of hidden sector particles that interact only gravitationally with the Standard Model particles.  In Ref.~\cite{Calmet:2008kx}, it has been suggested that the emergence of TeV scale quantum gravity may be interpreted as renormalization of the effective gravity coupling due to hidden sector particles, based on the one-loop computation \cite{Larsen:1995ax}.  Common to these models \cite{Dvali:2002mw, Dvali:2007hz, Dvali:2007ud, Calmet:2007uq, Calmet:2008kx, Calmet:2008vn} is the existence of order $(M_{\rm Pl}/TeV)^2 \sim 10^{32}$ hidden particles at or below the TeV scale, that can only be probed by the gravity interactions.

The effective Lagrangian for this type of theory can be expressed by the effective Fierz-Pauli Lagrangian combined with the interaction term of the graviton to the Standard Model particles:
\begin{equation}
{\cal L}_{\rm int}=-\frac{1}{{\overline M}(\mu_\star)} \, T^{\mu\nu} {\cal T}_{\mu\nu},
\label{eq.LagrMassless}
\end{equation}
where $T^{\mu\nu}$ is the massless graviton and ${\cal T}^{\mu\nu}$ is the SM energy-momentum tensor given by Eq.~(\ref{tensor}). The graviton coupling has the same form as that of Einstein's theory of general relativity, except for the effective Planck mass $\overline{M}(\mu_\star)$ which is assumed to be in the TeV range at energy scale $\mu_\star \sim$ TeV. 
One can note the similarity between Eq.~(\ref{eq.LagrMassless}) and the Eq.~(\ref{IL}). As a consequence, similar Feynman rules are expected. A detailed set of Feynman rules for the extra dimensional theories can be found in Refs.~\cite{Han:1999kl, Giudice:1999tg}, and for the MGM Model in Ref.~\cite{Calmet:2009fu}.

Presently, the MGM model is less constrained than ADD or RS models. It has been shown that the cosmic rays experiment AGASA sets a bound of 550 GeV on the effective Planck mass $\overline{M} (\mu_\star)$ in 4-dimensions \cite{Calmet:2008ye}. However, more recently results on monojet in addition to graviton emission \cite{Calmet:2009fu, Calmet:2009qo} predict this limit to be at most $\mathcal{O}$(5 TeV) at the LHC (14 TeV).

% --------------------------------------------------------------------
\section{Graviton implementation into MadGraph}
\label{implem}

Fortran subroutines to calculate helicity amplitudes with massive gravitons were added to the {\sc HELAS} (HELicity Amplitude Subroutines) library \cite{Murayama:1992gi, Hagiwara:1990dw} recently~\cite{Hagiwara:2008zr}. Thanks to these new subroutines which encode the Lorentz structure of the new couplings, the ADD and RS models could be implemented in \madgraph~after a few modifications. The implementation in \mgme, however, was not complete due to a few technical limitations.

First, \madgraph~version 4 can only generate Feynman diagrams with up to 4-point vertices, the diagrams involving the 5-point vertex (4 gluons-graviton) have therefore  to be added by hand.
Although it was fine for previous studies on graviton production with up to 2 jets~\cite{Karg:2009dq, Hagiwara:2008tw}, it is inconvenient for graviton production with higher multiplicity jets. 

Secondly, gravitons are densely distributed in the ADD model (see Eq.~(\ref{rho})) and a special treatment for the
propagator is called for.

Third, for the MGM model it is necessary to have the off-shell graviton current subroutines modified due to differences on the graviton propagator and polarization summation for the massless and massive cases. 

We present hereby the details on how each of the above issues has been tackled to achieve spin-2 graviton generation within \mgme~in a full automatic way.

\subsection{Five-point vertices}

\begin{figure}[t]
    \begin{center}
	\includegraphics[width=12cm]{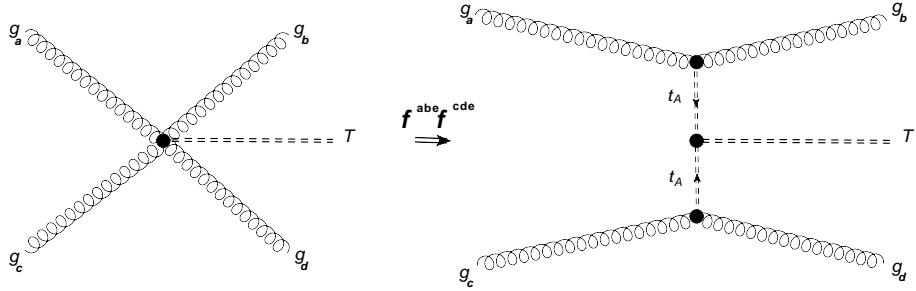}
	\end{center}
	\caption{Reduction of the portion of the 5 point vertex with color term $f^{abe}f^{cde}$into 3-point vertices by introducing auxiliary tensor particle $t_A$. $T$ represents the graviton.}
	\label{5point}
\end{figure}

The interactions (\ref{IL}) and (\ref{eq.LagrMassless}) entail a five-point vertex, {\tt ggggT}:

\begin{align}\label{g5t}
 {\cal L}_{\tt ggggT}
  = &-{\tt GT\,GC}^2\, f^{abe}f^{cde}\, T^{\mu\nu*} \times\Big[\frac{1}{4}\eta_{\mu\nu}
   g^{a,\rho*}g^{b,\sigma*}g^{c\ast}_\rho g^{d\ast}_\sigma
  -g^{b,\rho*}g^{a*}_\mu g^{c*}_{\nu}g^{d*}_\rho\Big],
\end{align}
with the coupling constants {\tt GT} = {\tt $-\frac{1}{\Lambda}$} and {\tt GC} = $g_s$. In  {\sc HELAS} such a vertex can be encoded as {\tt GGGGTX} computing the portion of the amplitude of the {\tt ggggT} vertex from four {\tt G}luon polarization vectors and a {\tt T}ensor boson wavefunction corresponding to the color structure $f^{abe}f^{cde}$, 
see figure~\ref{5point} the diagram on the left-hand side. 
In order to implement this vertex in \madgraph~we reinterpret it as scattering in the $t,u$ and $s$  channels of non-propagating octet tensor boson $t_A$ as an auxiliary particle. With $t_A$ we can reduce the portion of the 5-point vertex with color structure $f^{abe}f^{cde}$ into 3-point vertices: {\tt $ggt_A$} and {\tt $t_At_AT$}, which can then be generated by \madgraph~automatically, figure~\ref{5point} the diagram on the righ-hand side. Note we assign a flow to $t_A$, \ie, we treat $t_A$ different from its antiparticle in \madgraph~, and forbid the interaction with the antiparticle to avoid appearance of additional diagrams for $gg\rightarrow gg$.

After introducing particle $t_A$, we add several new {\sc HELAS} subroutines as specified on table 1, by matching the amplitude of the right side in figure~\ref{5point} with the portion of the left side of the same color structure, \ie, Eq.~(\ref{g5tb}). The interested reader will find a detailed description in Appendix \ref{5pointHELAS}.

\begin{table}
	\begin{center}
    \begin{tabular}{ | c | p{12cm} |}
    \hline
    Sub-routine & Summary \\ \hline
    VVTCXX & computes the amplitude of the {\tt gg$t_A$} vertex. \\ \hline
    JVTCXX &  computes the off-shell vector current {\tt J} made from the interactions of a {\tt V}ector gluon and an auxiliary {\tt T}ensor boson by the {\tt gg$t_A$} vertex. \\ \hline
    UVVCXX &  computes the off-shell tensor current {\tt U} for the auxiliary tensor $t_A$, made from two flowing-out {\tt V}ector gluons by the {\tt gg$t_A$} vertex. \\ \hline
    TTTXXX & computes the amplitude of the {\tt $t_A$ $t_A$T} vertex.\\ \hline
    UTTAXX & computes the off-shell non-propagating tensor current {\tt U}, made from the flowing-out graviton {\tt T}ensor and the auxiliary {\tt T}ensor by the {\tt $t_A$ $t_A$T} vertex.\\ \hline
    UTTBXX & computes the off-shell graviton tensor current {\tt U}, made from the two flowing-out auxiliary {\tt T}ensors by the {\tt $t_A$ $t_A$T} vertex.  \\
    \hline
    \end{tabular}
    \caption{Summary of new vertices introduced for the implementation of the 5-point vertex in the {\sc HELAS} structure.}
	\end{center}
    \label{Table5pHELAS}   
\end{table}

The  new {\sc HELAS} subroutines have been tested by using the QCD gauge invariance and the general coordinate transformation invariance of the helicity amplitudes.  Finally, we mention that the above technique has also been applied for a new implementation of the Higgs effective theory (HEFT) in \madgraph, see details  in Appendix~\ref{heftb}.

\subsection{Summation over the graviton mass spectrum}
\label{gmi}
The 4-dimensional graviton can be approximated as an infinite tower of KK graviton modes within the ADD theory. A MC simulation has therefore to take into account an extra phase-space integration to perform: the amplitude integration over the graviton mass density function (Eq.~(\ref{rho})). 

For virtual graviton exchange, the tower of KK graviton modes leads to the summation of their propagators, which is ultraviolet (UV) divergent since ADD is only an effective theory. One usually introduces an UV cutoff for the highest KK modes and replaces the summation by factors in some effective scale \cite{Han:1999kl}. This can be easily implemented in \madgraph~by setting the graviton mass to zero and the extra dimension scale to be the effective one.

Nevertheless, it is more complicated for the real graviton production case, for which we need to actually perform the integration described above.  The integration is carried out by the introduction of two massless colorless auxiliary particles: a scalar {\tt $x_1$}, and a pseudo-tensor {\tt $x_2$}. \madgraph~is then modified to include a new particle type {\tt P}, which is only involved in the interaction with graviton: {\tt $x_1$ $x_2$T}. Instead of generating process such as $pp\rightarrow T+X$ in \madgraph, one can generate $pp\rightarrow x_1x_2+X$. The phase spaces are related by
\begin{align}
 d\Phi(x_1x_2+X) =\frac{dm^2_T}{16\pi^2}d\Phi(T+X)\,.
 \label{ps}
\end{align}
Using Eq.~(\ref{ps}), we can perform the mass integration with the help of the new {\sc HELAS} subroutines for the {\tt $x_1$ $x_2$T} vertex summarized in table 2. Details on each new sub-routine are reported in the Appendix \ref{MassIntegration}.

\begin{table}
	\begin{center}
    \begin{tabular}{ | c | p{12cm} |}
    \hline
    Sub-routine & Summary \\ \hline
    PXXXXX & stores the helicity and momentum of the auxiliary particles $x_1$ and $x_2$. \\ \hline
    TPSXXX & computes the amplitude of $x_1 x_2  {\tt T}$ vertex from a set of parameters specified by the theory. \\ \hline
    UPSXXX & computes an off-shell graviton tensor current {\tt U}, made from the flowing-out auxiliary {\tt S}calar and {\tt P}esudo-particles, by the {\tt $x_1$ $x_2$T} vertex.\\
    \hline
    \end{tabular}
    \caption{Summary of new vertices introduced for the implementation of the mass integration in \mgme~within the framework of the ADD model.}
	\end{center}
    \label{TableMassIntegration}   
\end{table}

To conclude we mention that the new {\sc HELAS} subroutines and the new implementation of the propagators have been tested by comparing with previous implementation~\cite{Hagiwara:2008zr} always finding excellent agreement.

\subsection{Implementation of the massless graviton model}
\label{imgm}
A well-known property of spin-2 particles in the framework of Einstein's general relativity is the so called van Dam-Veltman discontinuity \cite{vanDam:1970vg}, which states that the massless propagator for a graviton cannot be obtained by
the massive one through a smooth limit. Such discontinuity  is due to the fact that the massive spin-2 particles have five states of polarization, while massless ones have only two of them. This can be observed on part of the the massive/massless graviton propagators \cite{Han:1999kl,Calmet:2009fu}:
\begin{eqnarray} \label{Propagators}
B^{\rm massive}_{\mu\nu,\alpha\beta} & \equiv & \sum_{\lambda = \pm2, \pm1, 0} \epsilon_{\mu\nu}(p,\lambda) \epsilon_{\alpha\beta}(p,\lambda)^ * =\nonumber \\
& = & \frac{1}{2}(\eta_{\mu\alpha} \eta_{\nu\beta}+\eta_{\mu\beta} \eta_{\nu\alpha} - \eta_{\mu\nu} \eta_{\alpha\beta}) - \frac{1}{2 m_T^2}(\eta _{\mu\alpha}p_\nu p_\beta +\eta _{\nu\beta}p_\mu p_\alpha + \nonumber \\
& & \eta _{\mu\beta}p_\nu p_\alpha +\eta _{\nu\alpha}p_\mu p_\beta) + \frac{1}{6} ( \eta_{\mu\nu}
+\frac{2}{m_T^2}p_\mu p_\nu) (\eta_{\alpha\beta}+\frac{2}{m_T^2}p_\alpha p_\beta),  \\
\nonumber \\
B^{\rm massless}_{\mu\nu,\alpha\beta} & \equiv & \sum_{\lambda = \pm2} \epsilon_{\mu\nu}(p,\lambda)\epsilon_{\alpha\beta}(p,\lambda)^ * = \nonumber \\
& = & \frac{1}{2}(\hat{\eta}_{\mu\alpha}\hat{\eta}_{\nu\beta}+ \hat{\eta}_{\mu\beta} 
\hat{\eta}_{\nu\alpha} - \hat{\eta}_{\mu\nu} \hat{\eta}_{\alpha\beta}) \label{Prop2},
\end{eqnarray}
where  $p^{\mu} = (p^ 0, \vec{p})$ is the four-momentum of the graviton, $\hat{\eta}_{\mu\nu} = \eta_{\mu\nu}- \frac{p_{\mu} \bar{p}_{\nu} + \bar{p}_{\mu} p_{\nu}}{p \cdot \bar{p}}$ and $\bar{p}^{\mu} = (p^ 0, -\vec{p})$. One can always check that the graviton propagators satisfy $B_{\mu\nu,\alpha\beta}\,p^ {\mu} = 0$, and lead to the correct number of spin degree of freedom by computing the traces:
 \begin{eqnarray}
 \eta^ {\mu\alpha} \eta^ {\nu\beta} B_{\mu\nu,\alpha\beta}^ {\rm massive}  =  5, \,\,\,\,\,
  \eta^ {\mu\alpha} \eta^ {\nu\beta} B_{\mu\nu,\alpha\beta}^ {\rm massless}  =  2 .
 \end{eqnarray}
Observe from Eqs.~(\ref{Propagators}) and (\ref{Prop2}) that $B^{\rm massive}_{\mu\nu,\alpha\beta}$ in the limit when $m_T \to 0$ is different than $B^{\rm massless}_{\mu\nu,\alpha\beta}$.

This property has to be taken into account for the  {\sc HELAS} routines that includes massless gravitons. The corresponding modifications have been  introduced for: i) the tensor wave function subroutine and ii) all the subroutines with an off-shell graviton.

The new subroutines have been tested through gauge invariance, and by comparing results of total cross sections for various $2 \to 2$ processes against the massive graviton case. The reason why we are allowed to perform the comparison is that although the discontinuity on the graviton propagator exists, it does not occur at the cross section level for processes relevant at colliders~\cite{Mirabelli:1999hc, Calmet:2009fu}.

Moreover, the implementation of the running of the Planck mass has been done in  \mgme~during the generation of the events: we have applied a minimum cut of ${\rm H}^{\rm min}_{\bot}>\overline{M} (\mu_\star) $ on the sum of the jets transverse momentum. In other words, a Heaviside step function-like behavior is added to the cross section, \ie, for collisions with $\sqrt{\hat s, (-t), (-u)} < \overline{M}(\mu_\star)$, gravity contributions are so weak that the cross section can be approximated to zero. 

% --------------------------------------------------------------------
\section{Parton shower merging with tree level matrix elements}
\label{matching}

As the parton center of mass energy increases, hadron collision events with large jet multiplicities become more probable. As a result, accurate simulations require to correctly account for the presence of QCD radiation, which might modify or alter
the leading order predictions for the relevant observables. 

In parton shower MC programs like \pythia~\cite{Sjostrand:2006zt} and \herwig~\cite{G.Corcella:2001jl} additional jets are usually obtained in the collinear and soft approximation. Hard and widely separated jets are therefore poorly described in
this approach.  On the other hand, tree-level fixed order calculations can provide reliable predictions in the hard
region, while failing in the collinear and soft limits.

To combine both descriptions and avoid double counting or gaps between samples with different multiplicity, an appropriate matching method is required. Several algorithms have been proposed over the years:
the CKKW method, based on shower veto and therefore on event re-weighting~\cite{Catani:2001ve,Krauss:2002ly}
and the MLM-based scheme, based on event rejection~\cite{Alwall:2007bh}.

For SM processes, in particular for the weak vector boson hadro-productions, matching schemes have been extensively applied and compared to the available data from the Tevatron~\cite{Krauss:2004gb}. Furthermore, studies have been performed for new physics in order express the significance of the matching, such as the notable reduction of the dependence on parton shower parameters~\cite{Alwall:2008qv}.

In this study, we use the $k_\bot$-MLM  matching scheme
implemented in \mgme \\ and interfaced to \pythia~for parton
shower and hadronization. Detailed comparisons of several event generators that include matching schemes have been performed in Refs.\cite{Mrenna:2004mb,Alwall:2007bh,Butterworth:2010ym}, showing a good overall agreement.

In the $k_\bot$ matching scheme, matrix element multi-parton events are produced with a minimum separation $k_\bot$ cutoff of $Q^{\rm ME}_{\rm min}$. For every event, the final-state partons are clustered according to the $k_\bot$ algorithm. Only clusterings corresponding to the Feynman diagrams provided by \madgraph~are allowed. In order to closely mimic the behavior of thIn the RS model, massive grae parton shower, the $k_\bot$ value for each clustering vertex corresponding to a QCD emission is used as renormalization scale for $\alpha_S$ in that vertex. For the central hard $2\to 1$ or $2\to 2$ process, the transverse mass $m^2_\bot = p^2_\bot + m^2$ of the particle(s) produced in the central process is used as a factorization scale, as well as a renormalization scale.

Subsequently, this event is passed to the parton shower MC simulator. Before hadronizing or decaying, the final partons are clustered into jets using the $k_\bot$ algorithm with a jet cutoff
of $Q^{\rm jet}_{\rm min} > Q^{\rm ME}_{\rm min}$. The jets are then compared to partons. They are considered to be matched if $k_\bot {\rm (parton, jet)}$ is smaller than the cutoff $Q^{\rm jet}_{\rm min}$. An impossibility of matching all partons with jets, results on the rejection of the event. For events with parton multiplicity smaller than the highest multiplicity, the number of jets must be equal to the number of partons.

\subsection{Choice of matching parameters and cuts}

As mentioned above, the aim of this paper is to study inclusive graviton production, \ie, $pp\, (\bar p)\to G+\,n\,\,{\rm jet(s)}$ at the LHC and Tevatron, within the three models: ADD, MGM and RS. 

For the ADD and MGM models, missing energy with hard jets will be the signal at hadron colliders to identify real emissions of gravitons. The main background is $pp\, (\bar p) \to Z\,+\,(n \textendash)j$ with Z boson decays into neutrino pairs. In the following, we consider matrix elements with parton multiplicity $n$ from 1 to 3 for the matching with parton shower for the signal processes.

To suppress SM backgrounds in the LHC graviton searches~\cite{Vacavant:2001sd}, we require
\begin{equation}\label{lhcset}
P_T^{\rm miss}> 500~{\rm GeV}\,.
\end{equation}
Jets are defined by the $k_{T}$ algorithm, with the resolution parameter set to $D=0.6$, and are required to satisfy $|\eta_j|<4.5$ and $P_T^j>50$~GeV.

Regarding the matching parameter at the LHC, the separation cutoff is set as:
\begin{eqnarray}
Q^{\rm ME}_{\rm min} & > &45\,{\rm GeV}, \\
Q^{\rm jet}_{\rm min} & > & 50\,{\rm GeV}\,.
\end{eqnarray}

At the Tevatron analysis we use the same settings as in the recent CDF study~\cite{Collaboration:2008fu, Collaboration:2006qa}, \ie, \
\begin{equation}\label{tevset}
  P_T^{\rm miss}> 120~{\rm GeV}\,,\quad
  P_T^{j}>150~{\rm GeV}\,,\quad {\rm and} \quad |\eta_j|<1\,.
\end{equation}
Here, jets are defined by the $k_{T}$ algorithm with $D=0.7$, and are required to satisfy $|\eta_j|<3.6$ and $P_T^j>20$~GeV. A second jet with $P_T>60$~GeV is vetoed. The separation cutoffs used for the matched analyses are:
\begin{eqnarray}
Q^{\rm ME}_{\rm min} & > &20\,{\rm GeV}, \\
Q^{\rm jet}_{\rm min} & > & 30\,{\rm GeV}.
\end{eqnarray}

In the RS model, massive gravitons can decay quickly into visible particles within the detector. Thus we first generate a full inclusive sample for $G$ productions at the hadron colliders, for which we take into account the matrix elements with parton multiplicity $n$ from 0 to 2. We then let $G$ in the inclusive sample decay into $e^+e^-$ or $\mu^+\mu^-$, keeping spin correlations exact.

To allow a thorough comparison with the NLO QCD calculations on graviton mono-jet productions of Ref.~\cite{Karg:2009dq}, and in analogy with the analysis of the ADD and MGM models, we generate a semi-inclusive sample for $G$-jet productions in the RS model with the same cuts and settings as for the ADD and MGM models. For simplicity, we do not ask for $G$ decay here, as the aim is to validate the matching method against NLO QCD calculations.

% --------------------------------------------------------------------

\section{Results}
\label{numres}

Let us fix now the model parameters. Considering the up-to-date experimental constrains as mentioned in Section~\ref{exdm}, for the ADD model, we choose $\Lambda = 5$ TeV for the LHC, $\Lambda = 1$ TeV for the Tevatron, with $\delta = 2,\:4,\:6$. For the RS model, we take $m_{1} = 1$ TeV and $100$ GeV, with $\Lambda = 3$ TeV. For the MGM model, we set $\overline{M} (\mu_\star) \sim \mu_\star = 1$ TeV and $2$ TeV.

In the event generation with \mgme~,  CTEQ6L1 PDF~\cite{Pumplin:2002cq} is employed which also fixes the values of $\alpha_S$ at the Z mass, while the choices on renormalization and factorization scales are defined to be the transverse mass of the graviton as explained in Sec.~\ref{matching}. Correspondingly, for the LO/NLO calculations, CTEQ6L1/6M PDFs are employed together with the corresponding values for the strong coupling $\alpha_S$. The renormalization and factorization scales are set to the transverse momentum of the graviton $P_T^G$, and we will also show the uncertainties by varying the scales in the range between $P_T^G/2$ and $2P_T^G$.

One relevant remark concerns the range of validity of the effective theories considered here. As they all assume
a  linearized Einstein gravity, they are valid only when the scales involved in the hard scattering process do not exceed the fundamental scale $M_S$, in which case a quantum gravity description is needed. 

In the following we present the matrix element-parton shower matched results and the LO/NLO ones, for the differential distributions:

\begin{itemize}
	\item \emph{Pseudo-rapidity of the leading jet ($\eta^{j}$)};
	\item \emph{Pseudo-rapidity ($\eta^{G}$)} of the graviton;
	\item \emph{Transverse missing $P_{T}$} for ADD and MGM models, or \emph{graviton $P_{T}$} for RS model; 
	\item \emph{Leading and Second jet $P_{T}$};
	\item \emph {$H_{T}$ of the jets}, which is defined as the sum of all the jet $P_{T}$ in each event:
\begin{equation}
H_{T}=\sum_{j}|P_{T}^{j}|.
\end{equation}
\end{itemize}
\newpage

\subsection{Inclusive sample for the RS graviton productions}

\begin{table}
\begin{center}
    \begin{tabular}{ | c  l | c  c | }
    \hline
    \multicolumn{2} {|c|} { Model } & $\kappa$-factor & Normalization factor\\
     \hline \hline
     & $m_{1}=1$ TeV & 1.65  & 1.73 \\ 
     RS&  $m_{1}=500$ GeV & 1.75  & 1.82 \\ 
    \hline
    \end{tabular}
    \caption{$\kappa$-factor $\sigma_{(NLO)}/\sigma_{(LO)}$ and normalization factor $\sigma_{(NLO)}/\sigma_{(MLM)}$ for inclusive $G$ production in the RS model at the LHC, from $p p \rightarrow G$ NLO QCD calculations and parton shower matching with the matrix elements of $p p \rightarrow G + n\,\,{\rm jet(s)}$ with $n=0, 1, 2$, $\Lambda=3$ TeV,  $m_{1}=0.5$ and $1$ TeV for $\sqrt{s} = 7$ and $14$ TeV LHC, respectively.}
    \label{KfactorFullInclusive}  
\end{center}
\end{table}

\begin{table}
\begin{center}
    \begin{tabular}{ | c  l | c  c |}
    \hline
     \multicolumn{2} {|c|} {} & \multicolumn {2}{|c|}{Normalization factor}\\
    \multicolumn{2} {|c|} {Model} & LHC &  Tevatron \\
	\hline
     \hline
     & $\delta=2$ & 2.05 & 2.40  \\ 
     ADD&  $\delta=4$ & 2.34 & 2.47   \\ 
     &  $\delta=6$ & 2.49 & 1.92   \\ \hline
     & $\overline{M}(\mu_\star)=1$  TeV& 1.56 & -   \\ 
     MGM& $\overline{M}(\mu_\star)=2$  TeV & 1.64  & -   \\ \hline
     & $m_{1}=1$ TeV & 1.99  & 1.95  \\ 
     RS&  $m_{1}=100$ GeV & 1.81  & 1.73   \\ 
    \hline
    \end{tabular}
    \caption{ Normalization factor $\sigma_{(NLO)}/\sigma_{(MLM)}$ for semi-inclusive G production in different models at the LHC and Tevatron. }
    \label{KfactorTable}  
\end{center}
\end{table}

\begin{figure}[H]
	\includegraphics[width=8cm]{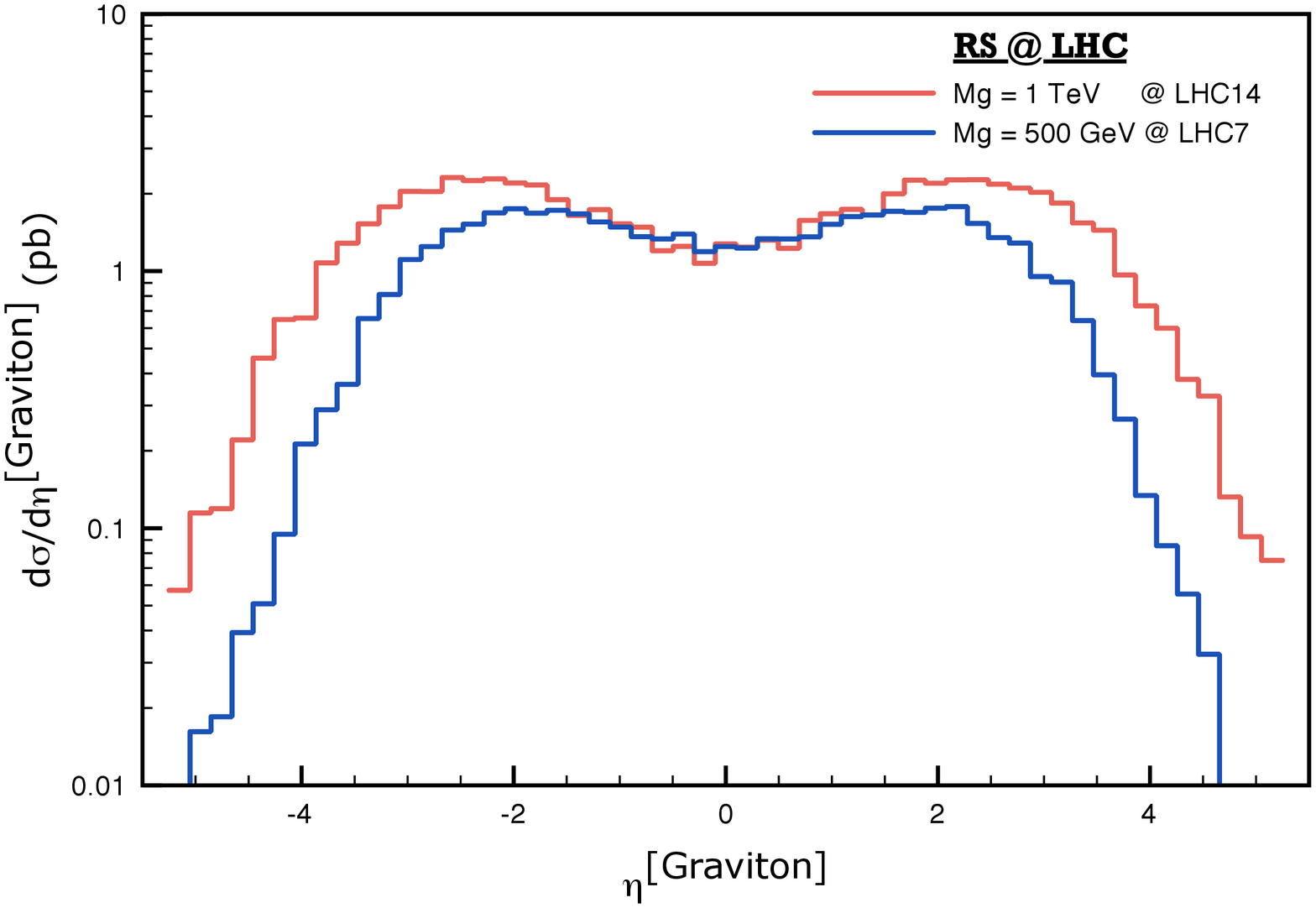}
	\includegraphics[width=8cm]{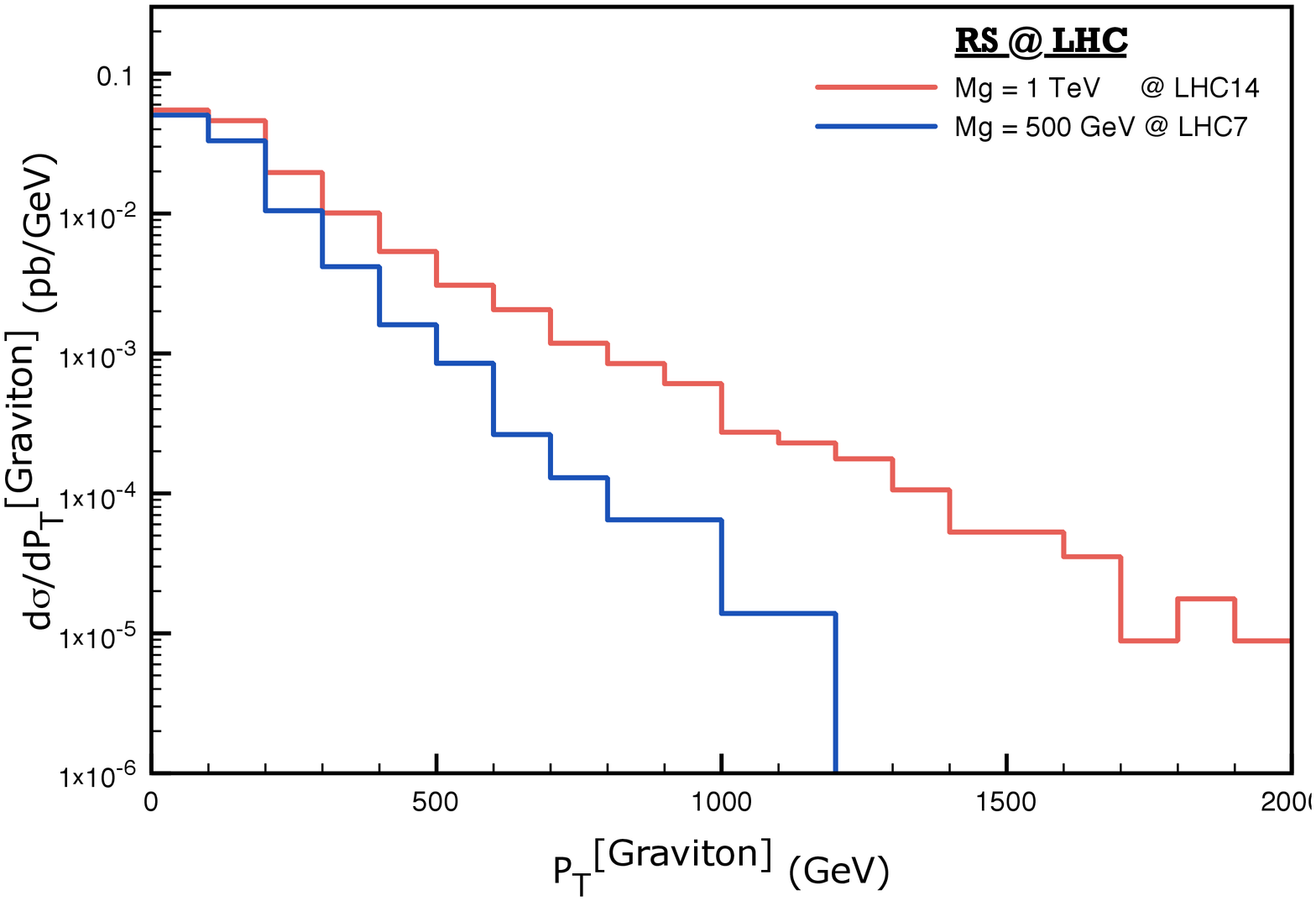}
	\includegraphics[width=8cm]{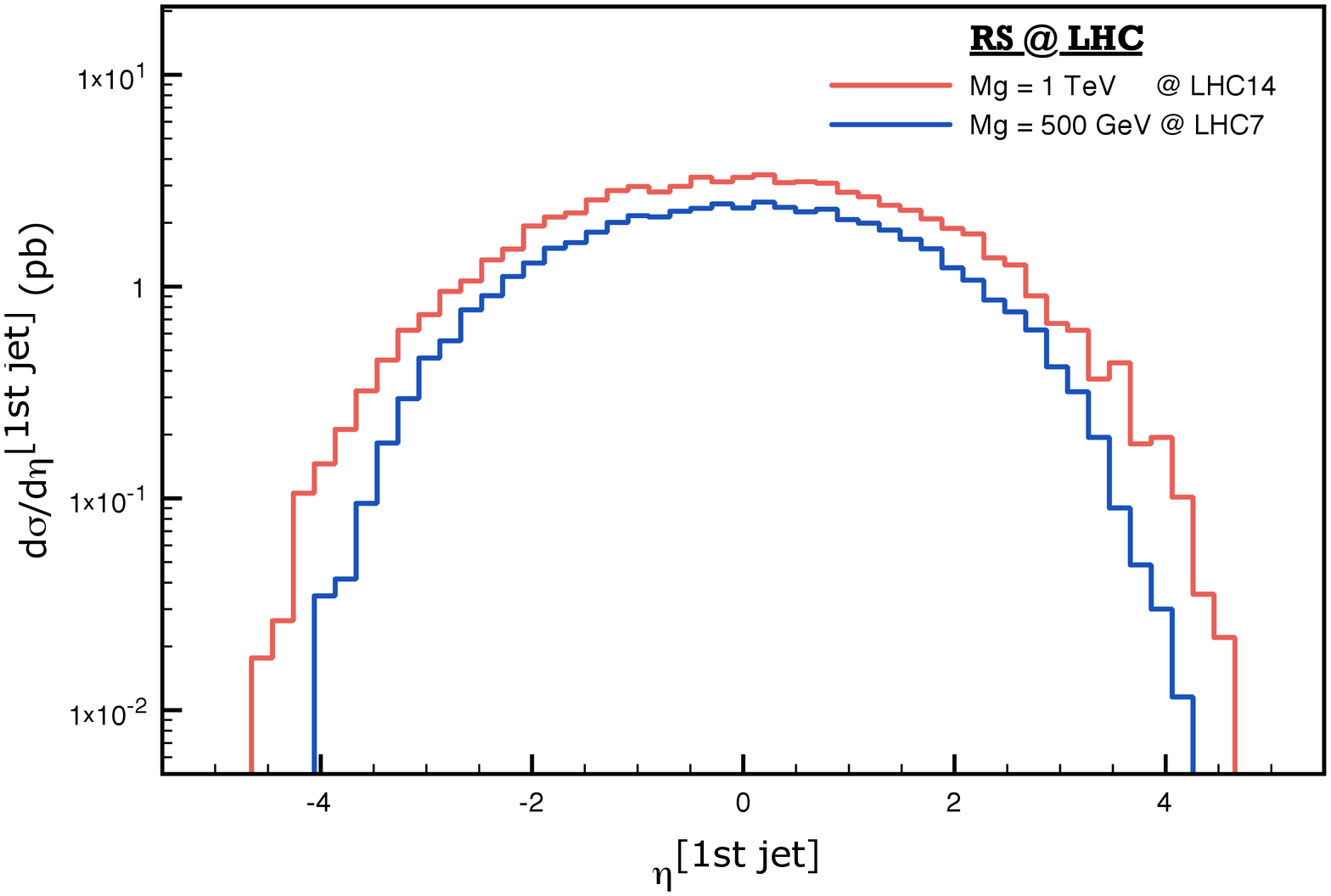}
	\includegraphics[width=8cm]{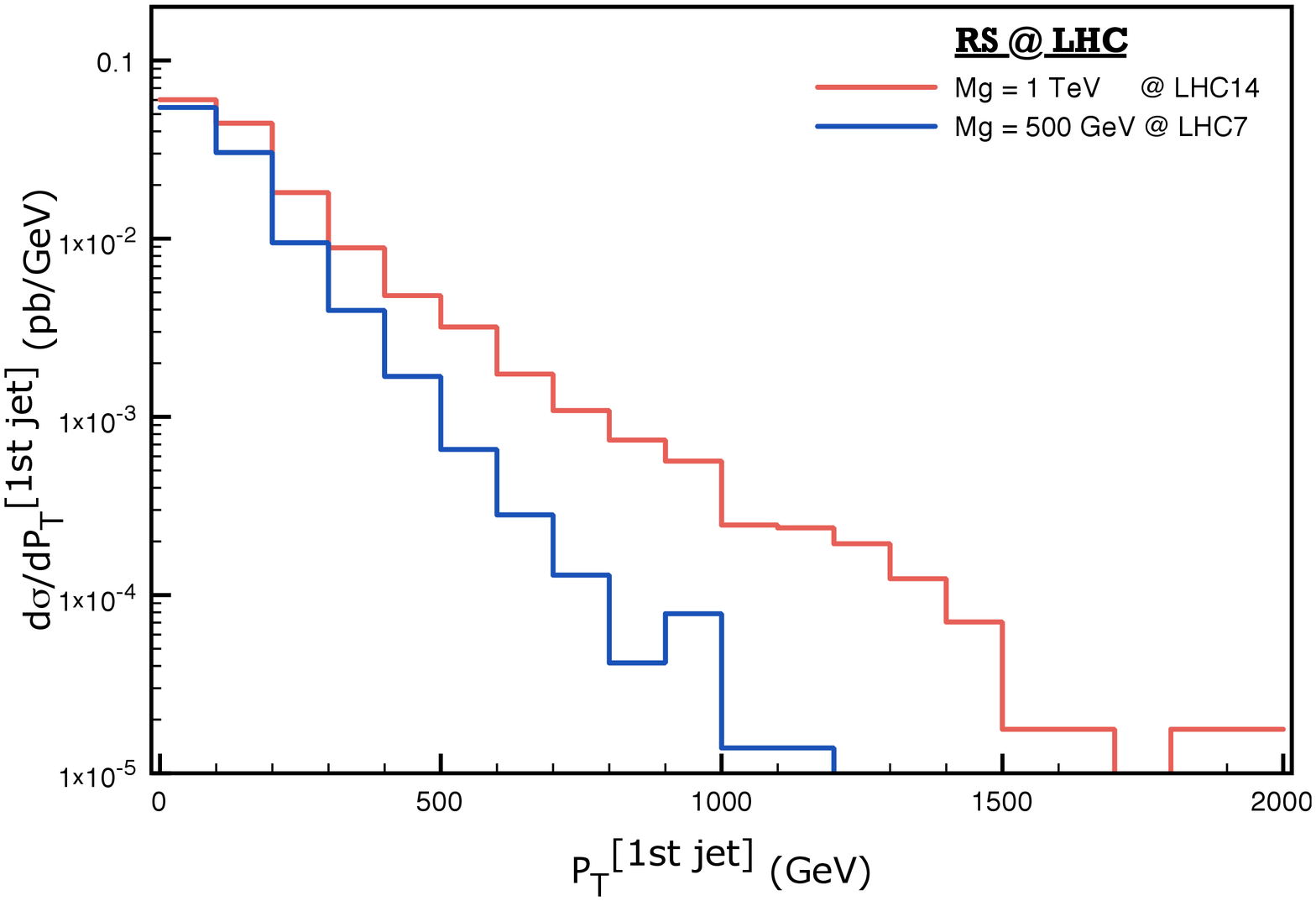}
	\includegraphics[width=8cm]{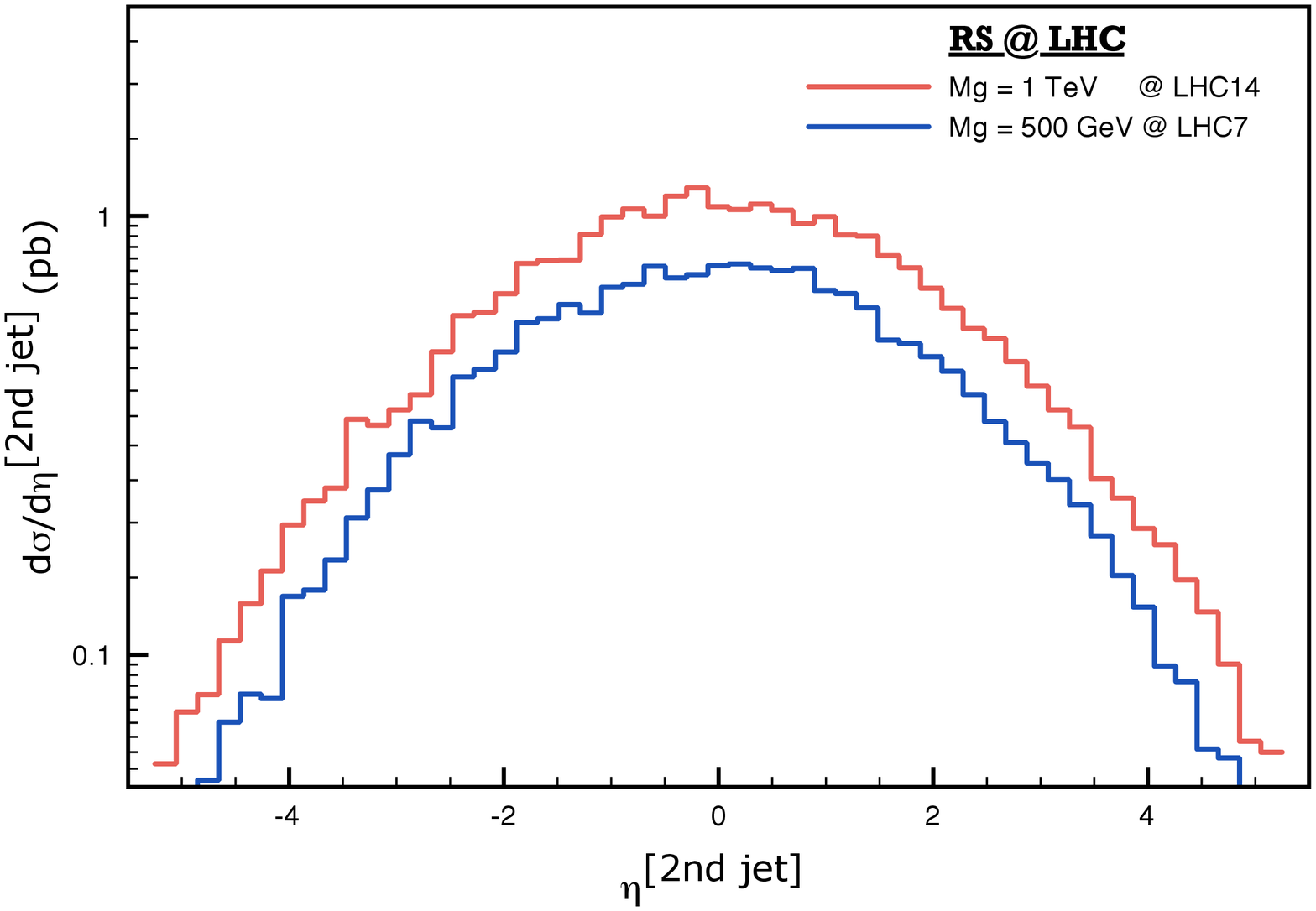}
	\includegraphics[width=8cm]{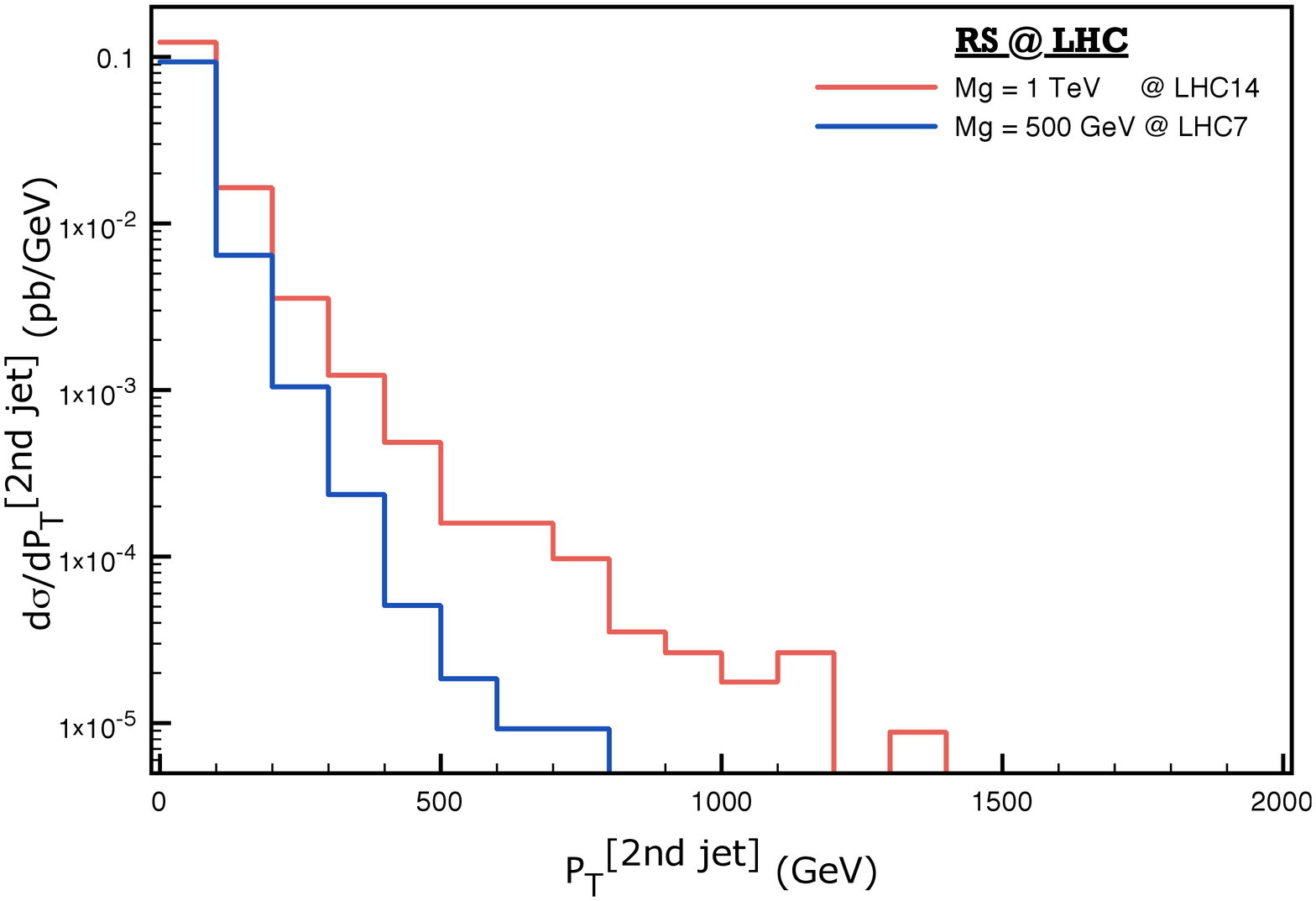}
	\caption{$k_\bot$-MLM matching results for a full inclusive sample of the RS model with $\Lambda=3$ TeV. The matched results are given by the red and blue curves for $m_{1}=1$ TeV at the LHC with $\sqrt{s}=14$ TeV, and $m_{1}= 500$ GeV at the LHC with $\sqrt{s}=7$ TeV, respectively. No cuts were applied in these samples.}
	\label{RSFull_LHC}
\end{figure}

\begin{figure}[H]
	\includegraphics[width=8cm]{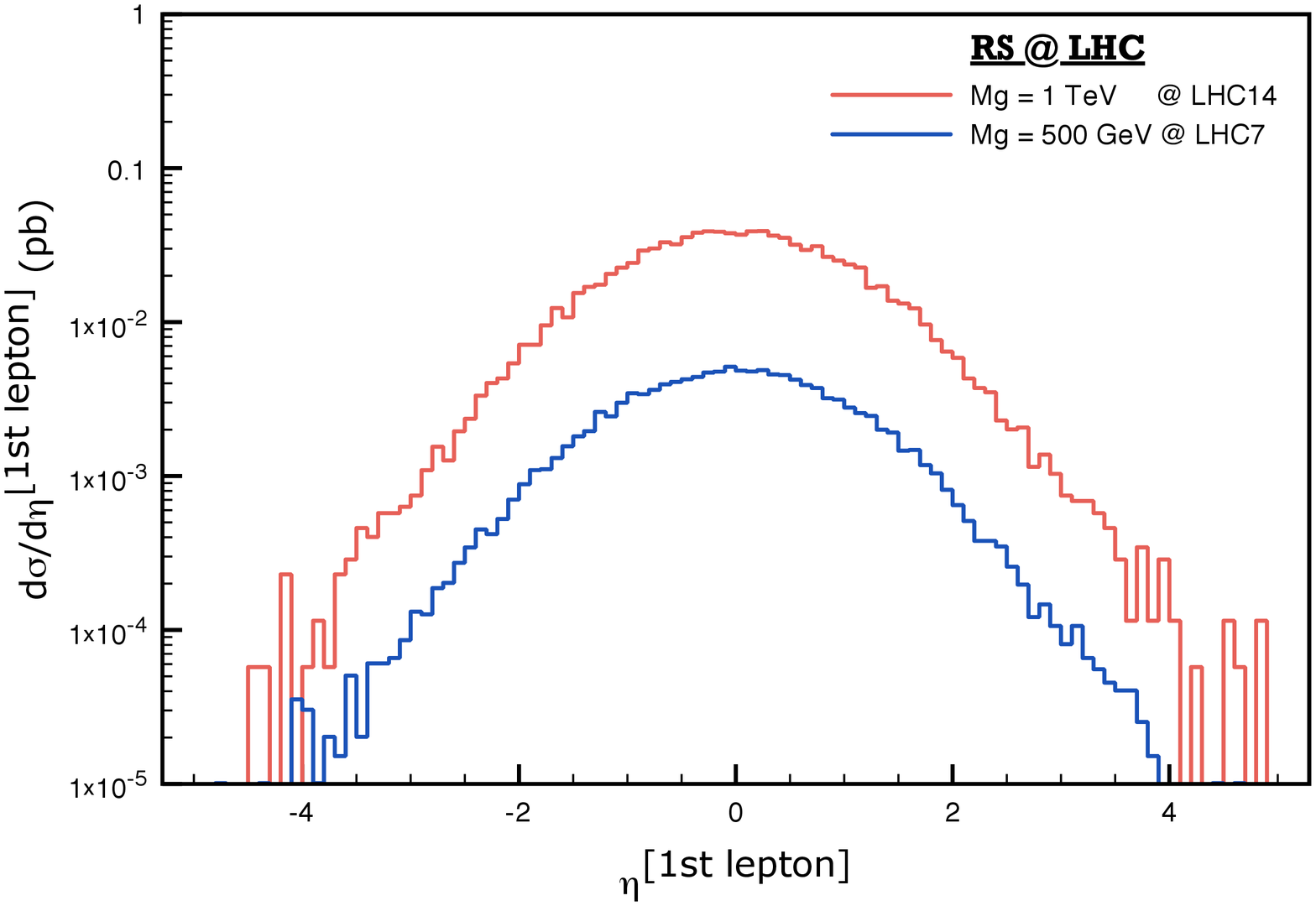}
	\includegraphics[width=8cm]{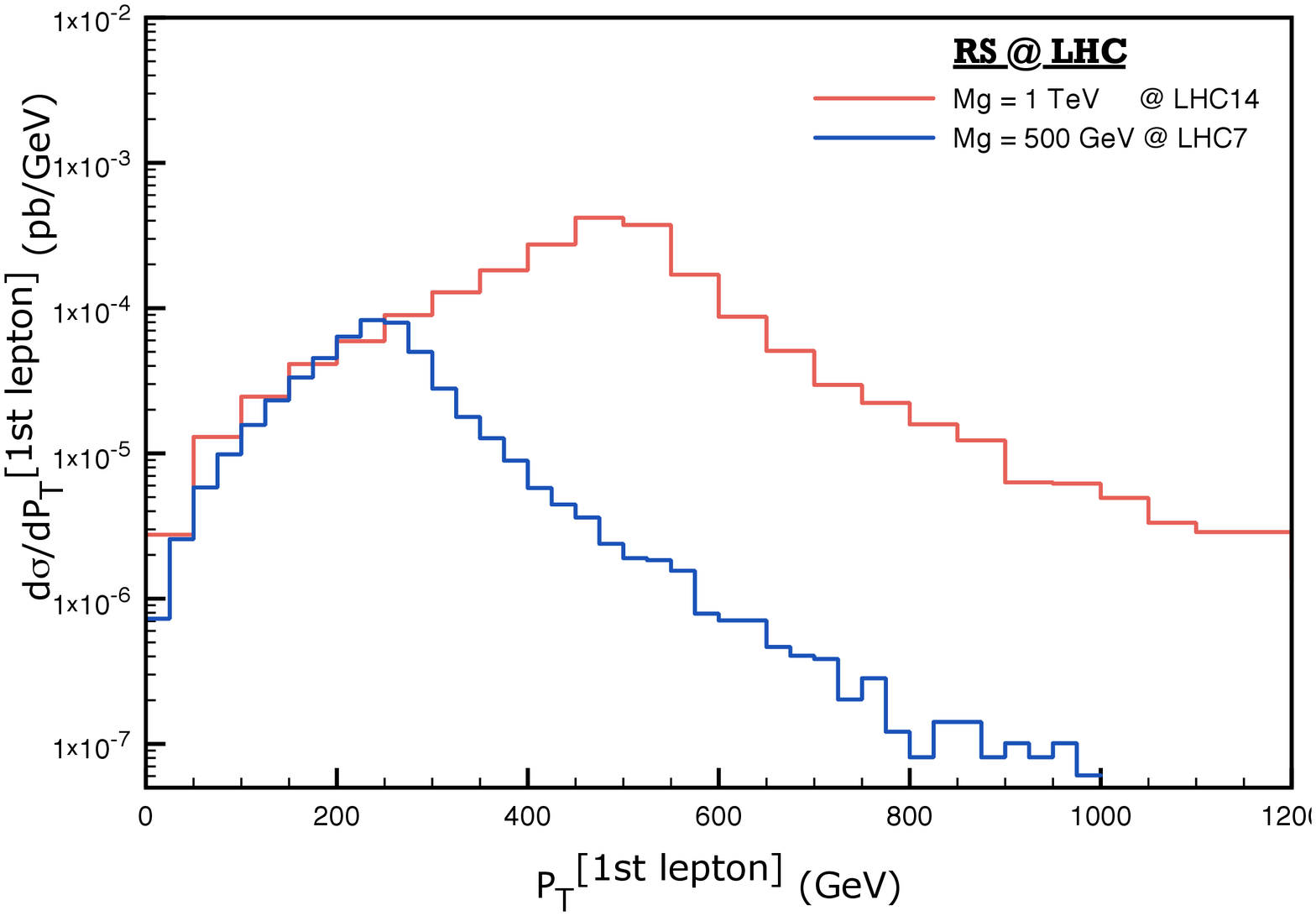}
	\includegraphics[width=8cm]{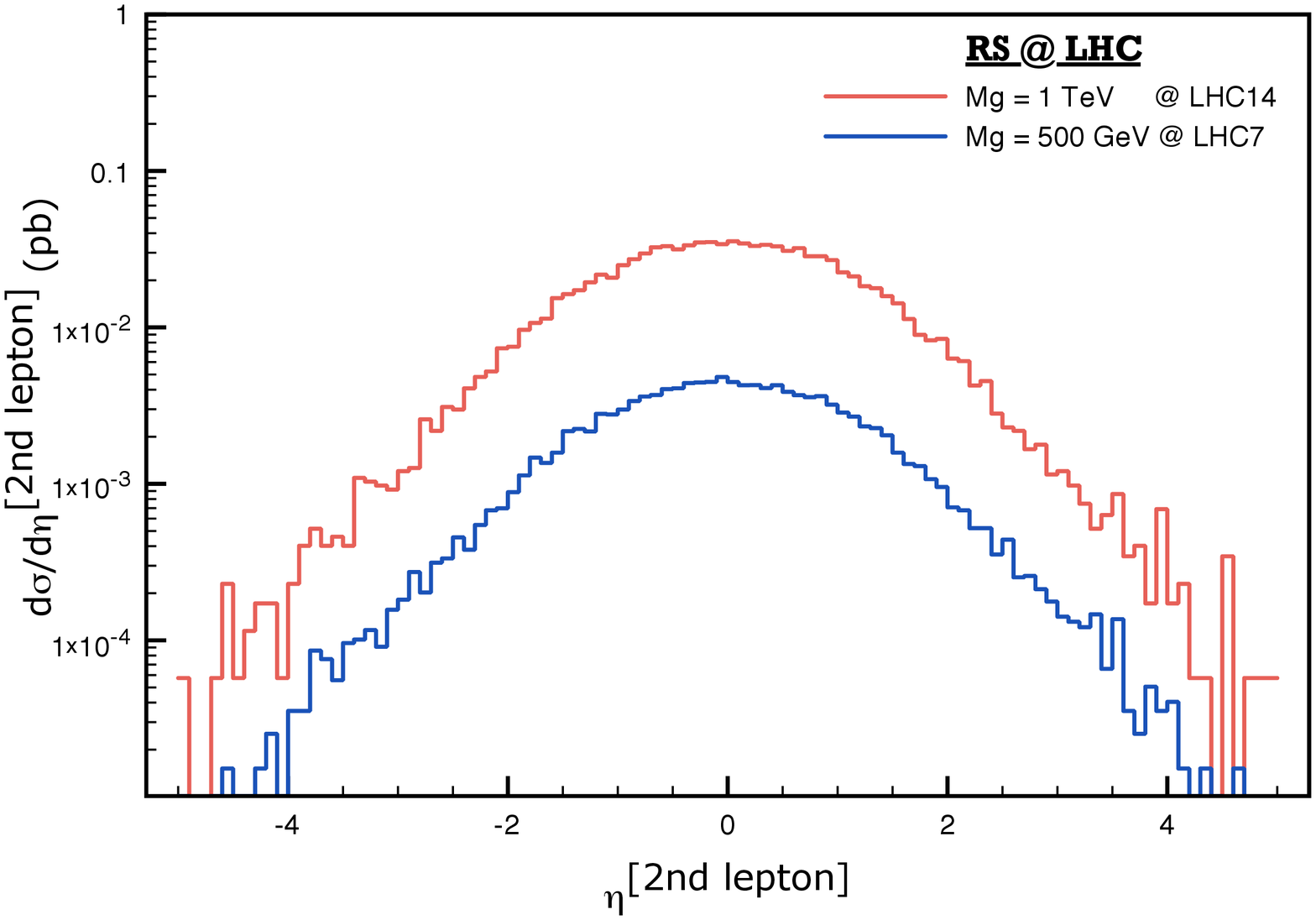}
	\includegraphics[width=8cm]{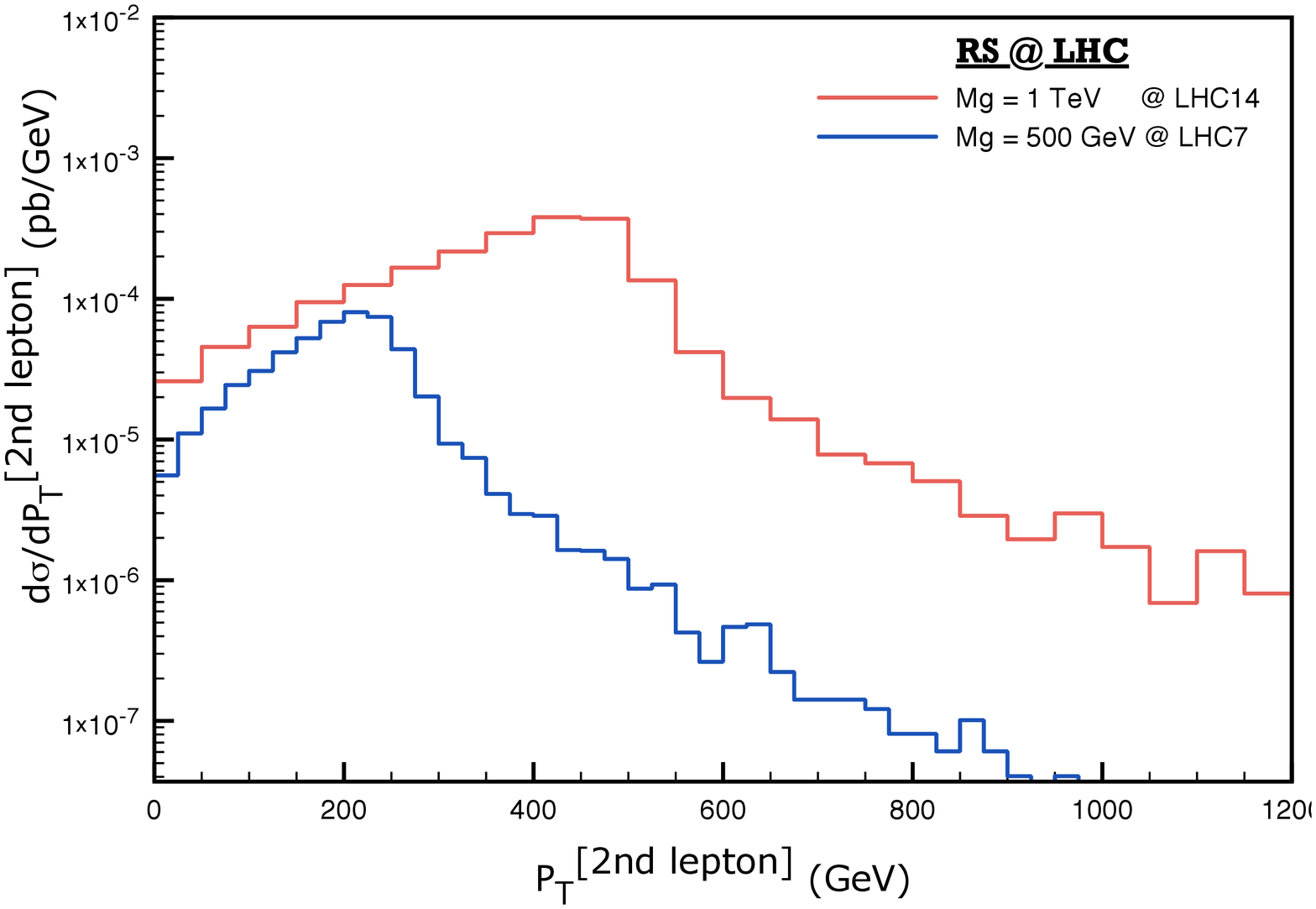}
	\caption{$k_\bot$-MLM matching results for a full inclusive sample of the RS model with $\Lambda=3$ TeV and the graviton decayed into a pair of leptons. The red and blue distributions are related to the mass of the first mode of the graviton and the collision energy. The red curves are for $m_{1}=1$ TeV at the LHC with $\sqrt{s}=14$ TeV, and the blue ones for $m_{1}= 500$ GeV at the LHC with $\sqrt{s}=7$ TeV. No cuts were applied in these samples.}
	\label{RSFullDecayed_LHC}
\end{figure}

Let us start the discussion with the results for the RS model where we first consider a full inclusive graviton production with graviton decay. The NLO QCD corrections for dilepton production in RS model via graviton production have been calculated in \cite{Mathews:2005cr, Li:2006nx} and compared to the LO expectation. Table~\ref{KfactorFullInclusive}  collects the results in terms of rather important $\kappa$-factors (as given by $\sigma_{(NLO)}/\sigma_{(LO)}$) which have
therefore to be included to achieve a satisfactory normalization of our fully inclusive samples, figures~\ref{RSFull_LHC} and \ref{RSFullDecayed_LHC}. 

In figure~\ref{RSFull_LHC}, $k_\bot$-MLM matched results for  $p p \rightarrow G +  n\,\,{\rm jet(s)}$ with $n=1, 2$ are presented. The graviton is decayed subsequently into a pair of leptons (\ie, either $e^+ e^-$ or $\mu^+ \mu^-$) and the corresponding results are shown in figure~\ref{RSFullDecayed_LHC}.  Here normalization factor between the $k_\bot$-MLM matched and the NLO expected cross section are also computed. These results are shown on the third column of table~\ref{KfactorFullInclusive}. Observe that the values for the NLO/LO normalization factor are very close to the ones found for the NLO/MLM normalization factor. In fact, the cross section after matching analysis is expected to be similar to the one computed for  $p p \rightarrow G$ by LO calculation. That is because LO calculation already includes the production of extra radiation through the PDFs (parton distribution functions). However, as the matching calculation considers explicitly the production of extra jets in the final state, it modifies the behavior of the distributions being analyzed such as the missing $P_T$, pseudo-rapidity and $H_T$ distribution. As a result, we expect that matched sample, normalized to NLO calculation, will yield better predictions for observables which involve extra radiation then a pure parton shower approach. 

\subsection{Semi-inclusive sample for graviton-monojet studies}
\begin{figure}   [H]
	\includegraphics[width=8cm]{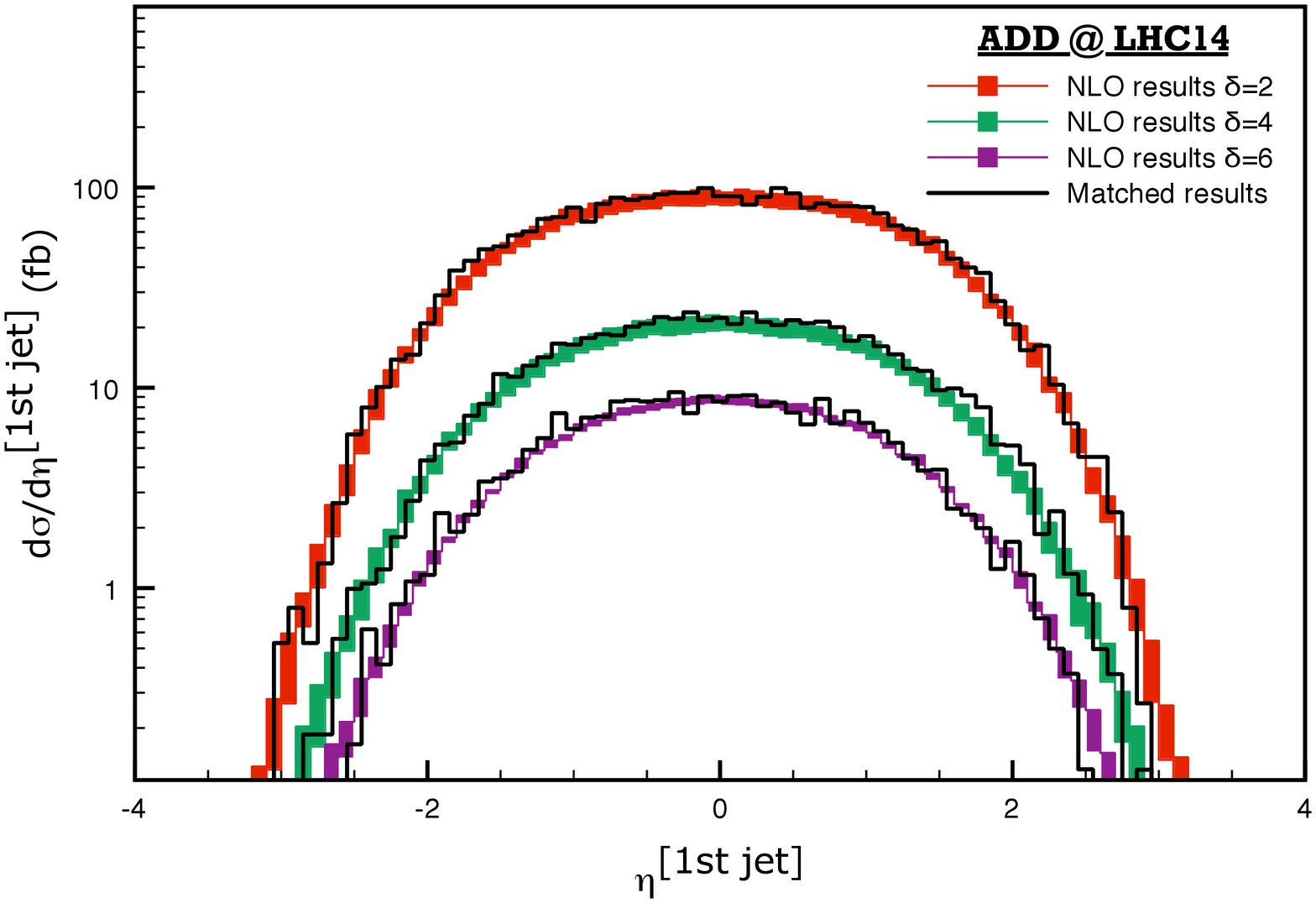}
	\includegraphics[width=8cm]{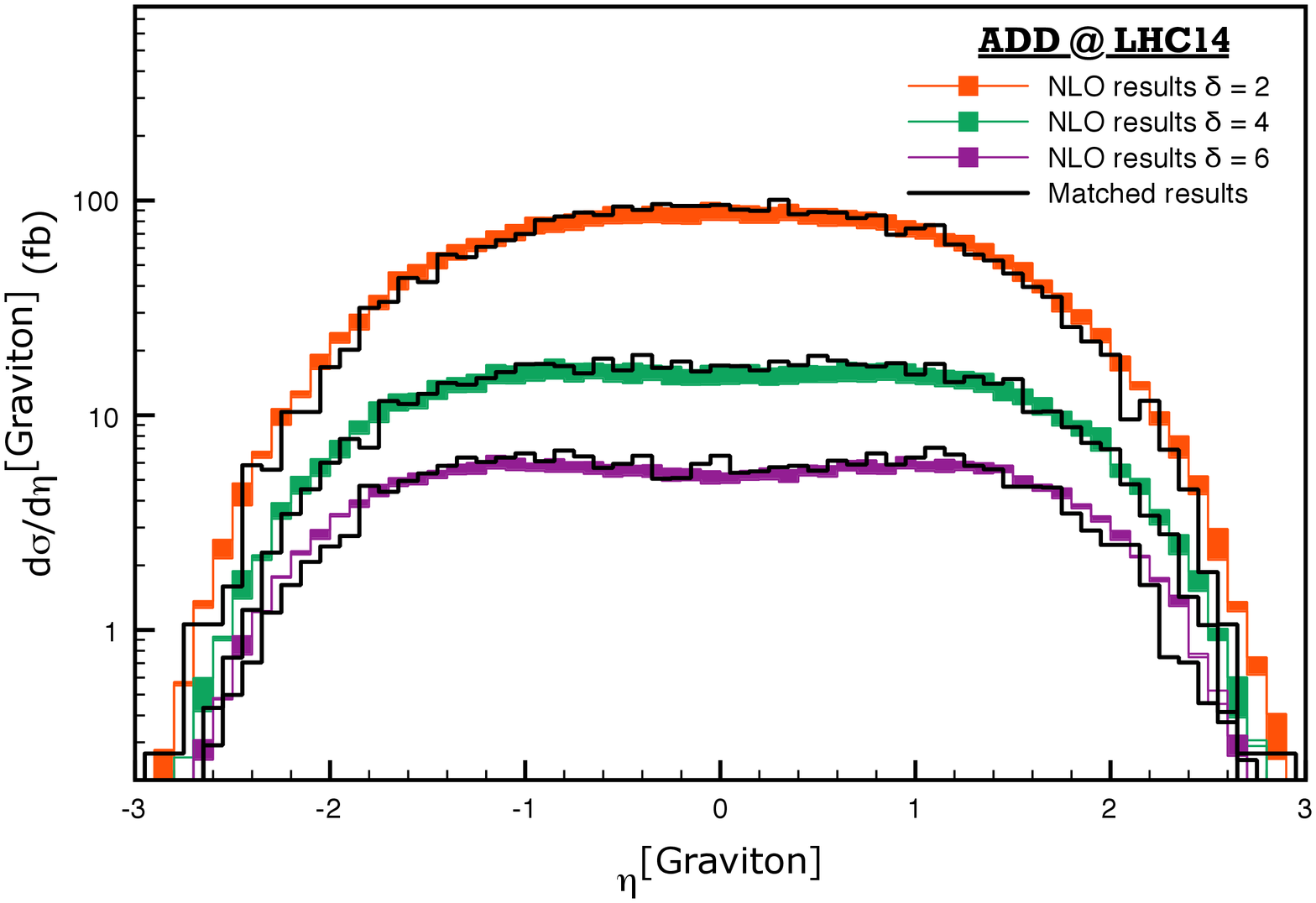}
	\includegraphics[width=8cm]{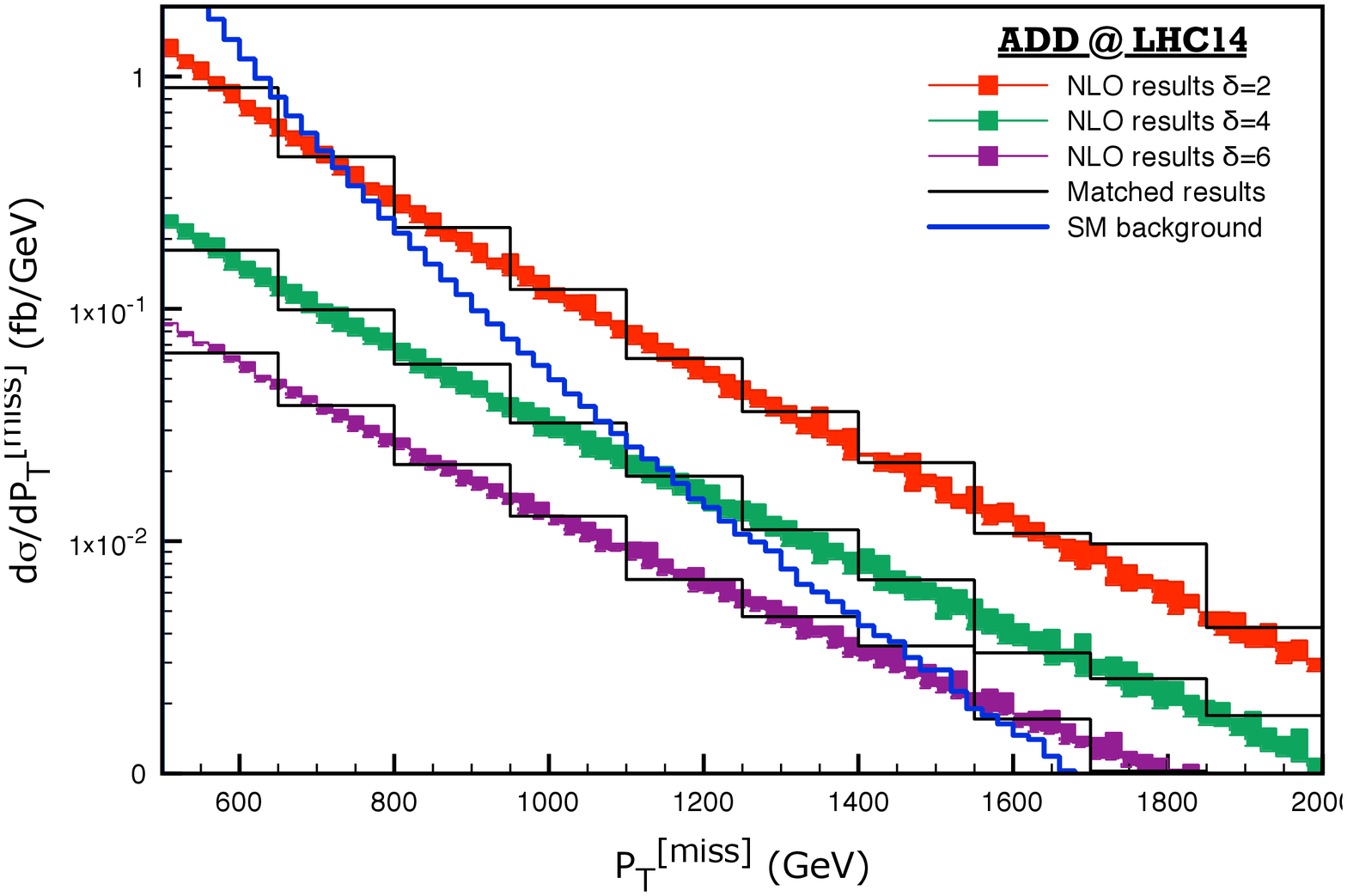}
	\includegraphics[width=8cm]{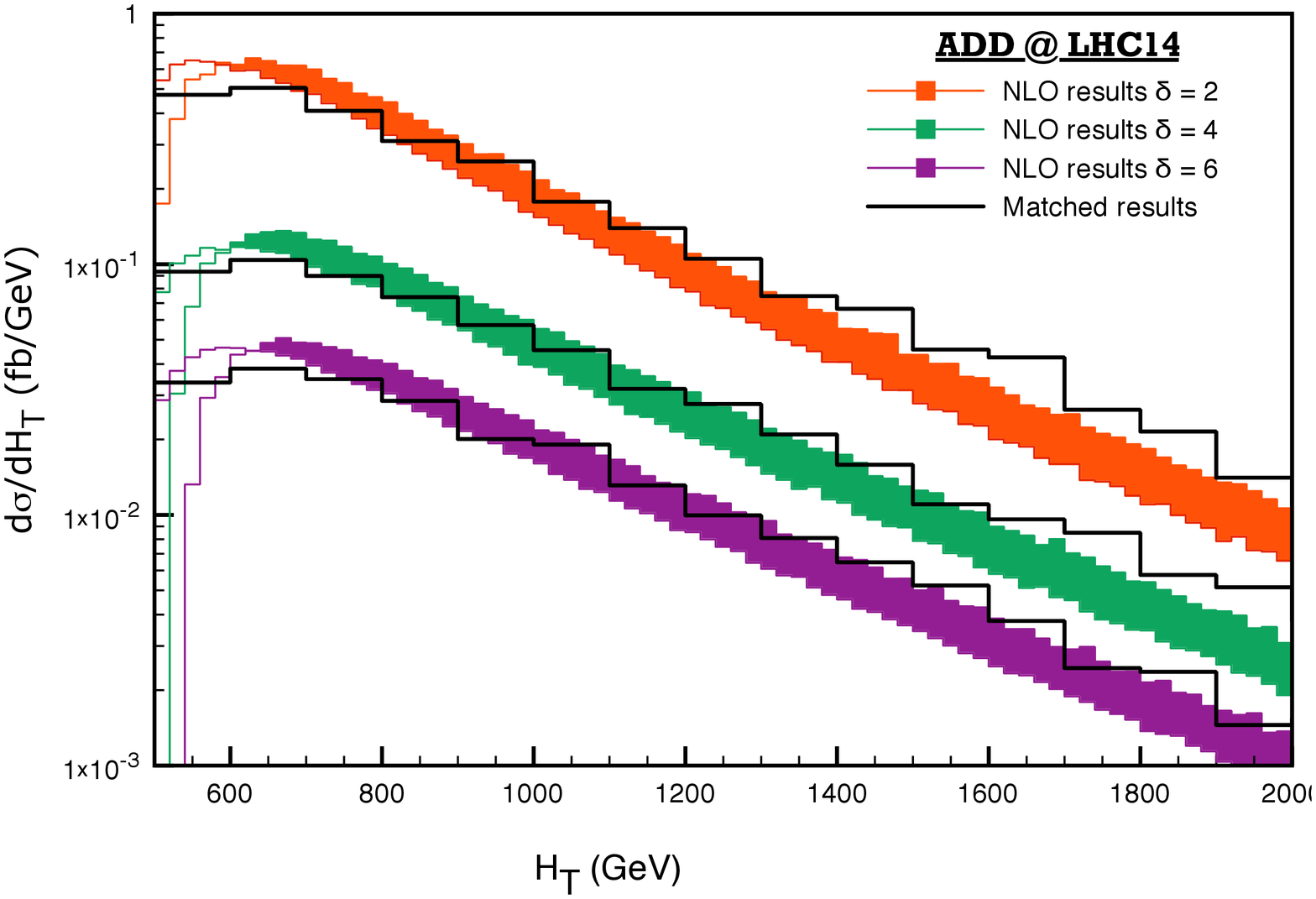}
	\includegraphics[width=8cm]{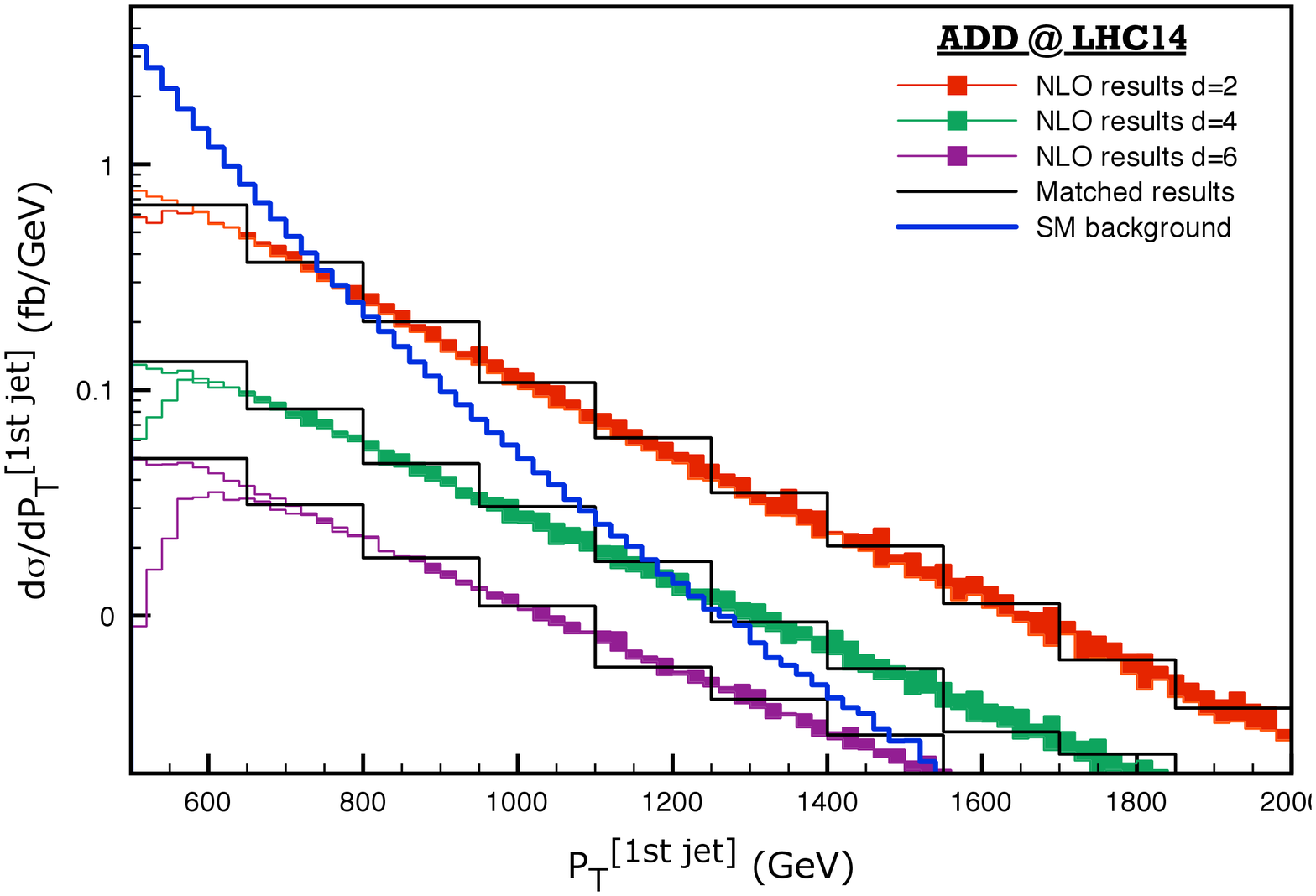}
	\includegraphics[width=8cm]{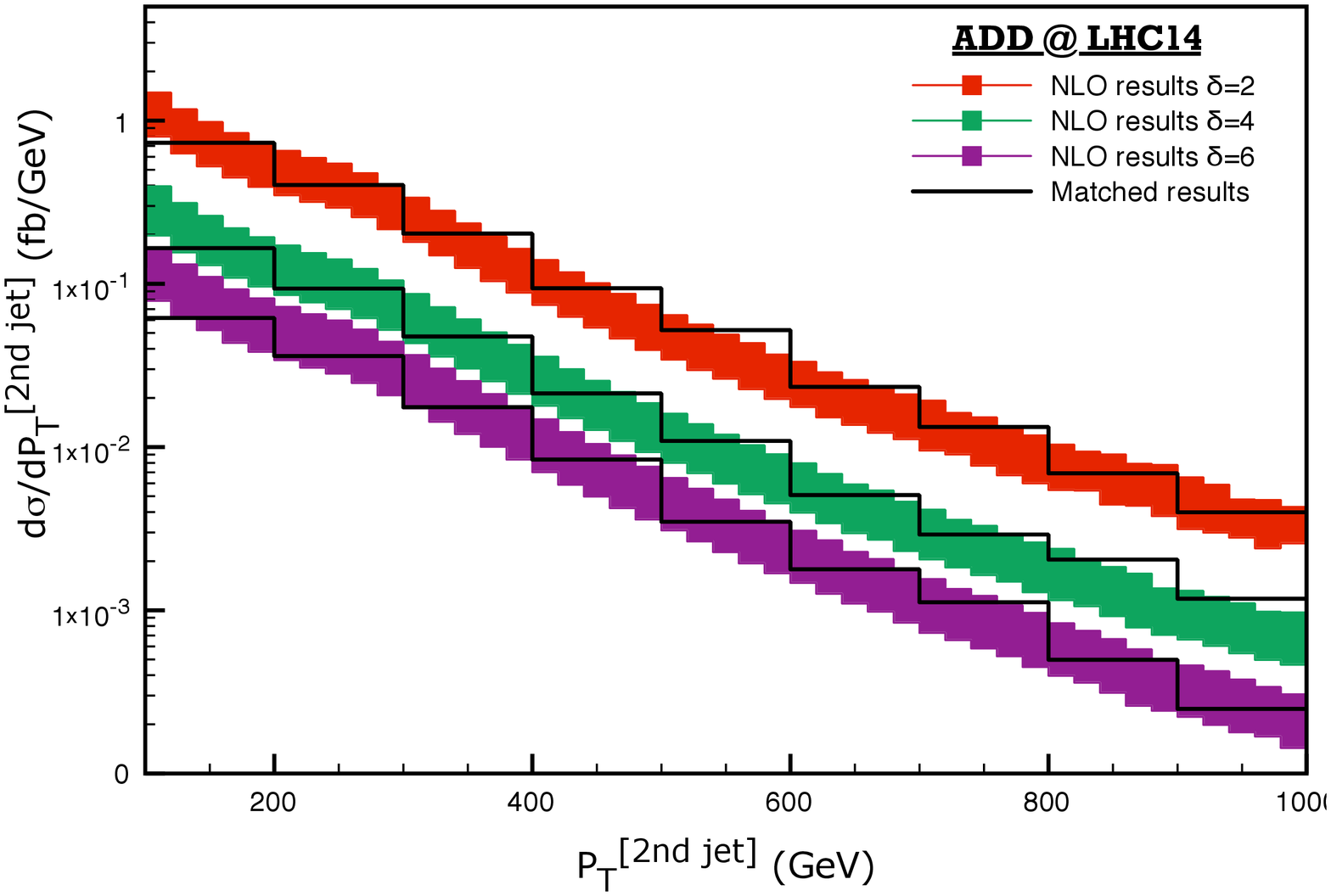}
	\caption{NLO/$k_\bot$-MLM matching comparison for the ADD model at the LHC ($\sqrt{s}=14$ TeV) with $\Lambda=5$ TeV. The NLO results are given by the red, green and purple bands for $\delta=2, 4, 6$, respectively. The matched results are given by the black curves with the same parameters as for the NLO ones. The dominant ${\rm Z} \to \nu\bar{\nu}$ background is also shown as a reference by the blue curve. The $k_\bot$-MLM matched curves are normalized by the NLO results, and the normalization factors can be found in table 4.}
	\label{ADD_LHC}
\end{figure}

\begin{figure}[H]
	\includegraphics[width=8cm]{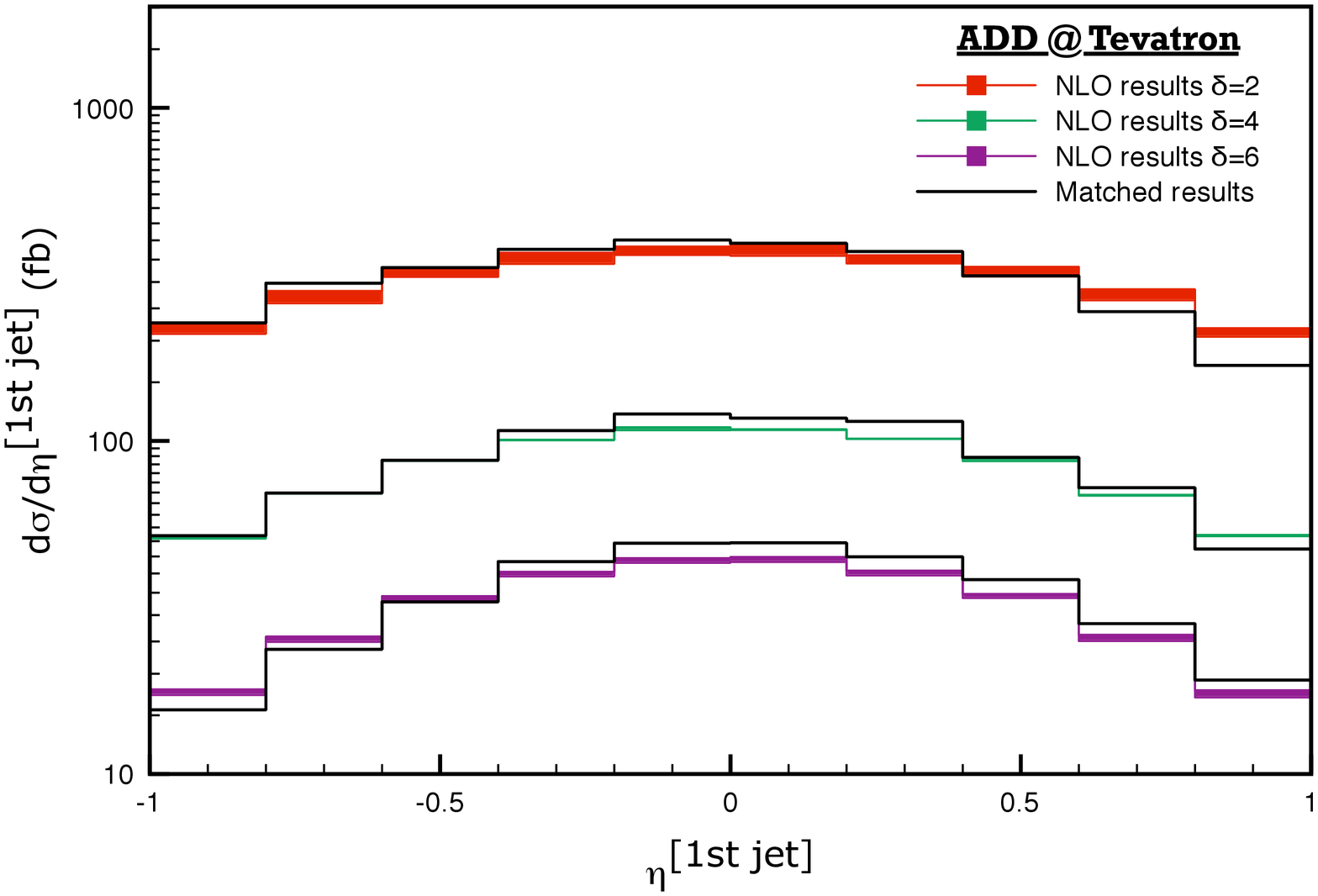}
	\includegraphics[width=8cm]{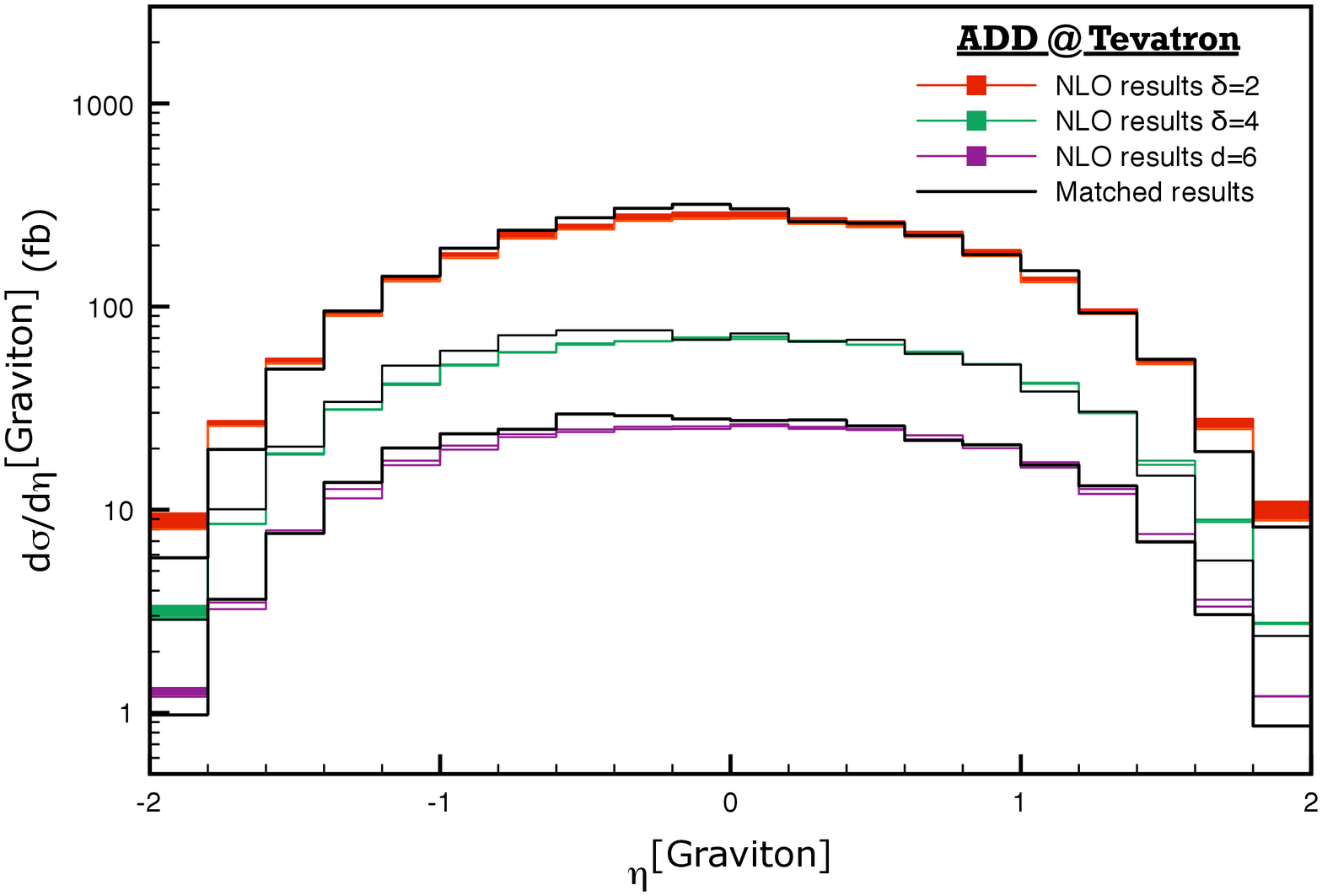}
	\includegraphics[width=8cm]{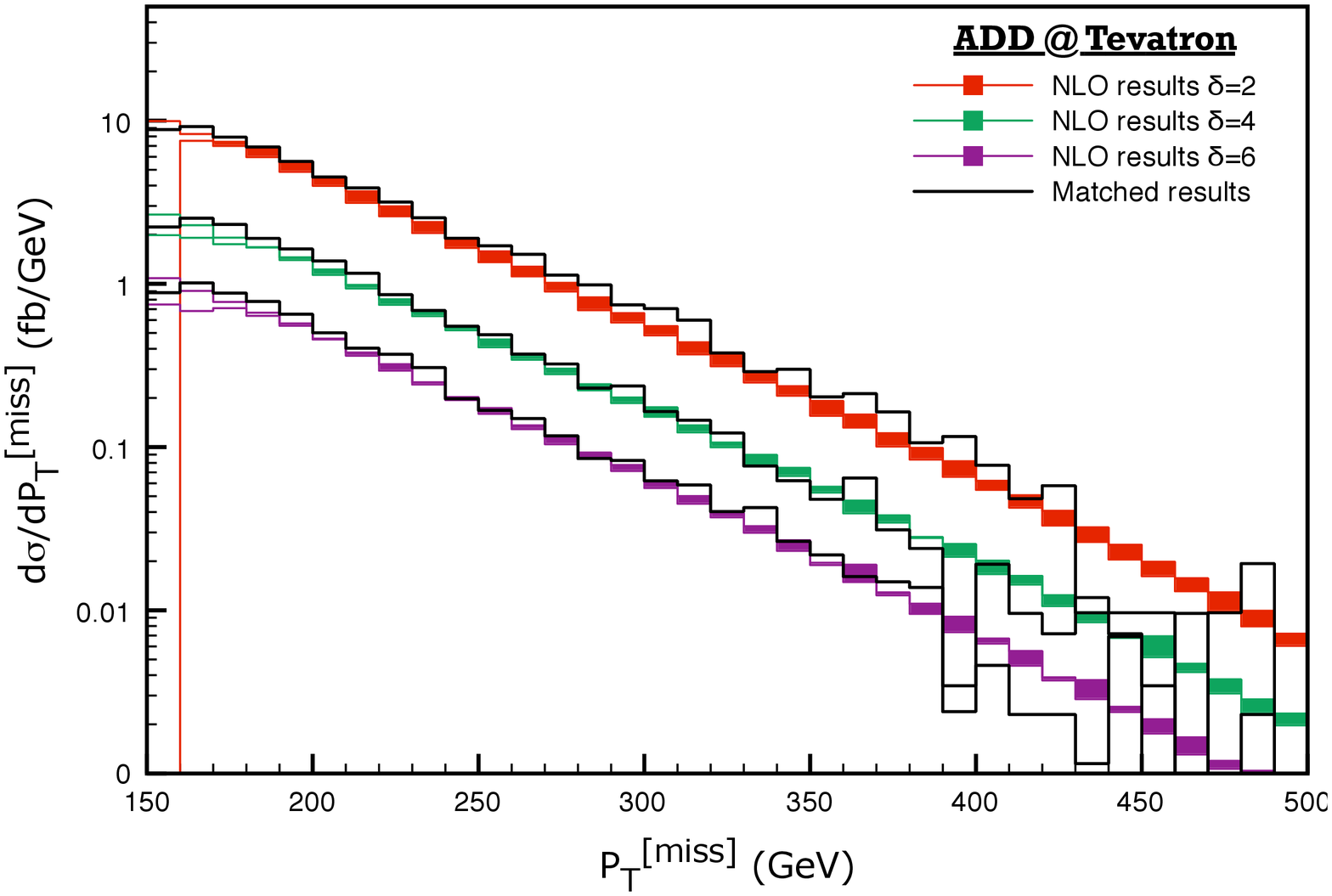}
	\includegraphics[width=8cm]{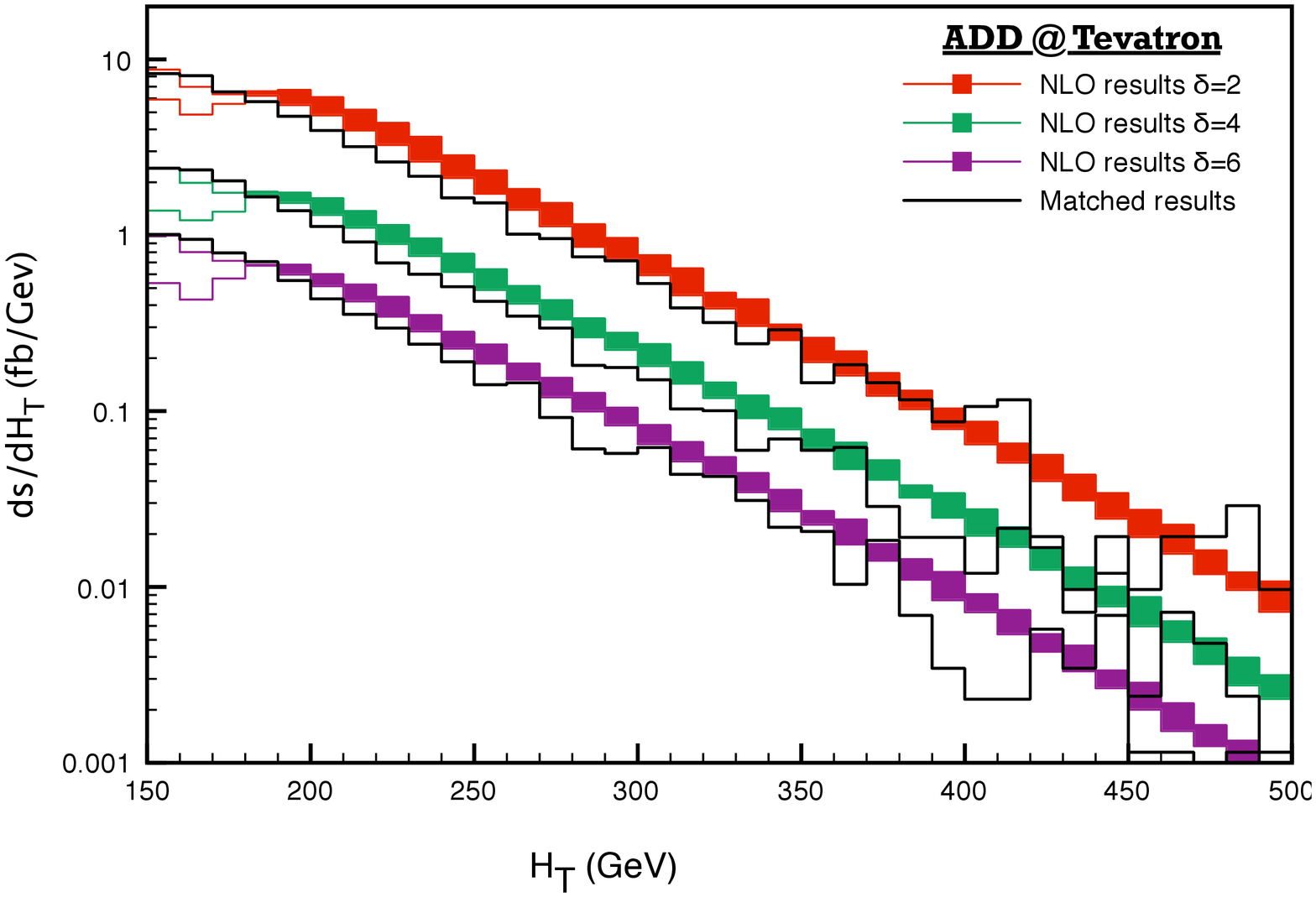}

	\begin{centering}
		\includegraphics[width=8cm]{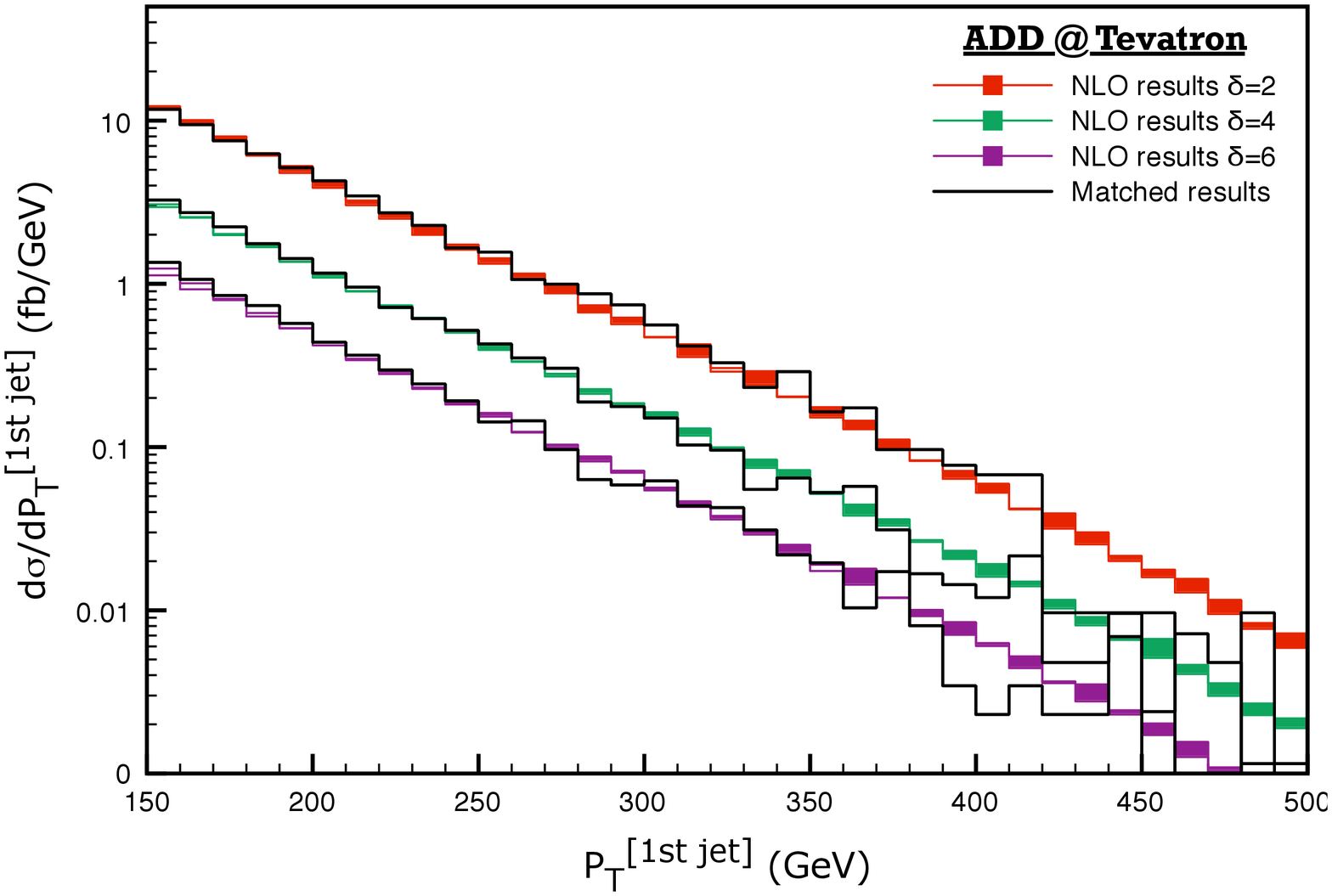}
	\par\end{centering}
	
	\caption{NLO/$k_\bot$-MLM matching comparison for the ADD model at the Tevatron ($\sqrt{s}=1.96$ TeV) with $\Lambda=1$ TeV for $\delta=2, 4, 6$. The matched curve is normalized by the NLO results and the normalization factors can be found in table 4.}
	\label{ADD_Tevatron}
\end{figure}

\begin{figure}[t]
	\includegraphics[width=8cm]{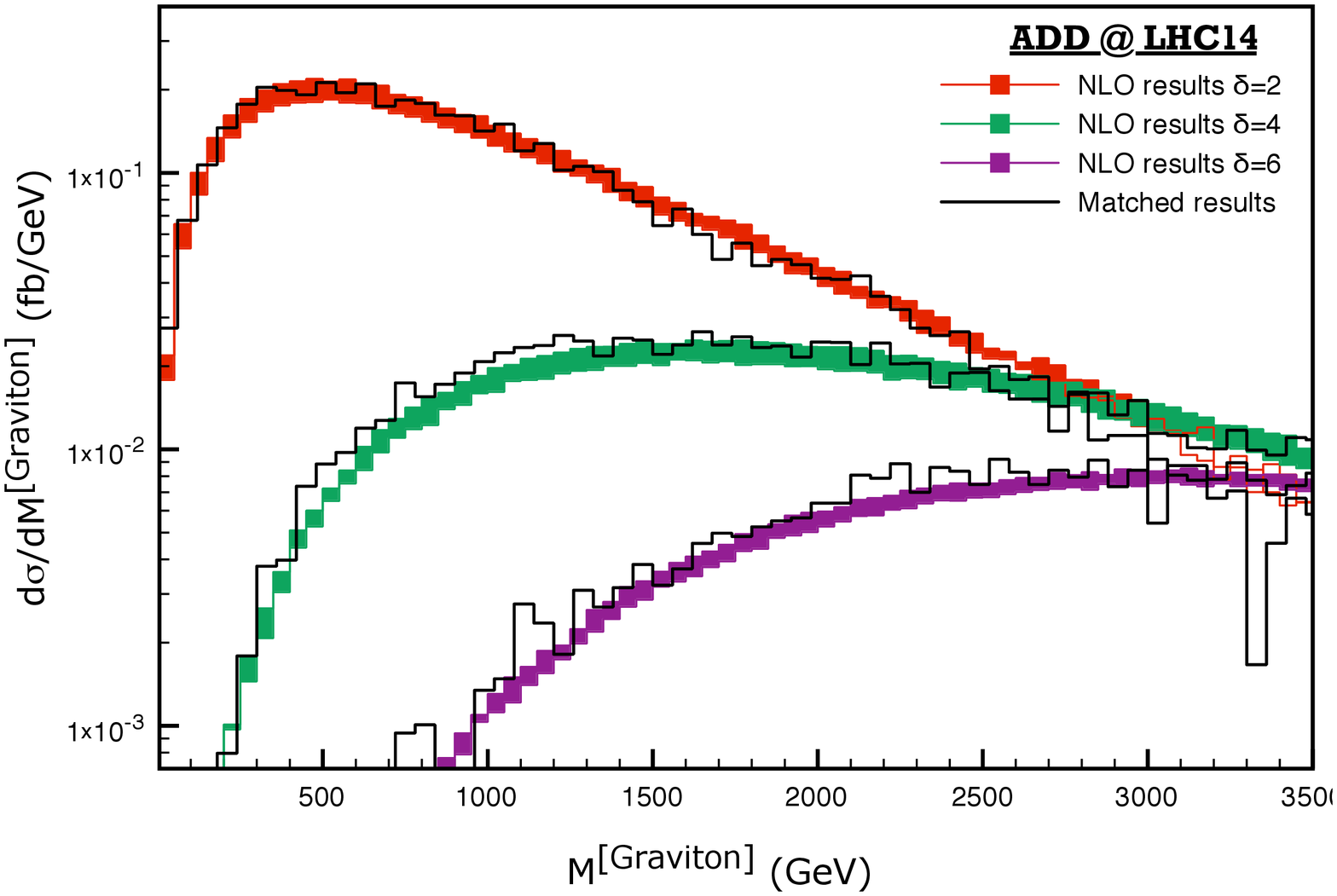}
	\includegraphics[width=8cm]{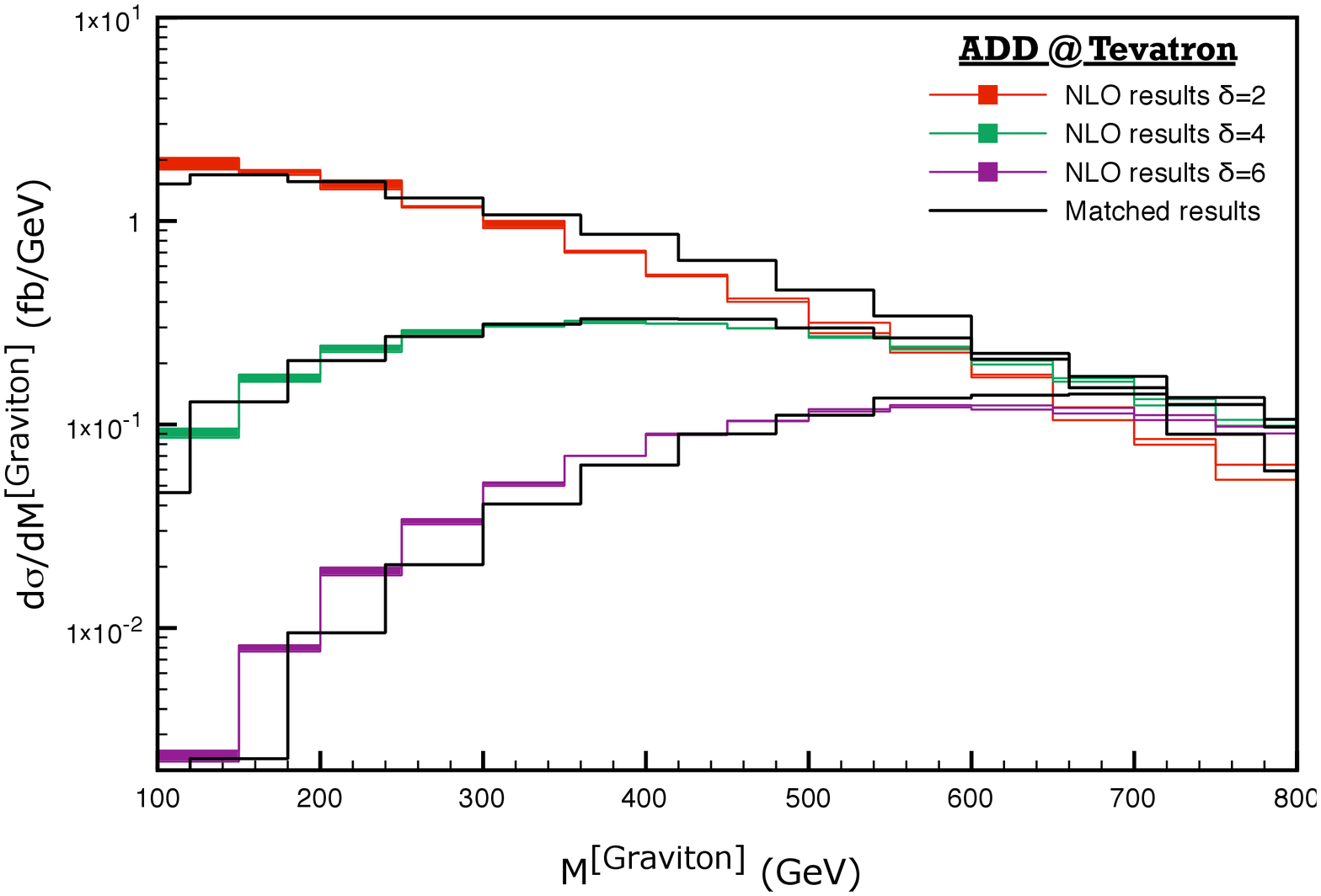}
	\caption{Graviton mass distribution at the LHC (left) and Tevatron (right) for the ADD model with $\delta=2, 4, 6$, respectively. The NLO results are shown by the red, green and purple bands, and the matched ones are shown by the black curves.}
	\label{MassDist_LHC}
\end{figure}

\begin{figure}[H]
	\includegraphics[width=7.5cm]{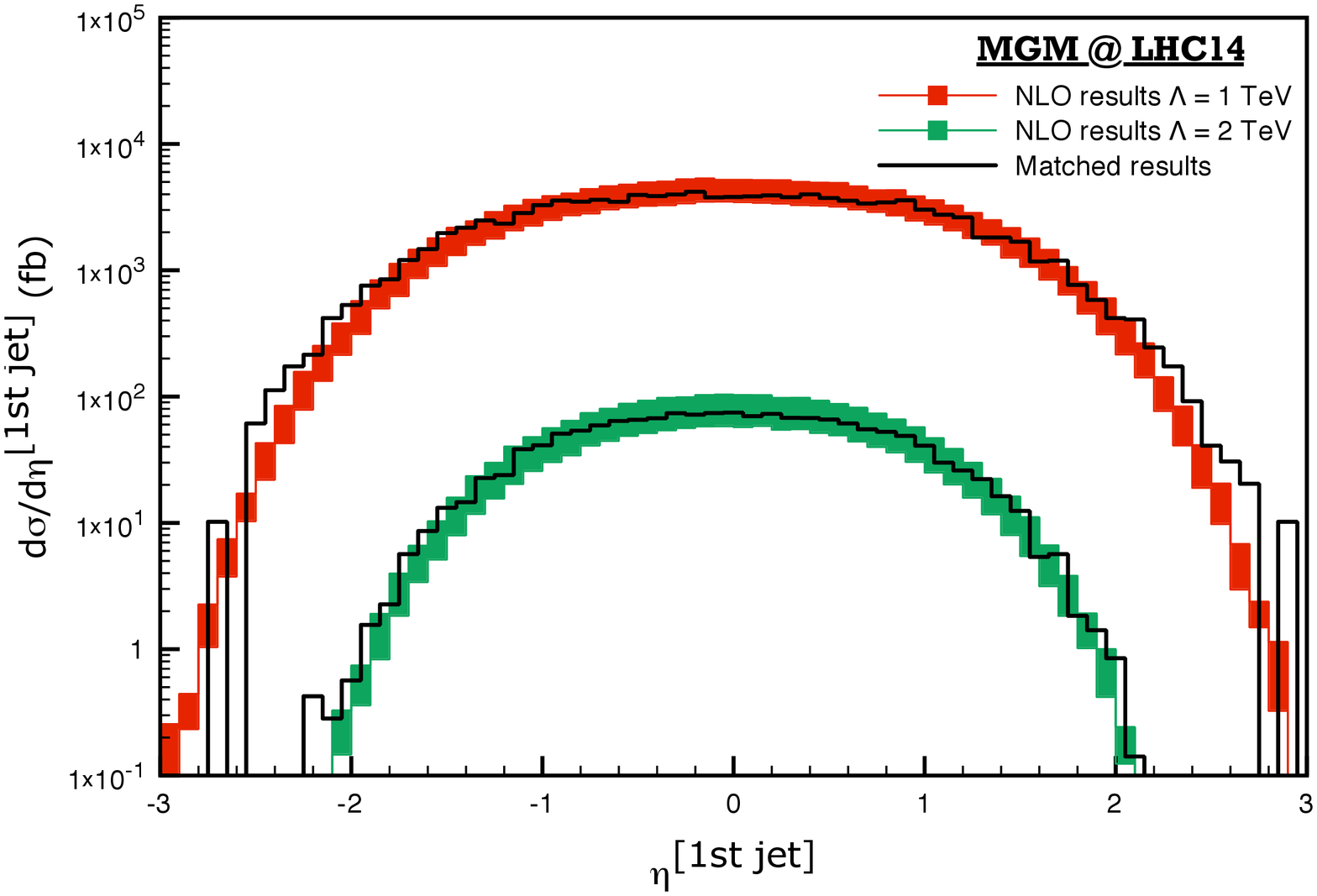}
	\includegraphics[width=7.5cm]{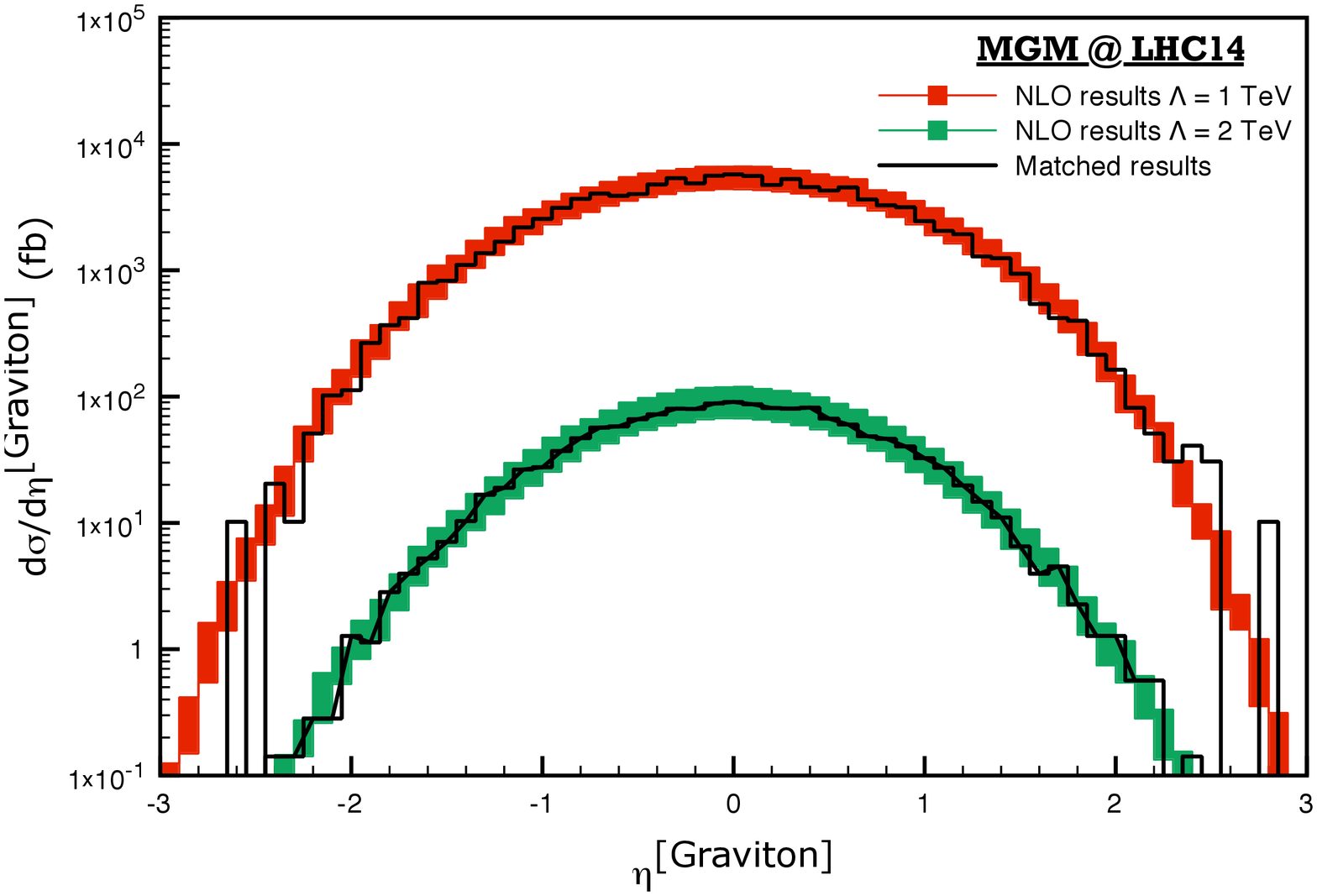}
	\includegraphics[width=7.5cm]{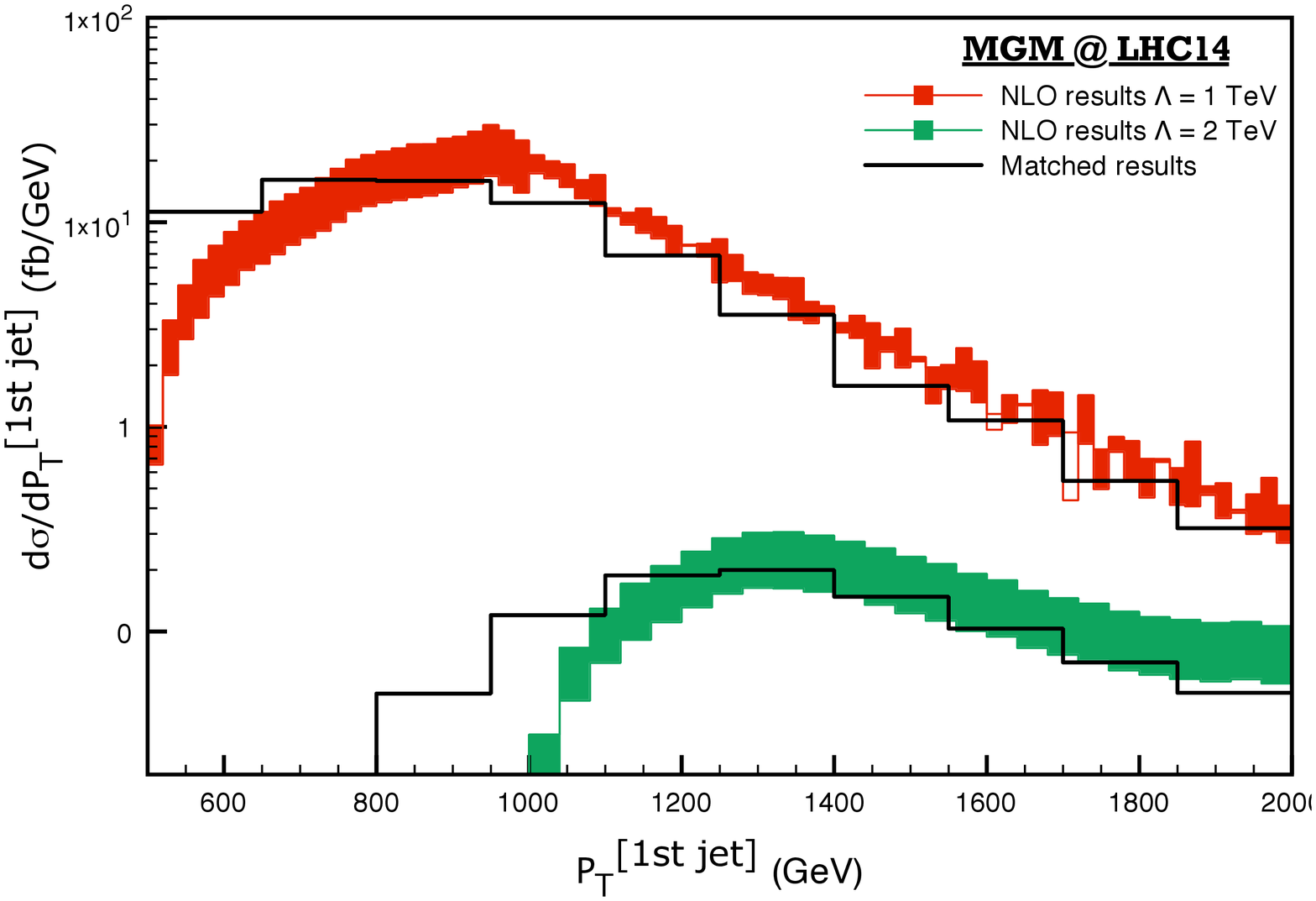}
	\includegraphics[width=7.5cm]{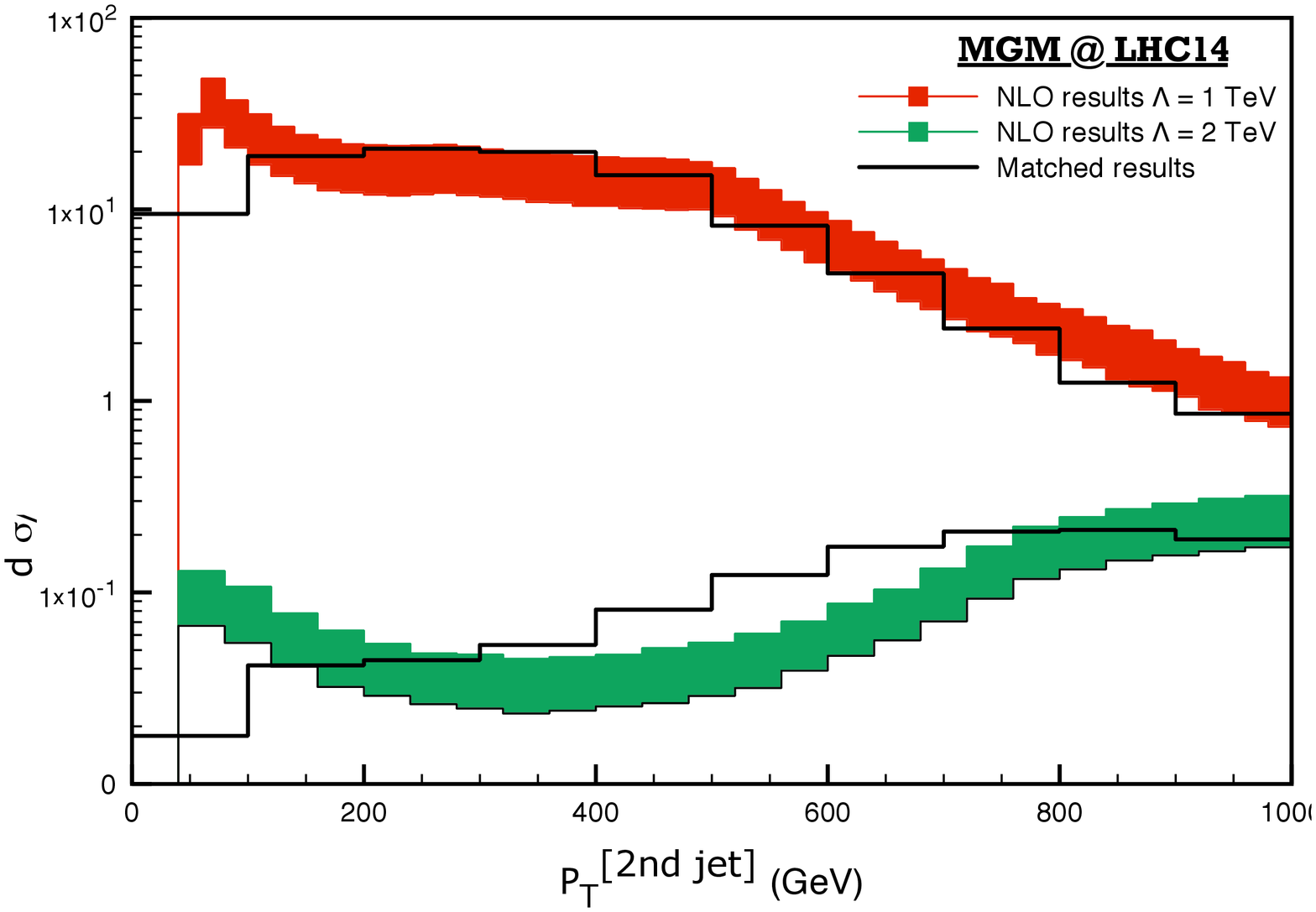}
	\includegraphics[width=7.5cm]{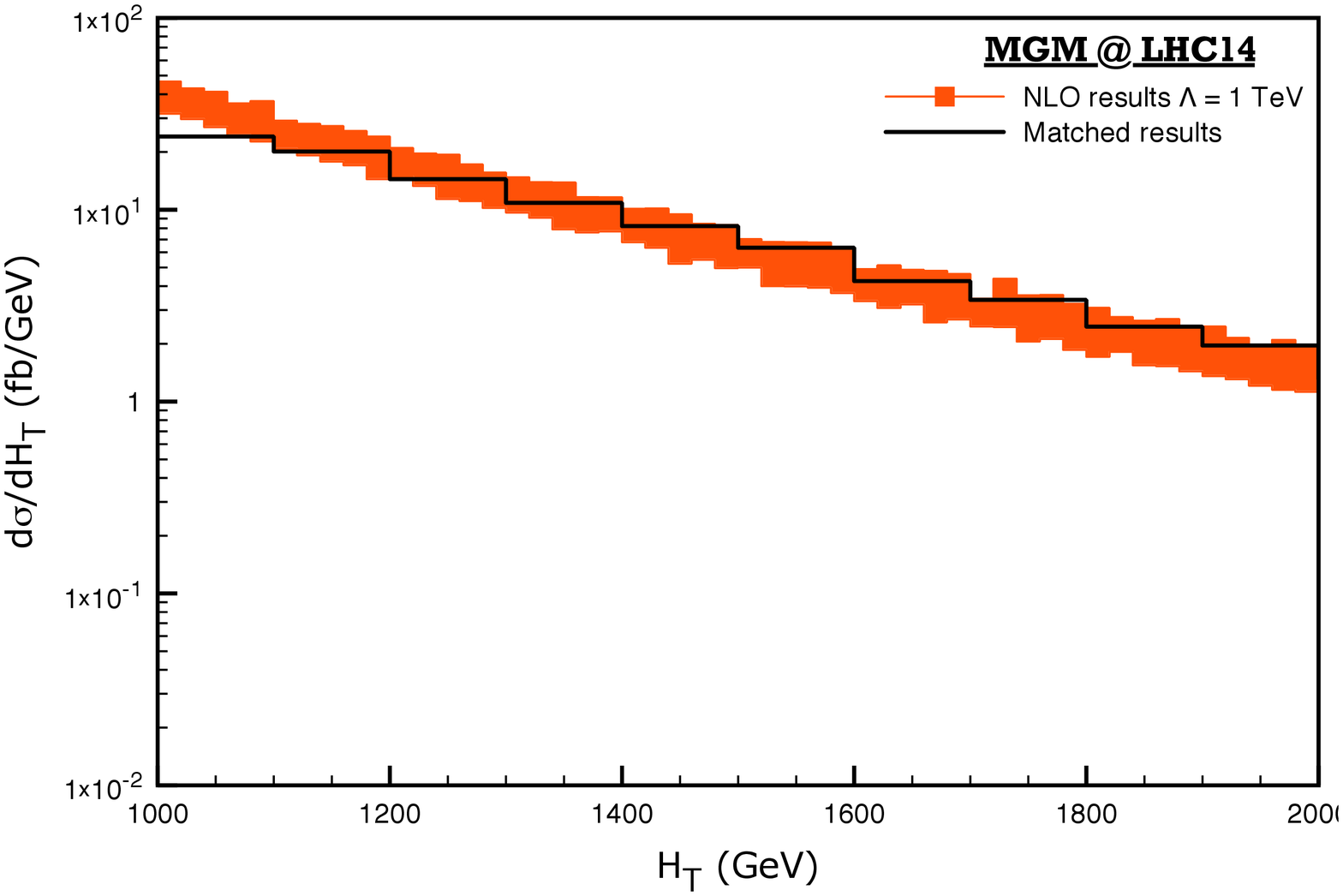} \hspace{0.9cm}  
	\includegraphics[width=7.5cm]{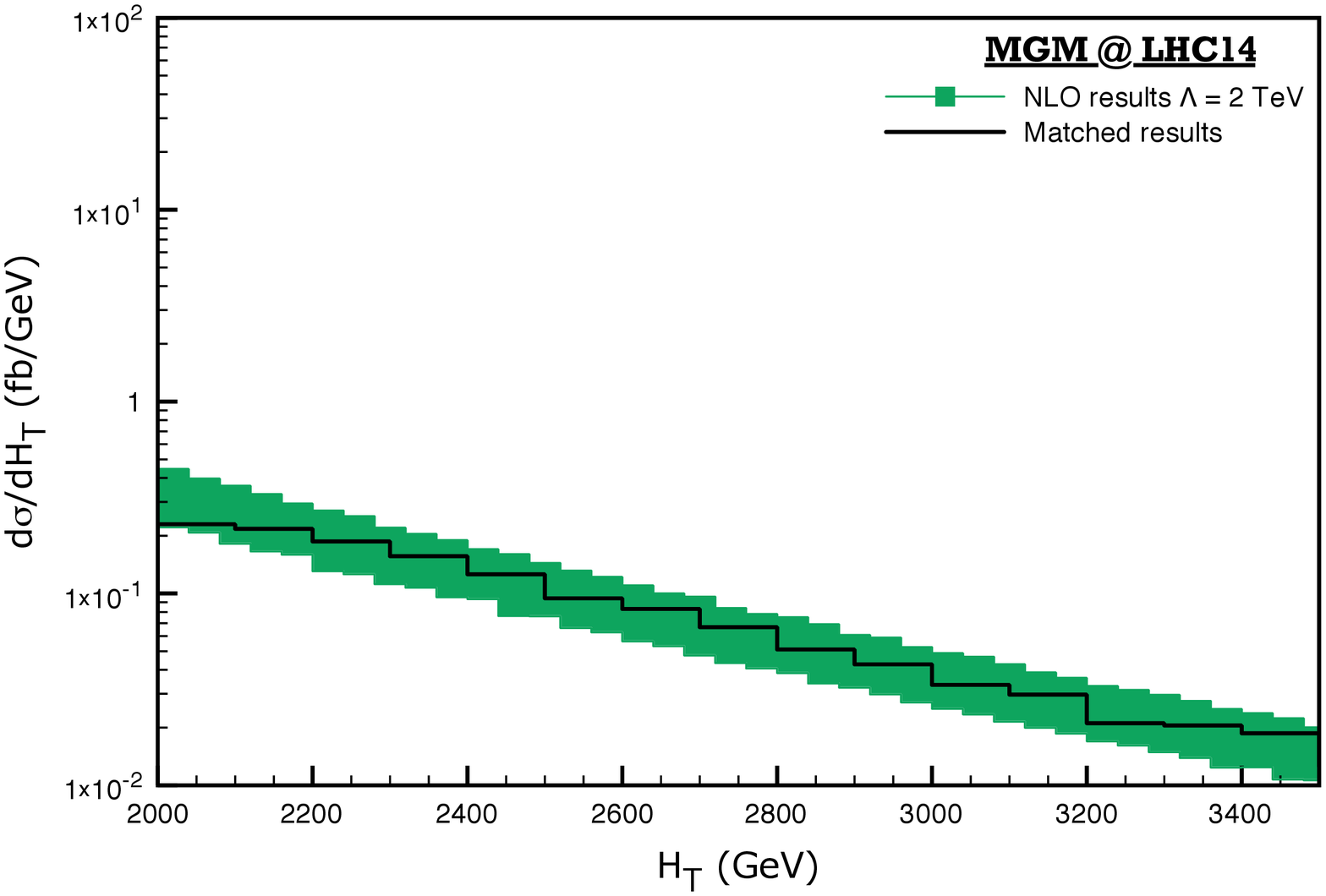}

	\begin{centering}
		\includegraphics[width=7.5cm]{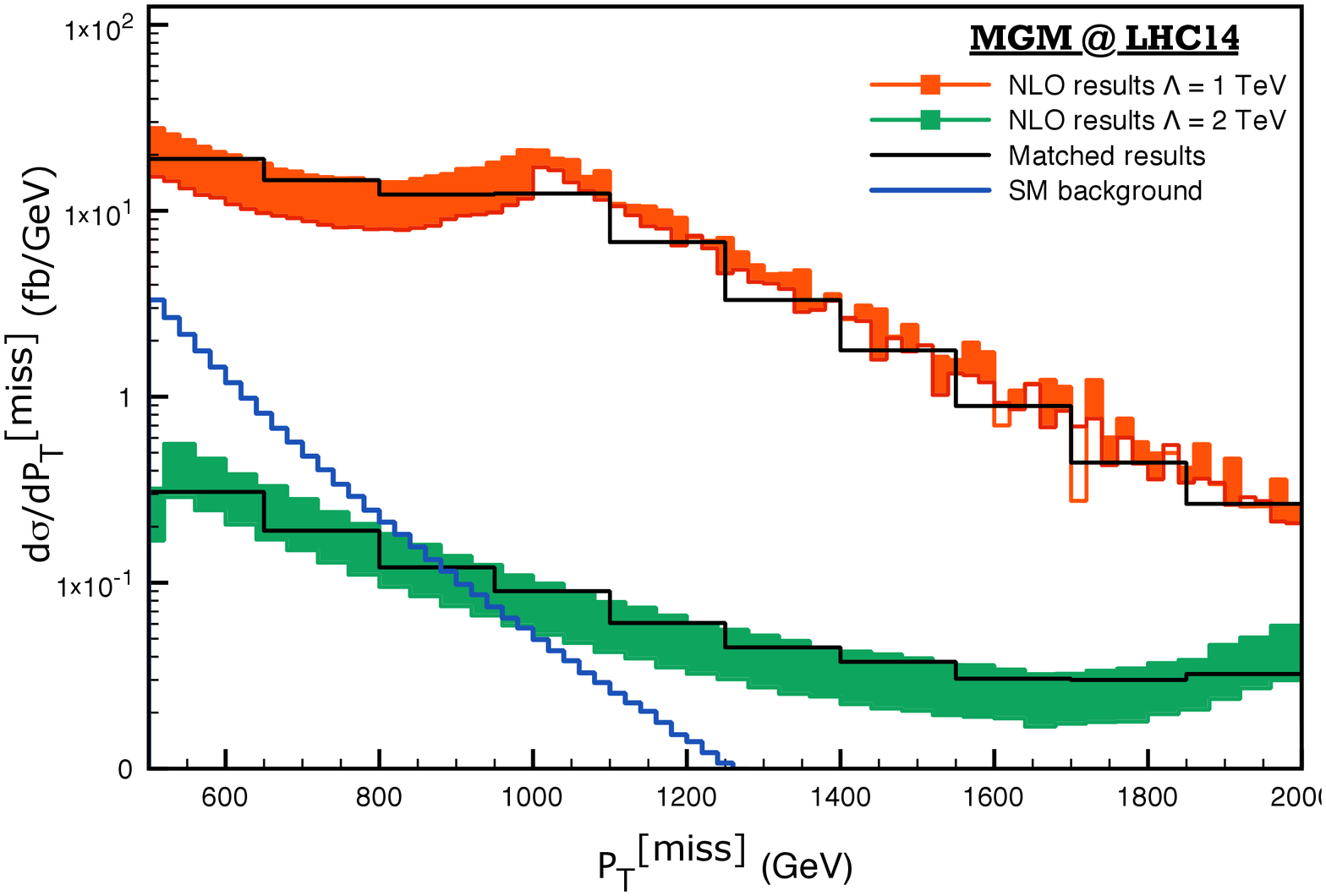} 
	\par\end{centering}

	\caption{ NLO/$k_\bot$-MLM matching comparison for the MGM model at the LHC. The NLO results are given by the red and green bands for $\overline{M}(\mu_\star) = 1,\, 2$  TeV, respectively. The matched results are given by the black curves with the same parameters as for the NLO ones. The dominant ${\rm Z} \to \nu\bar{\nu}$ background is also shown as a reference by the blue curve. The $k_\bot$-MLM matched curves are normalized by the NLO results, and the normalization factors can be found in table 4.}
	\label{MGM_LHC}
\end{figure}

\begin{figure}[H]
	\includegraphics[width=8cm]{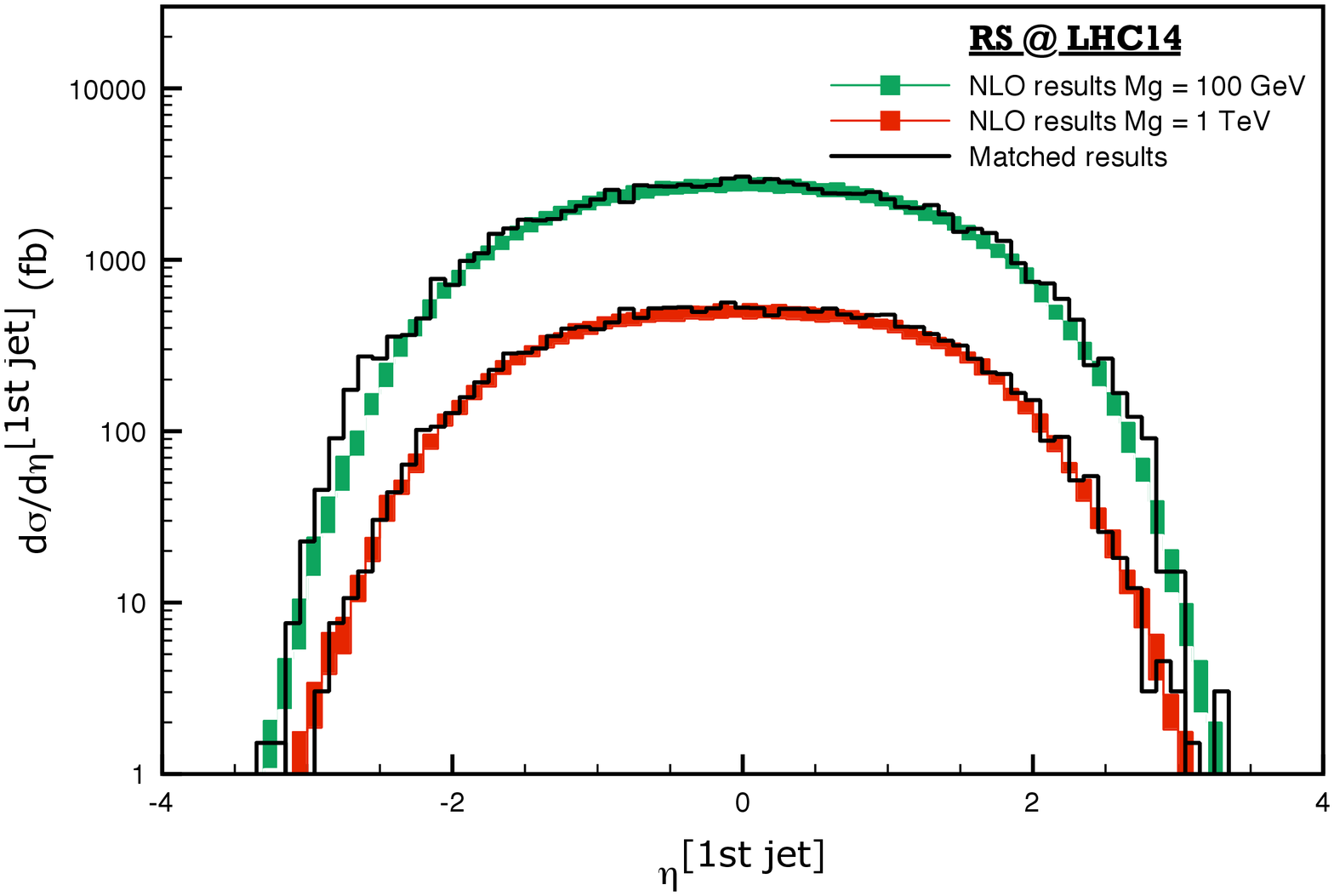}
	\includegraphics[width=8cm]{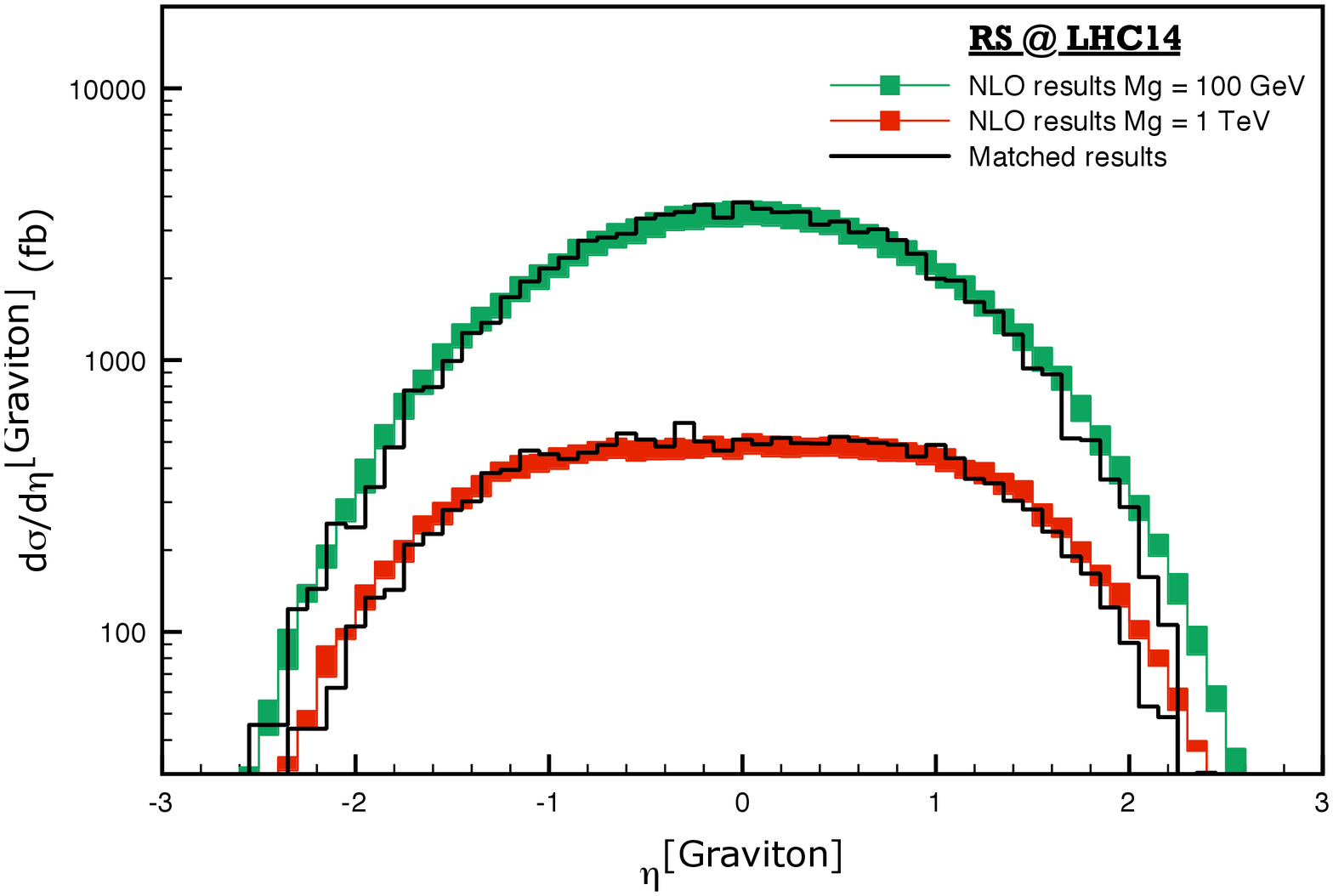}
	\includegraphics[width=8cm]{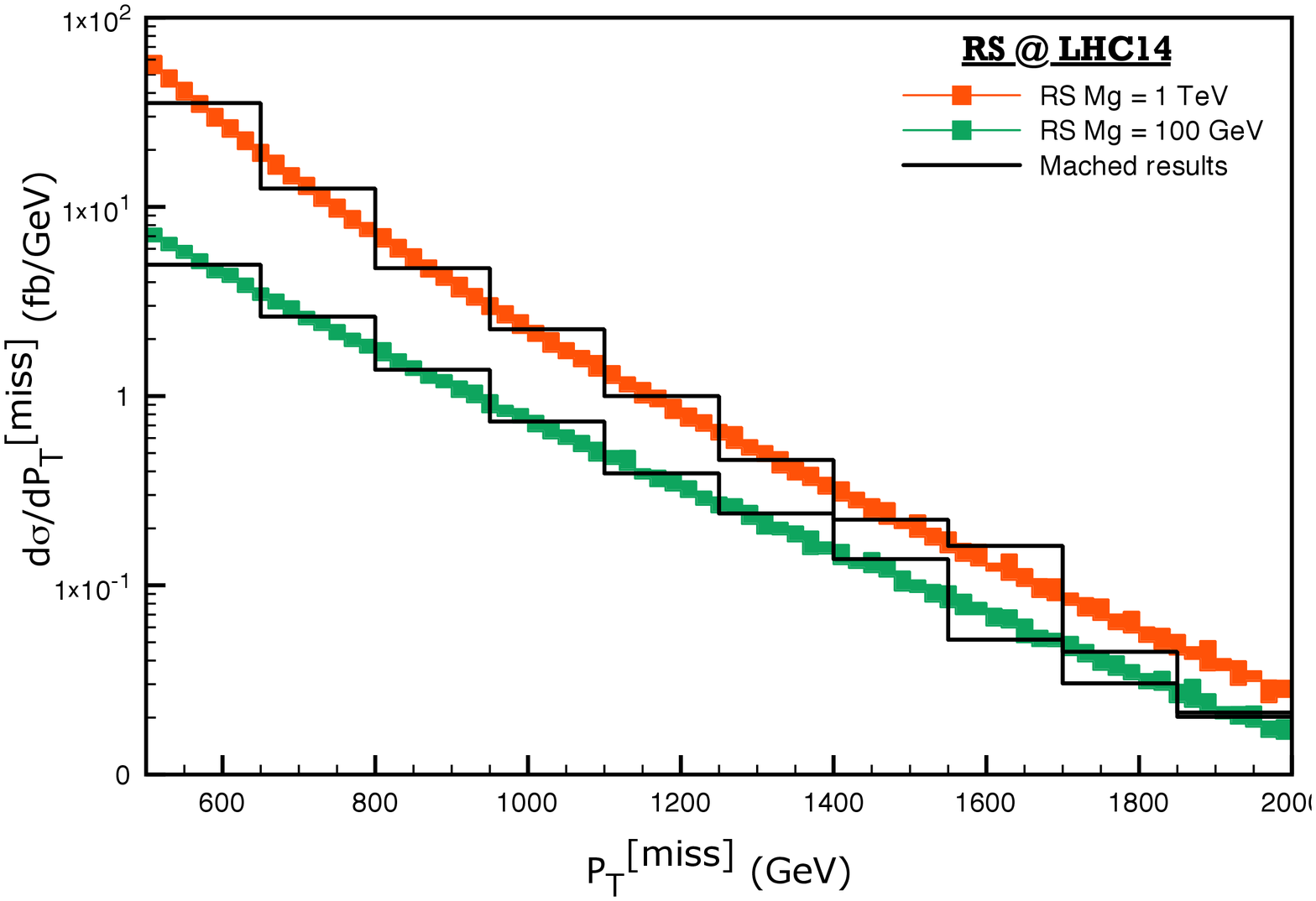}
	\includegraphics[width=8cm]{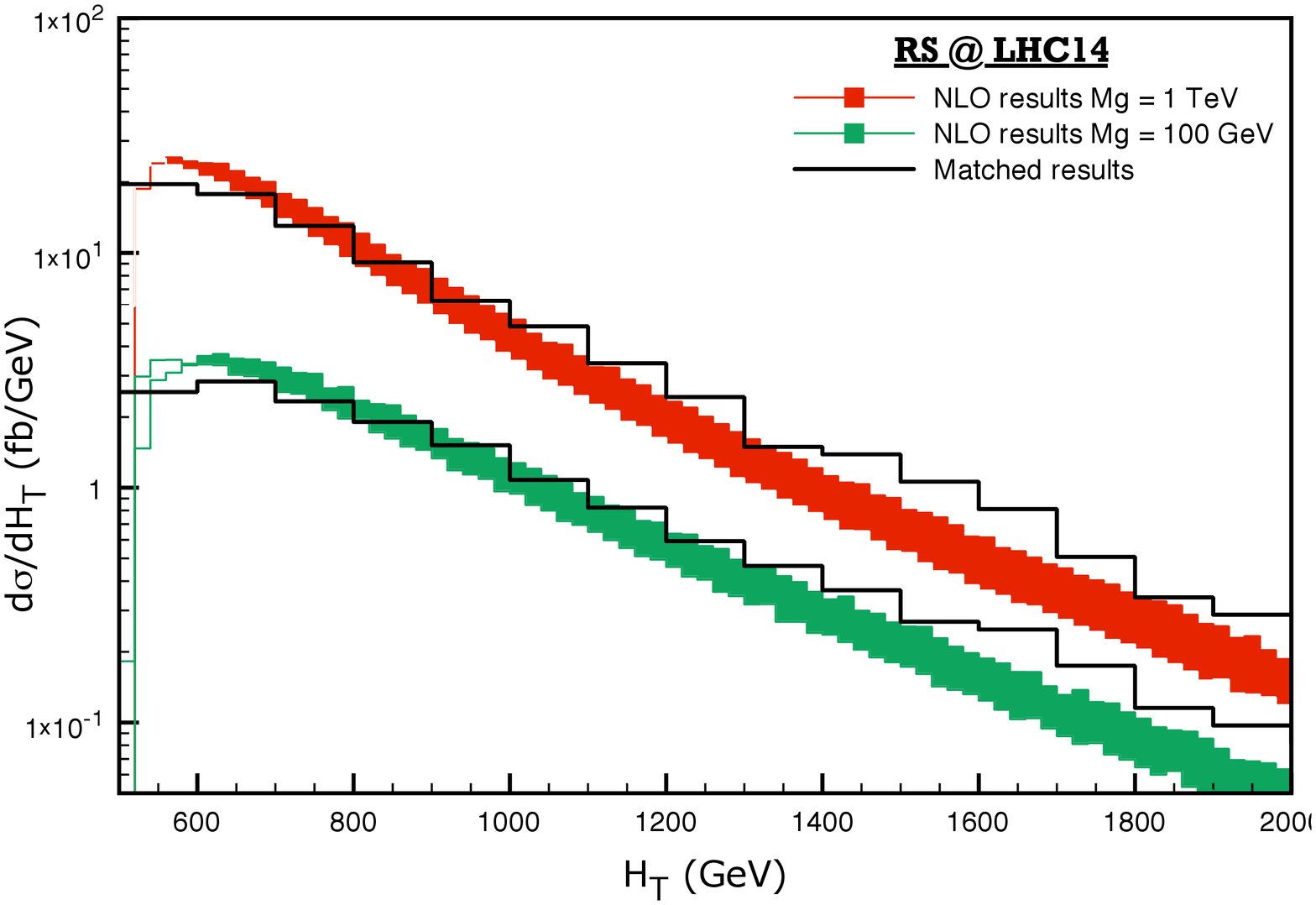}
	\includegraphics[width=8cm]{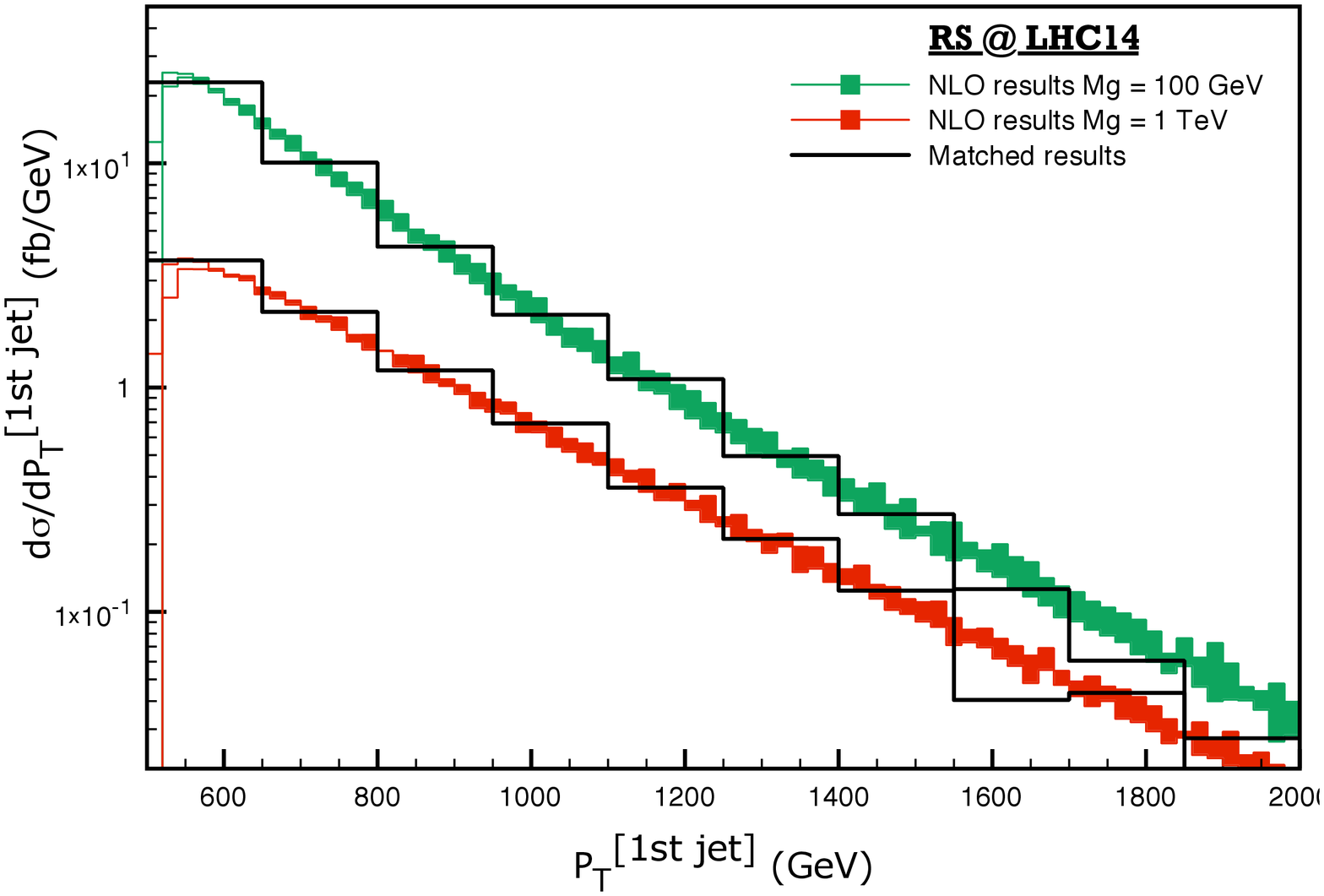}
	\includegraphics[width=8cm]{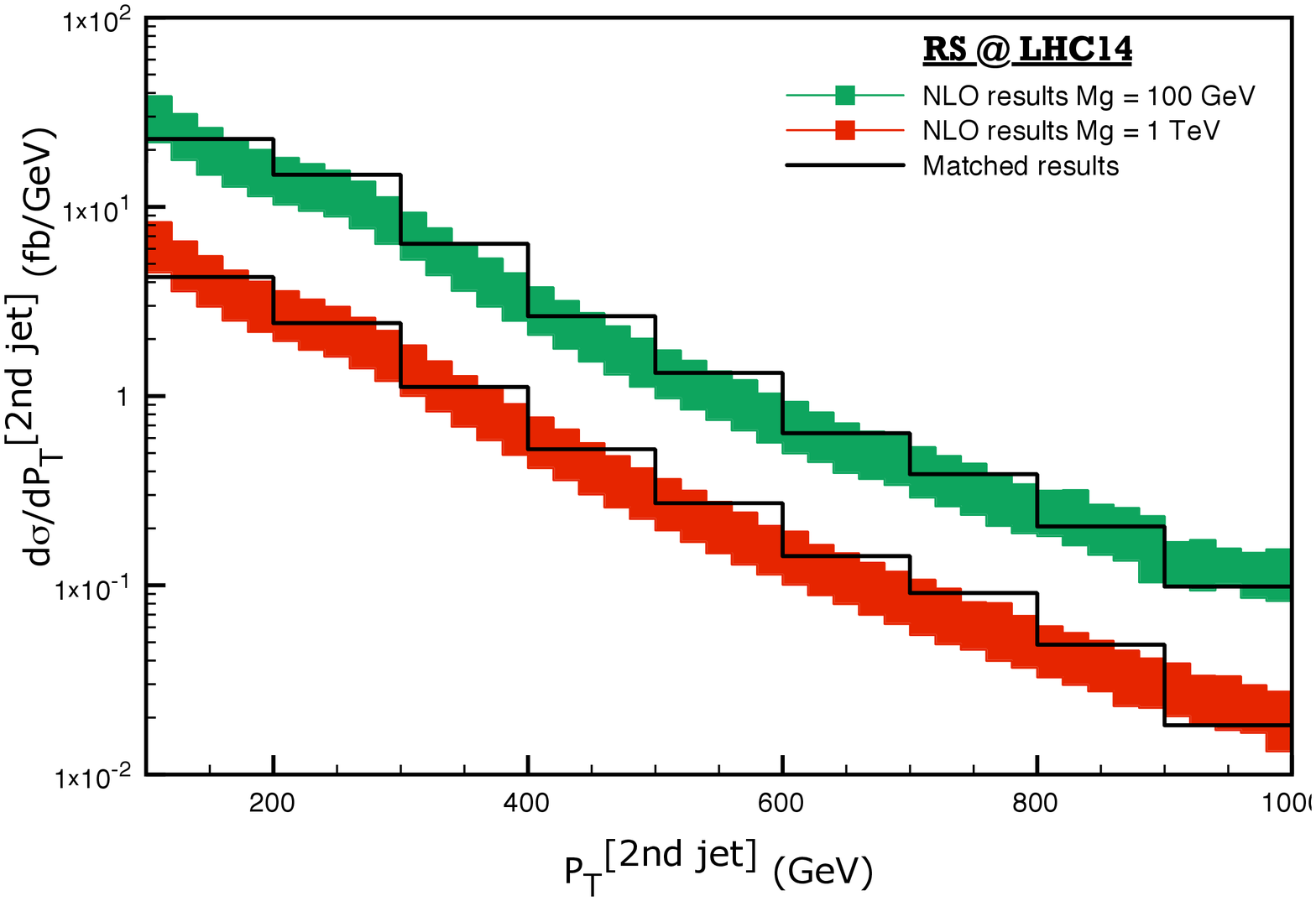}
	\caption{ NLO/$k_\bot$-MLM matching comparison for the RS model at the LHC with  $\Lambda=3$ TeV. The NLO results are given by the red and green bands for $m_{1}=1$ TeV and 100 GeV respectively. The matched results are given by the black curves with the same parameters as for the NLO ones. The $k_\bot$-MLM matched curves are normalized by the NLO results, and the normalization factors can be found in table 4.}
	\label{RS_LHC}
\end{figure}

\begin{figure}[H]
	\includegraphics[width=8cm]{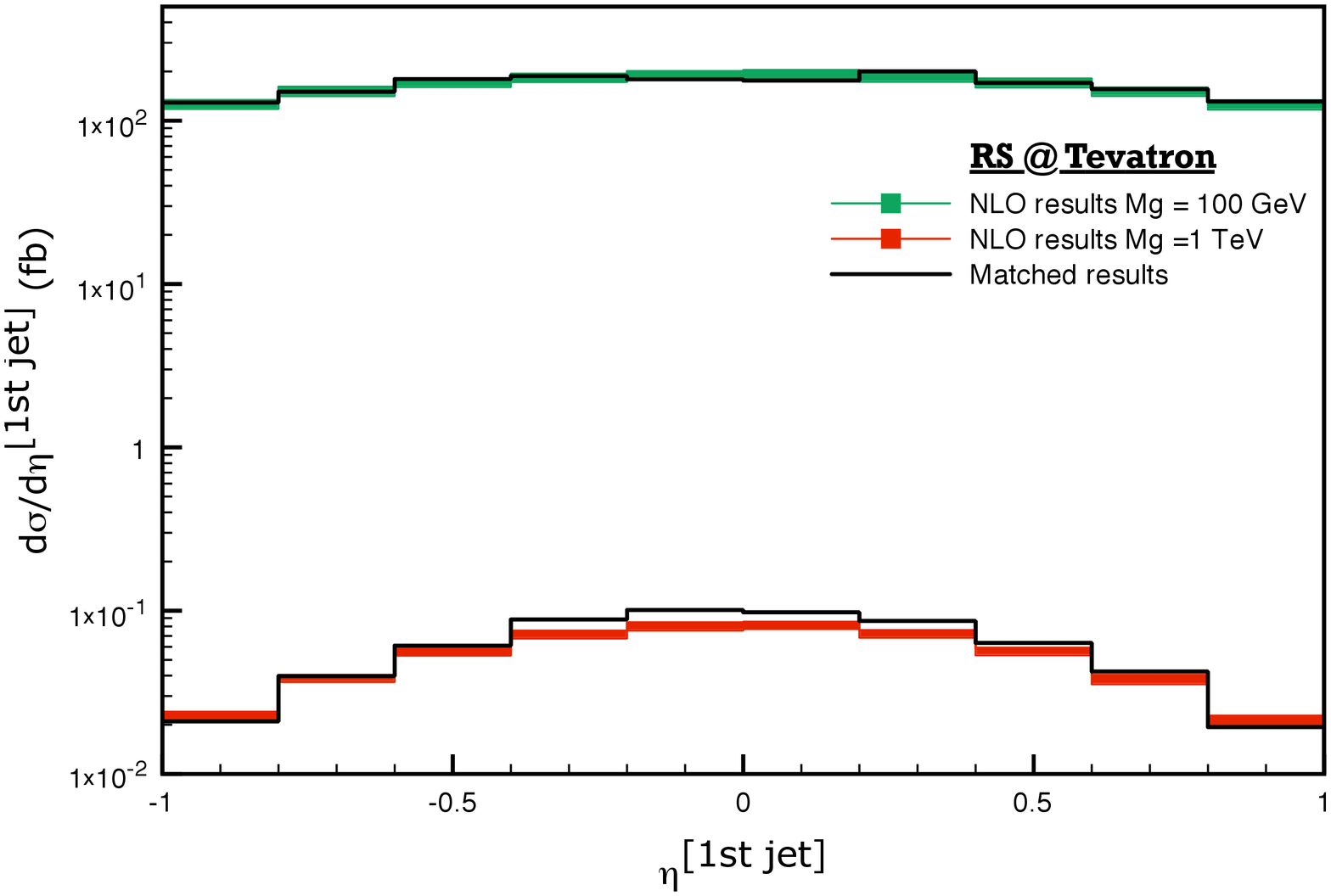}
	\includegraphics[width=8cm]{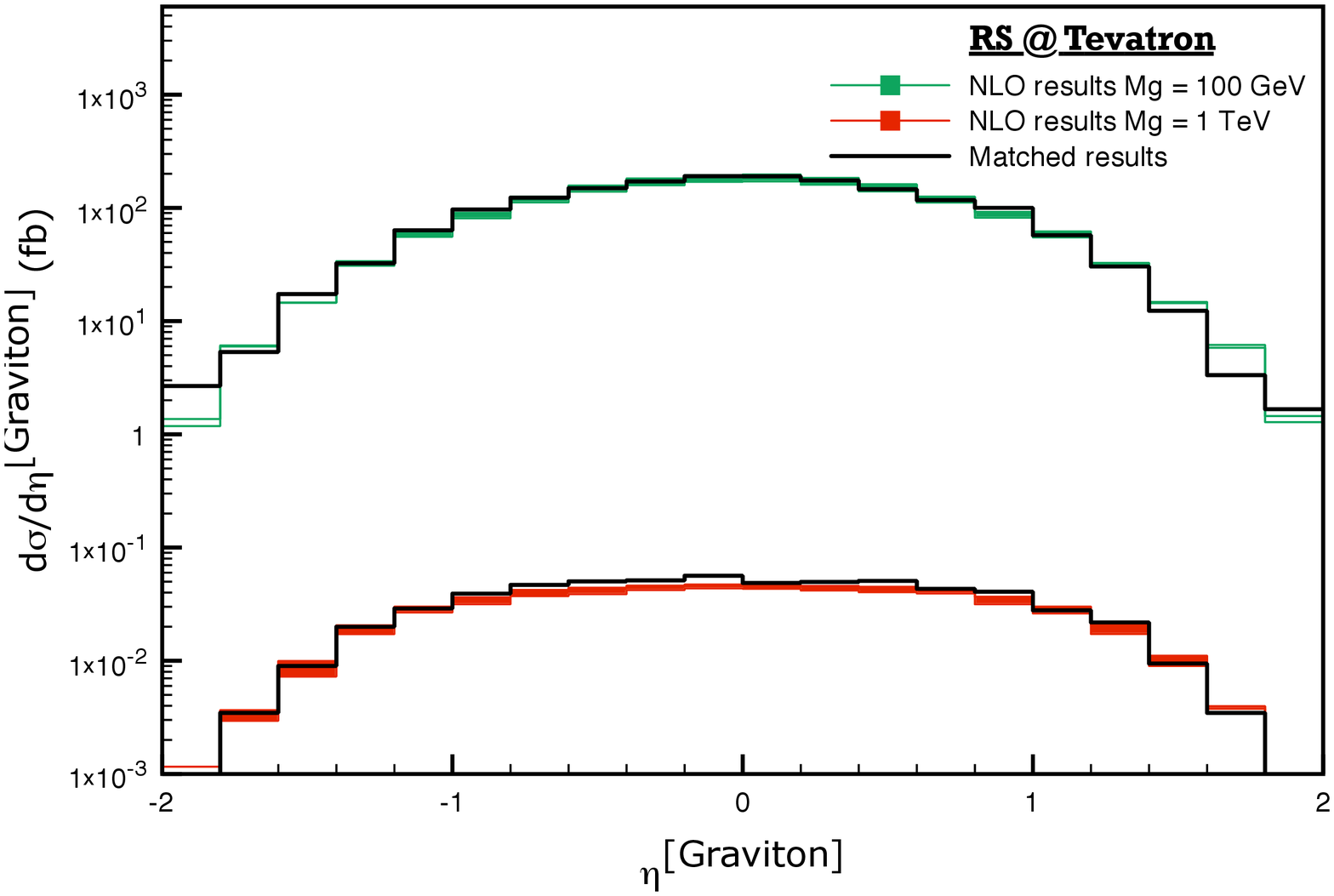}
	\includegraphics[width=8cm]{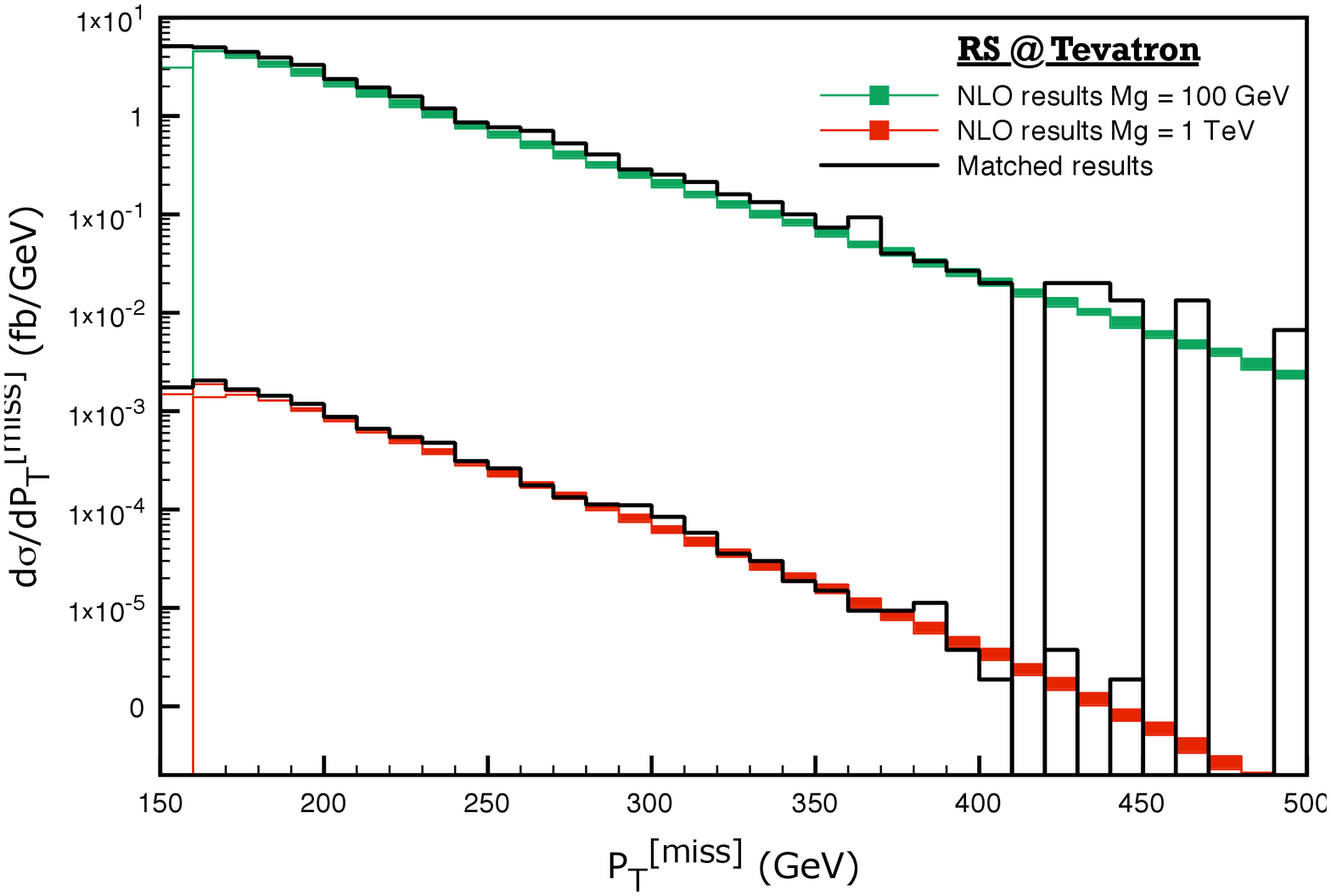}
	\includegraphics[width=8cm]{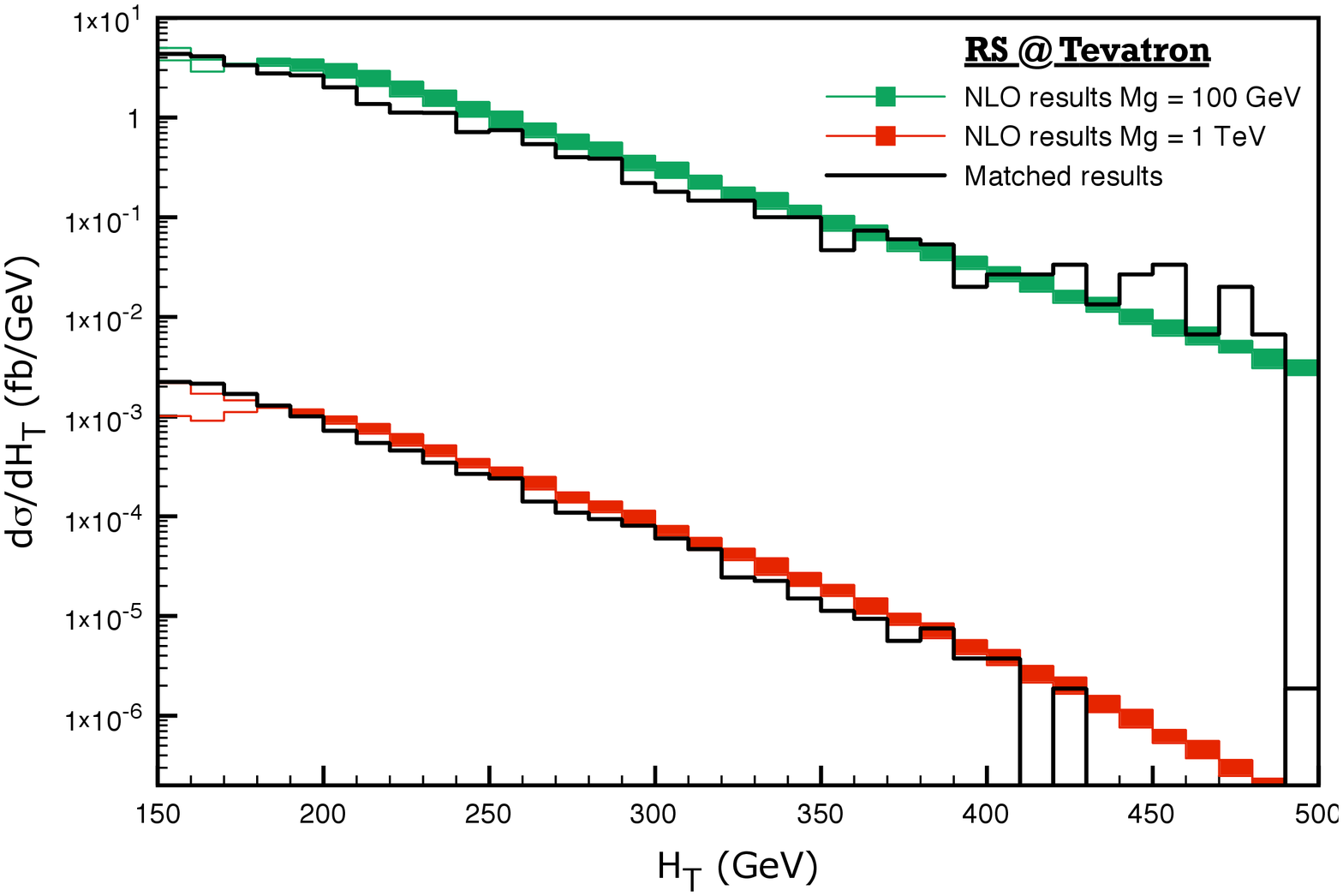}
	
	\begin{centering}
		\includegraphics[width=8cm]{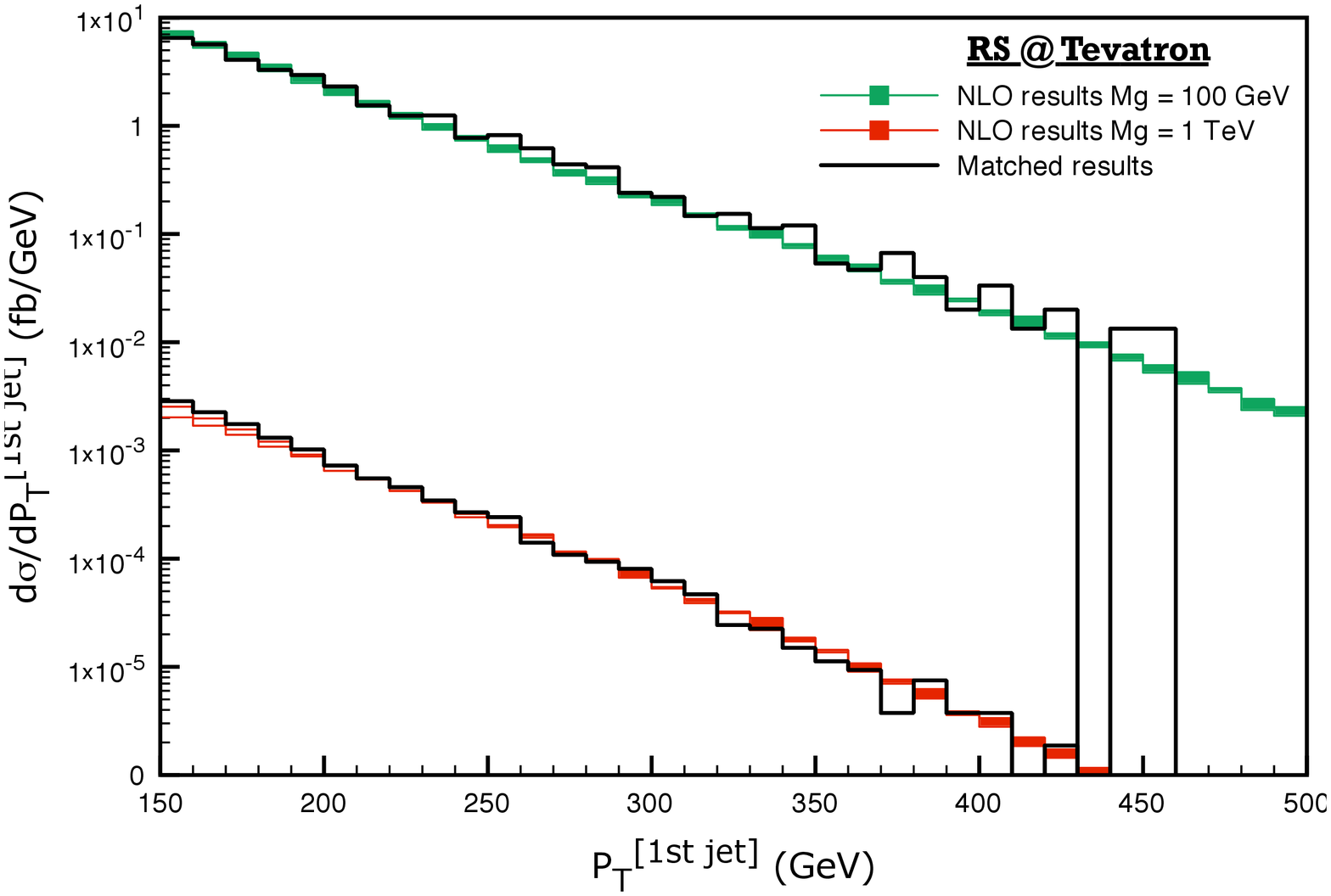}
	\par\end{centering}
	
	\caption{NLO/$k_\bot$-MLM matching comparison for the RS Model at the Tevatron with $\Lambda=3$ TeV, for $m_{1}=100$ TeV and 1 TeV, respectively. The $k_\bot$-MLM matched curve is normalized by the NLO results and the normalization factors can be found in table 4.}
	\label{RS_Tevatron}
\end{figure}

Figures~\ref{ADD_LHC}-\ref{RS_Tevatron} show all the differential distributions for the comparison between the NLO and $k_\bot$-MLM matched results. The NLO uncertainty bands are also presented by setting the renormalization and factorization scales to $\mu_F=\mu_R=\mu_0/2$, $\mu_0$ and $2\mu_0$ with $\mu_0=P_T^{G}$. The $k_\bot$-MLM matched results are normalized by the NLO ones with $\mu_F=\mu_R=\mu_0$,  as shows table~\ref{KfactorTable}. 

For ADD and MGM classes of models it is much more interesting to obtain a comparison between $G +  {\rm jet}$ at NLO and $G +  n\,\,{\rm jet(s)}$ with the $k_\bot$-MLM matching for $n=1,\,2,\,3$. In order to compare all the three classes of models, we also present the same analysis for the RS model.

We start by discussing the results for the ADD model. From figure~\ref{ADD_LHC}, one can immediately see that the NLO and $k_\bot$-MLM matched results for the LHC agree quite well on the shapes in pseudo-rapidity of the leading $P_{T}$ jet, pseudo-rapidity of the graviton and transverse missing $P_{T}$. For the leading jet $P_{T}$ distributions,  the agreement is satisfactory for large $P_{T}\gtrsim 800$~GeV region, while for $P_{T}\sim 500$~GeV, the NLO curves drop down too quickly and become unreliable. That is due to the cut chosen as in Eq. (\ref{lhcset}) which leads to inconsistency on leading jet $P_{T}$ lower bound at LO and NLO calculations, while $k_\bot$-MLM matching gives reliable results. The same argument also applies for $H_{T}$ distributions when $H_{T}\sim 500$~GeV. However, at large $H_{T}\gtrsim 1500$~GeV, the $k_\bot$-MLM matching tends to give harder distributions, because the contributions from the matrix elements of $G+3$ partons are also included. 

Figure~\ref{ADD_Tevatron} presents similar results for the Tevatron, and one can find, in general, a better agreement between NLO and $k_\bot$-MLM matched results. Tevatron's energy scale is much lower than the LHC and thus the contribution from more hard jet emission included in $k_\bot$-MLM matching does not play an important role within this case.

It is known (Sec.~\ref{gmi}) that the 4-dimensional graviton is represented by a infinite sum of KK gravitons in the ADD model. The mass of the effective graviton is therefore given by a distribution described by Eq. (\ref{rho}). In figure~\ref{MassDist_LHC} we show the results for the NLO and $k_\bot$-MLM matching comparison of the graviton mass distribution. One can easily observe the dependency on the number of extra dimension since the average graviton mass increases with it. Again, $k_\bot$-MLM matching results agree quite well with the NLO ones.

The results for the MGM model at the LHC are presented in figures~\ref{MGM_LHC}. The Tevatron plots are out of reach for $\overline{M}(\mu_\star) = 1,\, 2$ TeV, therefore the corresponding results are not shown. Moreover, we apply here the cut ${\rm H}^{\rm min}_{\bot}>\overline{M} (\mu_\star)$ as described in Sec.~\ref{imgm}. One can see a kink or turning points in the distributions of the leading or second jet $P_{T}$ and transverse missing $P_{T}$. The NLO and $k_\bot$-MLM matched results agree quite well on shapes for distributions of pseudo-rapidity of the leading $P_{T}$ jet and graviton, transverse missing $P_{T}$ and $H_{T}$ of the jets. For the second jet $P_{T}$ distribution, $k_\bot$-MLM matching gives a considerably different shape when compared to the NLO calculation, especially for $\overline{M}(\mu_\star) = 2$ TeV case.  For the leading jet $P_{T}$ distribution, the NLO and $k_\bot$-MLM matched results for the LHC agree well for large $P_{T}\gtrsim \overline{M} (\mu_\star)/2$, while for $P_{T} \lesssim\overline{M} (\mu_\star)/2$, the NLO curves are unreliable, again due to the inconsistency on leading jet $P_{T}$ lower bound at LO and NLO calculations.

Figures~\ref{RS_LHC}~-~\ref{RS_Tevatron} presents results for the RS model at the LHC and Tevatron similarly to the ADD figures showed in figures~\ref{ADD_LHC}~-~\ref{ADD_Tevatron}. Here, the agreement between the NLO and $k_\bot$-MLM matched results is again much better at the Tevatron than the LHC for the same reasons pointed above. Also, $k_\bot$-MLM matching tends to give harder $H_T$ distributions especially for large $H_{T}\gtrsim 1500$~GeV.

\subsection{Large $\kappa$-factors}

\begin{figure}[t]
	\includegraphics[width=8cm]{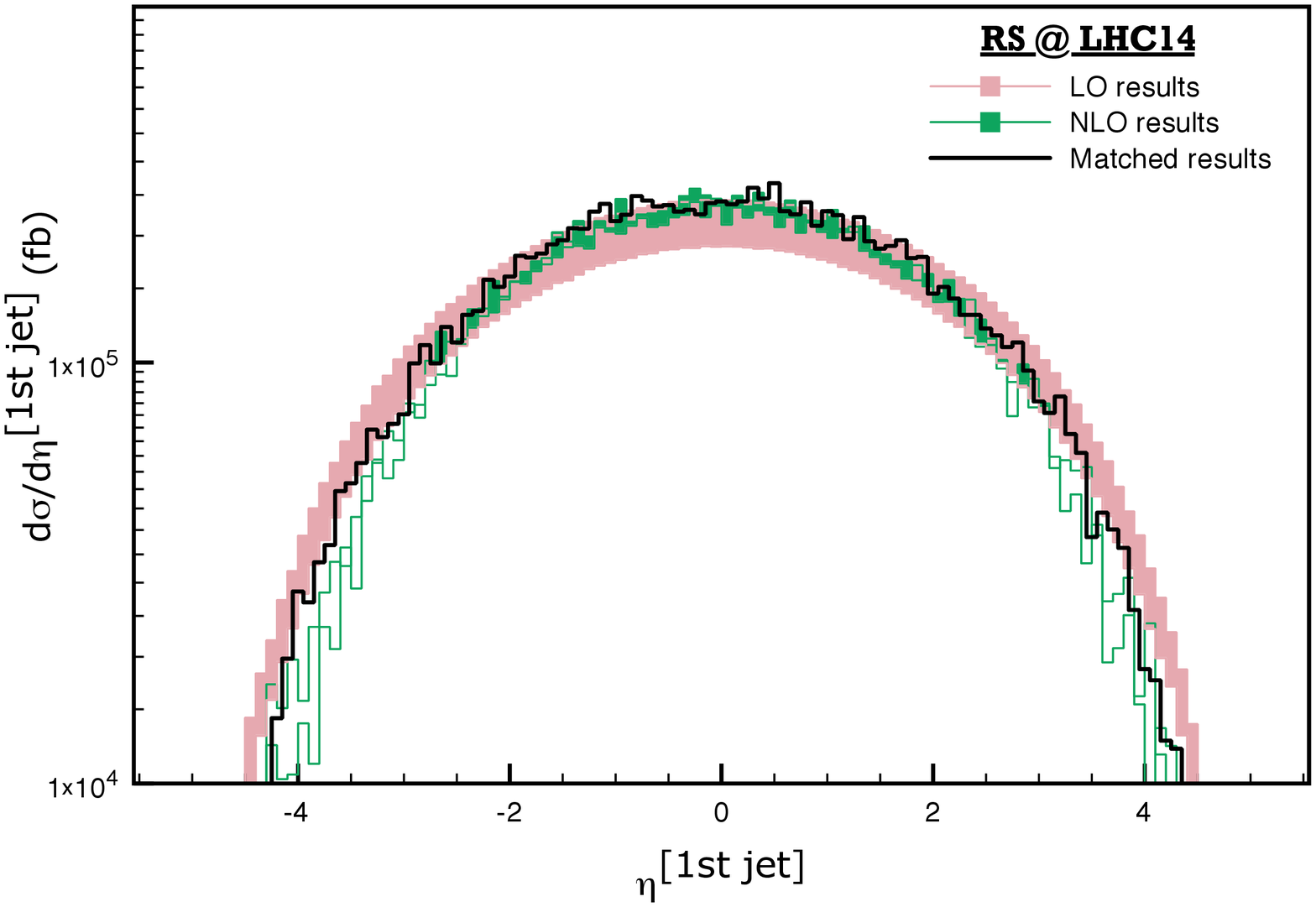}
	\includegraphics[width=8cm]{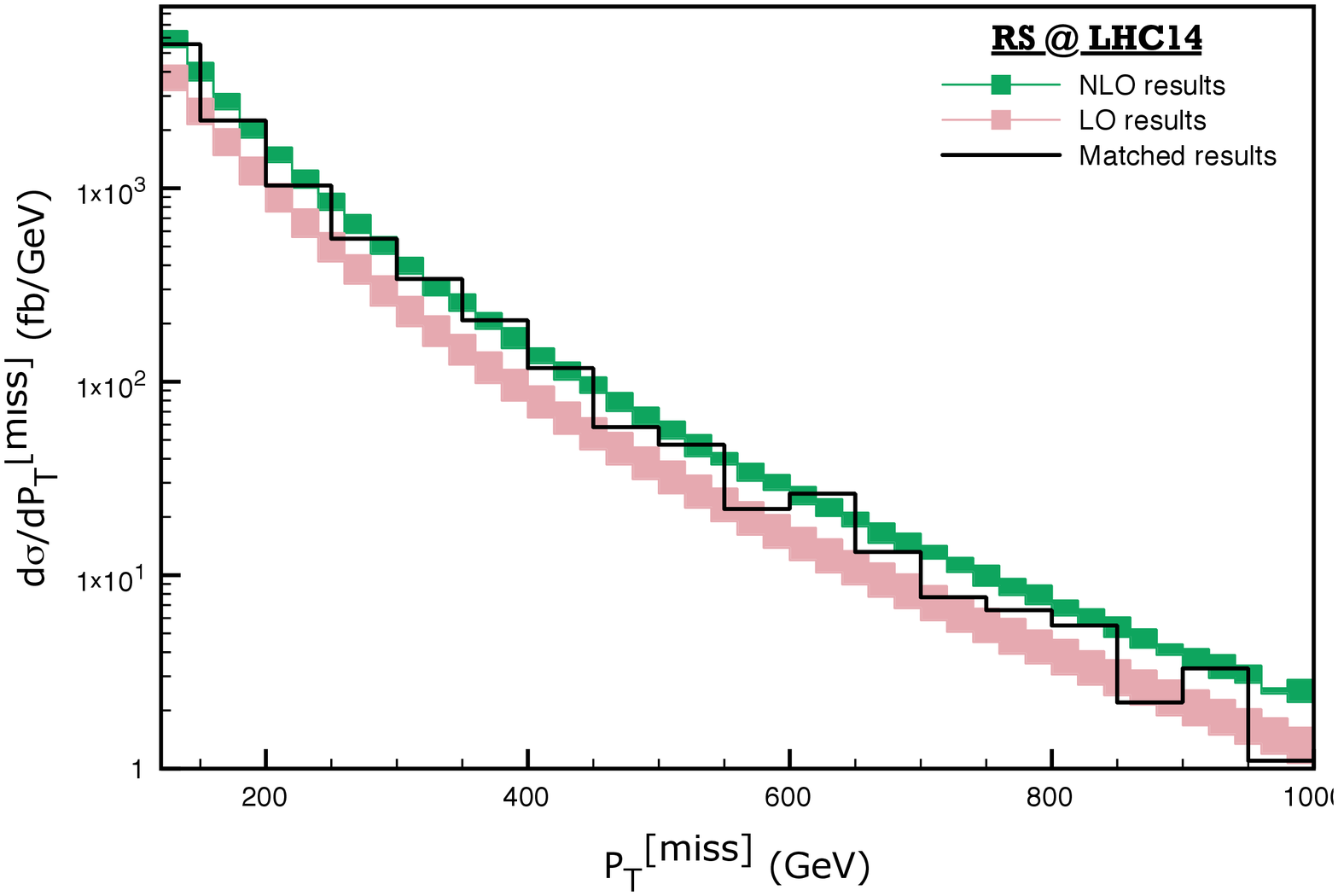}
	\includegraphics[width=8cm]{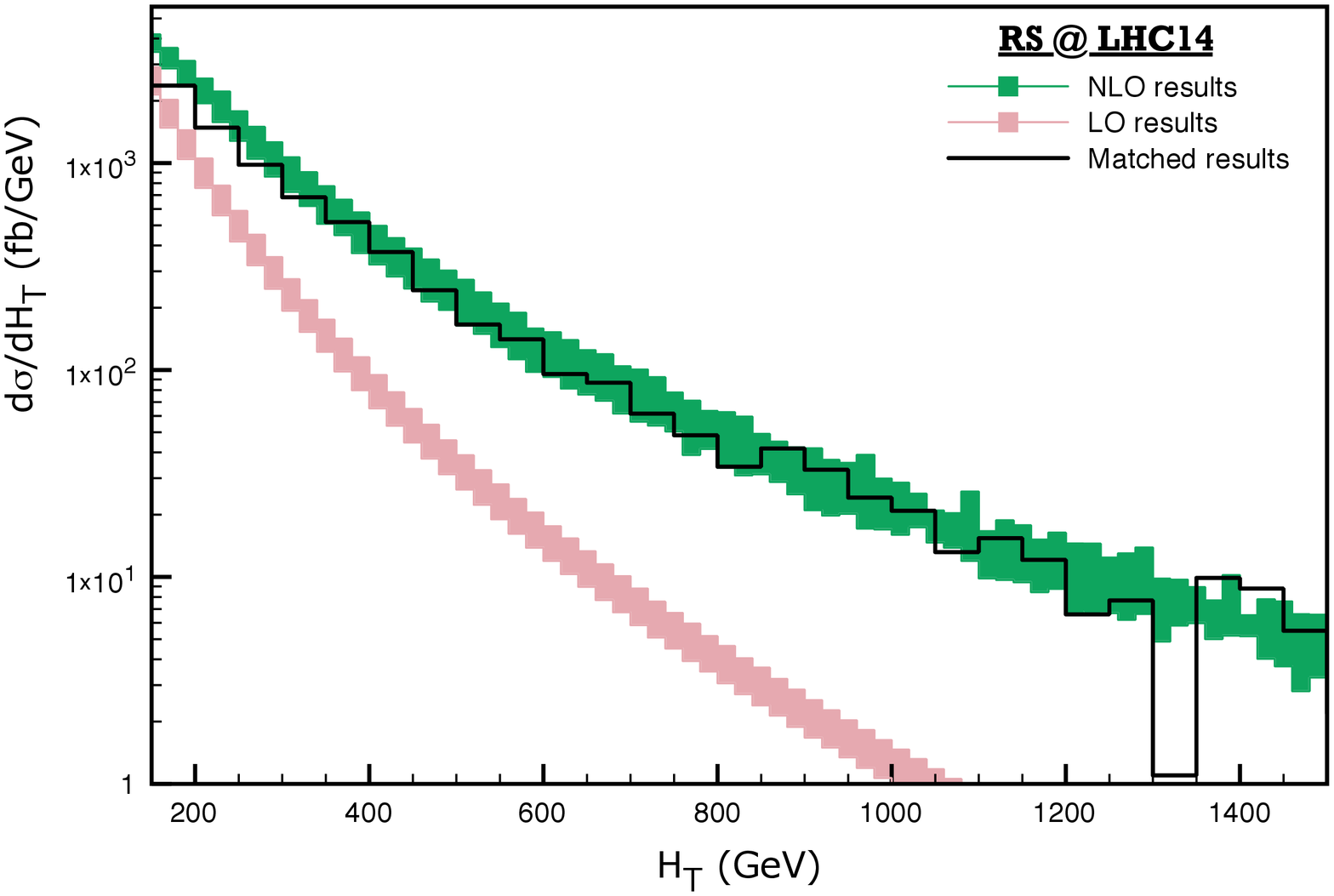}
	\includegraphics[width=8cm]{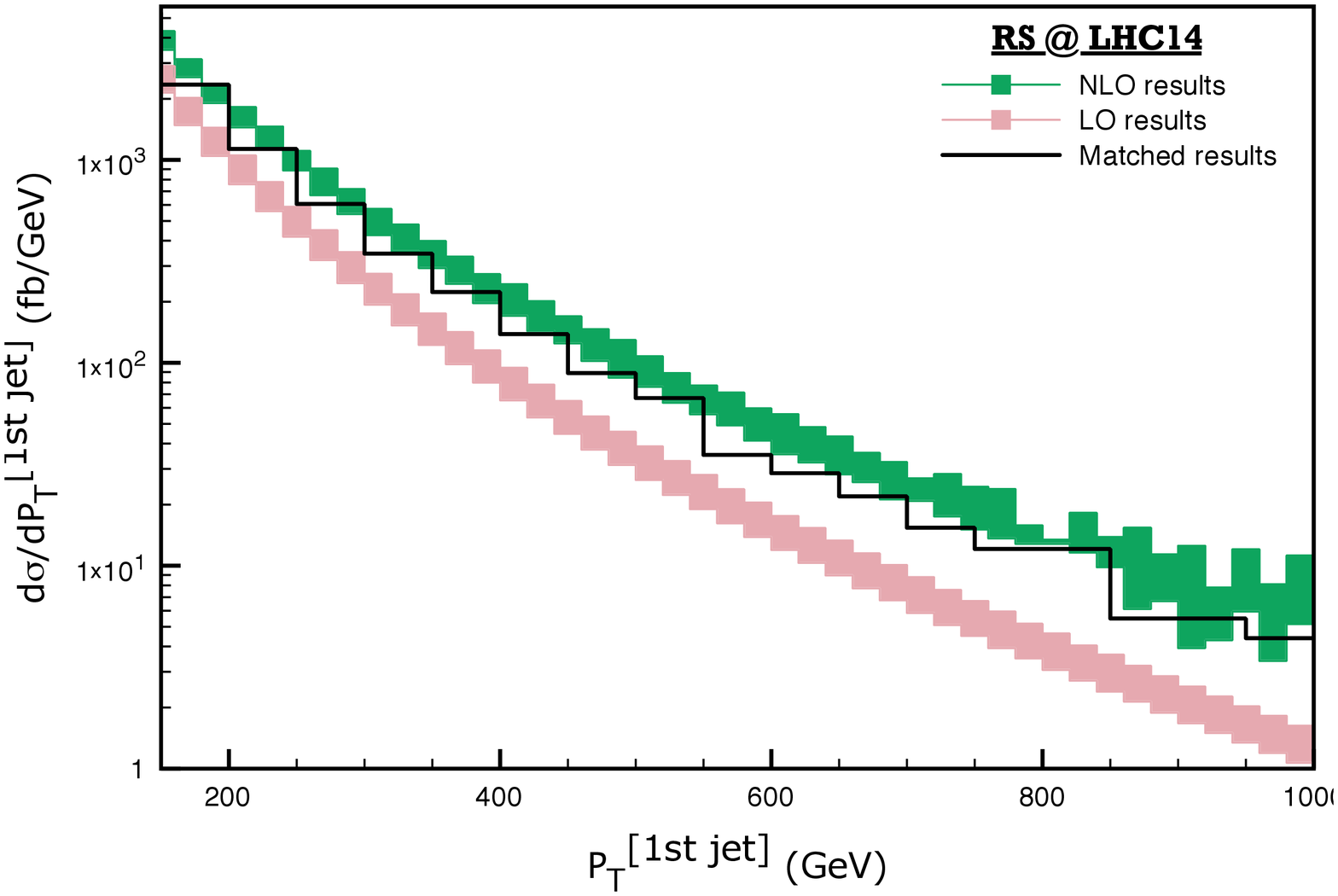}
	\caption{NLO/$k_\bot$-MLM matching comparison for the RS Model at the LHC with $m_{1}=100$ GeV and $\Lambda=3$ TeV. Different cuts from the ones specified above are used to show the interesting case with large $\kappa$-factors.}
	\label{RS_Kfactor_LHC}
\end{figure}%\newpage

In multi-jet final states, observables exist that can be subject to large NLO QCD corrections at TeV scales~\cite{Rubin:2010rq}. Examples include $H_T$ or the $P_T$ of the leading jet. Such observables are sensitive to the opening of new channels with new kinematic topologies at NLO, which are enhanced by parton distribution effects or by special kinematic configurations (leading to large logs). One is then lead to the impression that higher order corrections can be dominant and the perturbative series is not well-behaved. In fact, one can easily show that this is not the case for more inclusive quantities and this behaviour is due to the choice of the specific observable. In any case, the effects in some specific region  of the phase space can be large and need to be accounted for. One method was introduced Ref.~\cite{Rubin:2010rq}. Here we show that multi-parton samples with $k_\bot$-MLM matching provide a reliable description also of these kind of effects. In figure~\ref{RS_Kfactor_LHC} we collect some representative plots which show it. For simplicity, we consider the RS model at the LHC with $\Lambda=3$ TeV and $m_{1}=100$~GeV, with jet cuts $|\eta_j|<4.5$ and $P_T^j>50$~GeV, and graviton cut $P_T^G> 50~{\rm GeV}$. Comparing the LO and NLO curves, one can see the appearance of very large $\kappa$-factors, especially at large $P_T^G$, leading jet $P_{T}$ and $H_T$. The matched samples describe extremely well the NLO shapes. 
%
%--------------------------------------
\section{Summary and conclusions}
\label{sec:end}

In this paper we have presented  a detailed comparison between predictions from next-to-leading order QCD calculations and  the  tree level matrix-element matching with parton showers in the context of  three representative theories featuring a graviton: ADD, RS and MGM. This study validates and motivates the use of a fully automatized event generator such as \mgme~including spin-2 particles and multi-parton jet matching for LHC simulations. Let us summarize the main aspects of our approach and the key results.

First, we have presented the complete implementation of spin-2 particles in \\ \mgme~ event generator, which now includes: (i) new {\sc HELAS} subroutines to handle 5-point vertex diagrams; (ii)  proper  account  of summation over graviton excited modes; and (iii) modifications to handle massless spin-2 particles. As a result, processes with spin-2 particles can now be simulated in a fully automatic way.

Second, we have studied the simulation of graviton production in inclusive multi-jet samples. The $k_\bot$-MLM parton shower matching scheme has been employed for several benchmark models and points: for ADD we have chosen
$\Lambda = 5$ TeV for the LHC, $\Lambda = 1$ TeV for the Tevatron, with $\delta =2,\:4,\:6$; for the RS model, we have taken $m_{1} = 100$ GeV and $1$ TeV, and $\Lambda = 3$ TeV; and finally for MGM model, we defined $\overline{M}(\mu_\star) = 1$ and $2$ TeV. We have considered  $pp/p\bar p \rightarrow G \,+\,n$-jets in all the scenarios. In ADD and MGM models, the graviton appears as missing energy while  in RS model it may be observed through its decay products (here, as a an example, we have chosen a pair of leptons). The fully inclusive samples have been normalized to the corresponding NLO inclusive total cross sections for producing $G+ X$.  Key distributions for the most relevant observables involving one or more jets have  then been compared to those obtained by the NLO calculation $G + {\rm jet} + X$.

The overall agreeement in shape as well as in normalization between NLO observables and the corresponding ones obtained by \mgme~+ \pythia~is excellent. In particular the fact that a unique normalization factor (the one normalizing the overall inclusive sample) suffices to describe not only the shape but also the normalization of jet distributions makes the use of matched samples in experimental analyses straightforward and accurate. Finally, we note that observables which are sensitive to the number of jets, such as the $H_T$ are better described by the matched sample than a fixed-order calculation.

In conclusion we have implemented and validated the generation of fully inclusive samples with the correct leading-order jet multiplicity for spin-2 particle production at the Tevatron and the LHC. We look forward to their use in the current and forthcoming searches at hadron colliders.

\acknowledgments

We would like to thank  Tao Han for useful discussions on extra-dimensional theories, Simon de Visscher for his great help on the matching analysis framework, Claude Duhr for advice on implementing 5-point vertex in  \madgraph~version 4 and Jean-Marc G\'erard for discussion on the polarizations of the massless vs.~massive graviton. This work is supported in part by the FWO - Vlaanderen, project G.0235.05 and in part by the Belgian Federal Office for Scientific, Technical and Cultural Affairs through the 'Interuniversity Attraction Pole Program - Belgium Science Policy' P6/11-P. The work presented here has been in part supported by the IISN convention 4.4511.10, in part by Grant-in-Aid for Scientific Research 20348064 from Japan Society for Propution of Sciences and in part by the European Community's Marie-Curie Research Training Network HEPTOOLS under contract MRTN-CT-2006-03 3 and by the Concerted Research action (GOA) of the Vrije Universiteit Brussel.

%-----------------------------------------------------------------

\appendix

\section{New HELAS subroutines for the 5-point vertex implementation}
\label{5pointHELAS}

Recall that {\tt ggggT} vertices are obtained from the interaction Lagrangian among the graviton tensor and four gluons by Eq.~(\ref{g5t}). The subroutine {\tt GGGGTX} computes the portion of the amplitude of the {\tt ggggT} vertex from four {\tt G}luon polarization vectors and a {\tt T}ensor boson wavefunction corresponding to the color structure $f^{abe}f^{cde}$, which should be called as
\begin{align}
 {\tt CALL\ GGGGTX(VA,VB,VC,VD,TC,GC,GT\ ,\ VERTEX)}
\end{align}
The inputs {\tt VA(6)}, {\tt VB(6)} and {\tt VC(6)} are complex six-dimensional
arrays which contain the {\tt V}ector boson wavefunctions, and
their momenta as
\begin{align*}
 p_a^{\mu} &= (\Re e{\tt VA(5)},\Re e{\tt VA(6)},
               \Im m{\tt VA(6)},\Im m{\tt VA(5)}), \\
 p_b^{\mu} &= (\Re e{\tt VB(5)},\Re e{\tt VB(6)},
               \Im m{\tt VB(6)},\Im m{\tt VB(5)}), \\
 p_c^{\mu} &= (\Re e{\tt VC(5)},\Re e{\tt VC(6)},
               \Im m{\tt VC(6)},\Im m{\tt VC(5)}).
\end{align*}
The input {\tt TC(18)} is a complex 18-dimensional array which contains the wavefunction of the {\tt T}ensor boson:
\begin{align}
 {\tt T^{\mu+1,\nu+1}=TC(4\mu+\nu+1)},
\label{TtoTC}
\end{align}
and its four-momentum as
\begin{align*}
p^{\mu} &= (\Re e{\tt TC(17)},\Re e{\tt TC(18)},
            \Im m{\tt TC(17)},\Im m{\tt TC(17)}),
\end{align*}
The output {\tt VERTEX} is a complex number in units of GeV:
\begin{align}\label{g5tb}
 {\tt VERTEX}=-{\tt GT\,GC}^2\,T^{\mu\nu}\,
  G_{\mu\nu,\rho\lambda\sigma\delta}\,
 V_A^{\rho}V_B^{\sigma}V_C^{\lambda}V_D^{\delta}
\end{align}
with
\begin{align}\label{g5tc}
 G_{\mu\nu,\rho\sigma\lambda\delta}
  &=\eta_{\mu\nu}( \eta_{\rho\sigma}\eta_{\lambda\delta}
                  -\eta_{\rho\delta}\eta_{\sigma\lambda}) \nn\\
  &\quad+\big[ \eta_{\mu\rho}\eta_{\nu\delta}\eta_{\lambda\sigma}
            +\eta_{\mu\lambda}\eta_{\nu\sigma}\eta_{\rho\delta}
            -\eta_{\mu\rho}\eta_{\nu\sigma}\eta_{\lambda\delta} \nn\\
  &\qquad\ -\eta_{\mu\lambda}\eta_{\nu\delta}\eta_{\rho\sigma}
  +(\mu\leftrightarrow\nu)\big],
\end{align}
and we use the notation
\begin{align}
 V_A^{\mu} &= {\tt VA(\mu+1)}, \, V_B^{\mu} = {\tt VB(\mu+1)}, \nn\\
 V_C^{\mu} &= {\tt VC(\mu+1)}, \, V_D^{\mu} = {\tt VD(\mu+1)}.
\end{align}

In order to insert automatically 5-point vertices in \madgraph, we introduce non-propagating colorless tensor boson $t_A$ as an auxiliary particle. With $t_A$ the portion of the 5-point vertex with color structure $f^{abe}f^{cde}$ can be reduced automatically. After introducing particle $t_A$, several new {\sc HELAS} subroutines were added. They are summarized on table 1, and can be called on the following way: 

\subsubsection*{VVTCXX}

This subroutine computes the amplitude of the {\tt gg$t_A$} vertex,
\begin{align*}
 {\tt CALL\ VVTCXX(VA,VB,TC,GC,VMASS\ ,\ VERTEX)}
\end{align*}
{\tt VMASS} represents the vector gluon mass which is zero (although
it does not play any role here, we keep it as an input argument in
accordance with the subroutine {\tt VVTXXX}, for convenience of \madgraph). What we compute here is
\begin{align}
 {\tt VERTEX} = {\tt GC}\, \eta_{\mu\rho} \eta_{\nu\sigma} T^{\mu\nu}
 V_A^{\rho}V_B^{\sigma}.
\end{align}

\subsubsection*{JVTCXX}

This subroutine computes an off-shell vector current {\tt J} made from
the interactions of a {\tt V}ector gluon and an auxiliary {\tt T}ensor boson by the
{\tt gg$t_A$} vertex, and should be called as
\begin{align*}
 {\tt CALL\ JVTCXX(VC,TC,GC,VMASS,VWIDTH\ ,\ JVTC)}
\end{align*}
The input {\tt VC(6)} is the wavefunction and momentum of the gluon.
The output {\tt JVTC(6)} gives the off-shell vector current
multiplied by the gluon propagator, which is expressed as a complex
six-dimensional array:
\begin{align}
 &{\tt JVTC(\alpha+1)} = -\frac{GC}{q^2}\eta_{\alpha\mu} V^\alpha T^{\mu\nu},
\end{align}
and
\begin{align}
 {\tt JVTC(5)} &= {\tt V(5)}+{\tt TC(17)}, \\
 {\tt JVTC(6)} &= {\tt V(6)}+{\tt TC(18)}.
\end{align}
Here the momenta $q$ are
\begin{align*}
 q^{\mu} &= (\Re e{\tt JVTC(5)},\Re e{\tt JVTC(6)},
             \Im m{\tt JVTC(6)},\Im m{\tt JVTC(5)}).
\end{align*}

\subsubsection*{UVVCXX}

This subroutine computes an off-shell tensor current {\tt U}
for the auxiliary tensor $t_A$, made from two flowing-out {\tt V}ector gluons by the {\tt gg$t_A$}
vertex, and should be called as
\begin{align*}
 {\tt CALL\ UVVCXX(VA,VB,GC,VMASS,XM,XW\ ,\ UVVC)}
\end{align*}
The inputs {\tt XM} and {\tt XW} are two dummy arguments for which
we keep them in accordance with the subroutine {\tt UVVXXX}. The
output {\tt UVVC(18)} is a complex 18-dimensional array:
\begin{align}
 T^{\alpha\beta} &= {\tt GC}V_A^{\alpha}V_B^{\beta}
\end{align}
for the first 16 components of {\tt UVVC}, and
\begin{align}
 {\tt UVVC(17)} &= {\tt VA(5)}+{\tt VB(5)}, \\
 {\tt UVVC(18)} &= {\tt VA(6)}+{\tt VB(6)}.
\end{align}

\subsubsection*{TTTXXX}

This subroutine computes the amplitude of the {\tt $t_A$$t_A$T}
vertex, called as
\begin{align*}
{\tt CALL\ TTTXXX(TC,T1C,T2C,GT,\ VERTEX)}
\end{align*}
The inputs {\tt T1C(18)} and {\tt T2C(18)} are complex 18-dimensional arrays which contain
the wavefunction and momenta for the auxiliary tensors. What we compute here is
\begin{align}
 {\tt VERTEX} = {\tt GT}\, T^{\mu\nu} T_1^{\rho\lambda} T_2^{\sigma\delta}
 G_{\mu\nu,\rho\sigma\lambda\delta},
\end{align}
with $T_{1,2}$ defined from {\tt T1C(18)} and {\tt T2C(18)} as in
Eq.~(\ref{TtoTC}).

\subsubsection*{UTTAXX}

This subroutine computes an off-shell non-propagating tensor current {\tt U},
made from the flowing-out graviton {\tt T}ensor and the auxiliary {\tt T}ensor by the {\tt $t_A$$t_A$T}
vertex, and should be called as
\begin{align*}
 {\tt CALL\ UTTAXX(TC,T1C,GT,\ UTTA)}
\end{align*}
The output {\tt UTTA(18)} is a complex 18-dimensional array:
\begin{align}
T_A^{\alpha\beta} &= {\tt GT}\, \eta^{\sigma\alpha} \eta^{\delta\beta} T^{\mu\nu} T_1^{\rho\lambda} G_{\mu\nu,\rho\sigma\lambda\delta}
\end{align}
for the first 16 components of {\tt UTTA}, and
\begin{align}
 {\tt UTTA(17)} &= {\tt T(17)}+{\tt T1(17)}, \\
 {\tt UTTA(18)} &= {\tt T(18)}+{\tt T1(18)}.
\end{align}

\subsubsection*{UTTBXX}

This subroutine computes an off-shell graviton tensor current {\tt U},
made from the two flowing-out auxiliary {\tt T}ensors by the {\tt $t_A$ $t_A$T}
vertex, and should be called as
\begin{align*}
 {\tt CALL\ UTTBXX(T1C,T2C,GT, TMASS, TWIDTH, \ UTTB)}
\end{align*}
The inputs {\tt TMASS} and {\tt TWIDTH} are the graviton mass and
width, $m_T$ and $\Gamma_T$. The output {\tt UTTA(18)} is a complex
18-dimensional array:
\begin{align}
T^{\alpha\beta} &= {\tt GT}\frac{- B_{\mu\alpha,\nu\beta}}{q^2-m_T^2+im_T\Gamma_T}
\, T_1^{\rho\lambda} T_2^{\sigma\delta} G_{\mu\nu,\rho\sigma\lambda\delta}
\end{align}
for the first 16 components of {\tt UTTB}, and
\begin{align}
 {\tt UTTB(17)} &= {\tt T1(17)}+{\tt T2(17)}, \label{qUTT1}\\
 {\tt UTTB(18)} &= {\tt T1(17)}+{\tt T2(18)}. \label{qUTT2}
\end{align}
Here $q$ is the momentum of the off-shell tensor boson given in Eqs.~(\ref{qUTT1}) and
(\ref{qUTT2}) as
\begin{align*}
 q^{\mu} = (\Re e{\tt UTTB(17)},\Re e{\tt UTTB(18)},
            \Im m{\tt UTTB(18)},\Im m{\tt UTTB(17)}).
\end{align*}
And $B_{\mu\alpha,\nu\beta}$ is the polarization summation
tensor\cite{Hagiwara:2008zr}.

%----------------------------------------------------------------------------------------------------------------
\section{New subroutines for summation over continuous mass spectrum}
\label{MassIntegration}

In order to have the mass integration realized in \madgraph~within the framework of the ADD model, new auxiliary particles were introduced in Sec.~\ref{implem}. In consequence, three new sub-routines summarized in table 2 were added into the {\sc HELAS} structure. The details for the insertion of these sub-routines are the following:

\subsubsection*{ PXXXXX}

This subroutine stores the helicity and momentum of the auxiliary
spin-2 {\tt P}seudo-particle for further usage, and should be called
as
\begin{align*}
  {\tt CALL\ PXXXXX(P,XM,NHEL,NST\ ,\ PC)}
\end{align*}
The input {\tt P(0:3)} is a real four-dimensional array which
contains the four-momentum $p^{\mu}$ of the pseudo particle, {\tt
NHEL} (${\tt =\pm 2,\pm 1,0}$) specifies its helicity $\lambda$,
{\tt NST} specifies whether the boson is in the final state ({\tt
NST = 1}) or in the initial state ({\tt NST = -1}). {\tt XM} is its
mass but does not play any role here.

The output {\tt PC(18)} is a complex 18-dimensional array, among
which only the following matter, namely
\begin{align}\label{pxxxxx}
{\tt PC(\ 1)=  NHEL},
\end{align}
and
\begin{align}
  ({\tt PC(17)},\,{\tt PC(18)})
 ={\tt NST}\,({\tt P(0)}+i{\tt P(3)},\,{\tt P(1)}+i{\tt P(2)}).
\end{align}

\subsubsection*{TPSXXX}

This subroutine computes the amplitude of the {\tt $x_1$$x_2$T}
vertex, and should be called as
\begin{align*}
{\tt CALL\ TPSXXX(TC0,PC,SC,GP,TMASS0,TWIDTH,\ VERTEX)}
\end{align*}
{\tt SC(3)} is a complex three-dimensional array which contains the
momentum for $x_1$. {\tt GP}(1:2) here are not coupling constant,
but contain the ADD model inputs:
\begin{align}
&{\tt GP(\ 1)}=\Lambda + i \delta, \nn \\
&{\tt GP(\ 2)}=M_{low} + i M_{up} ,
\end{align}
$M_{low}$ and $M_{up}$ are the lower and upper limits for graviton
mass integration. {\tt TMASS0} ($m_{T0}$ for below) is only
temporarily used and got from the MG input file {\tt
param\_card.dat}, while the real graviton mass $m_{T}$ is defined as
the invariant mass of x1 and x2, \ie,
\begin{align}
m_T \equiv |q_1+q_2| ,
\end{align}
with $q_1$ and $q_2$ are the momenta of $x_1$ and $x_2$.

Also note the input {\tt TC0} is indeed calculated from the other
side of Feynman diagram, and contains the polarization summation
tensor $B_{\mu\nu\rho\sigma}$, which can be used to project out the
real graviton wave function, due to the relations of the 5 helicity
states of the tensor boson~\cite{Hagiwara:2008zr}:
\begin{align} \label{Bmassive}
B_{\mu\nu,\alpha\beta}(p)= \sum_{\lambda=\pm2,\,\pm1,0}
\epsilon_{\mu\nu}(p,\lambda)
  \epsilon_{\alpha\beta}(p,\lambda)^*,
\end{align}
and
\begin{align}
  \epsilon^{\mu\nu}(p,\lambda)\epsilon_{\mu\nu}(p,\lambda^\prime)^*
 =\delta_{\lambda\lambda^\prime},
\end{align}
with $p \equiv q_1+q_2$, and the helicity $\lambda$ is got from the
pseudo-particle's wavefunction (Eq.~(\ref{pxxxxx})).

Now we can get the true graviton tensor wave function by calling the
subroutine
\begin{align}
  {\tt CALL\ TXXXXX(P},m_T,{\tt \lambda, +1 ,\ TC)}
\end{align}

Finally, the output {\tt VERTEX} is
\begin{align} \label{TPS}
 {\tt VERTEX} =& \frac{8\pi^2}{m_T}\rho{(m_T)}
 (p^2-m_{T0}^2+im_{T0}\Gamma_T) T_{\mu\nu} T_0^{\mu\nu}\nn\\
 &\times  \theta(m_T-M_{low})\theta(M_{up}-m_T),
\end{align}
where $\theta$ represents the Heaviside step function. In
Eq.~(\ref{TPS}), we include the mass density factor(Eq.~(\ref{ps})),
the inverse graviton propagator, and the compensation factor for the
decay phase space.

\subsubsection*{UPSXXX}
This subroutine computes an off-shell graviton tensor current {\tt
U}, made from the flowing-out auxiliary {\tt S}calar and {\tt
P}esudo-particles, by the {\tt $x_1$$x_2$T} vertex, and should be
called as

\begin{align*}
{\tt CALL\ UPSXXX(PC,SC,GP,TMASS0,TWIDTH,\ UPS)}
\end{align*}
The output UPS is a complex 18-dimensional array, which is indeed
got simply by calling inside {\tt UPSXXX}
\begin{align}
  {\tt CALL\ TXXXXX(P},m_T,{\tt \lambda, +1 ,\ UPS)}.
\end{align}
and including the step functions in Eq.~(\ref{TPS}).

\section{New subroutines for the massless spin-2 particles }
\label{MasslessSpin2}

In order to have massless spin-2 particles also implemented in \mgme~, the tensor wave function subroutine was modified, and four subroutines for the off-shell tensor were added to the former  {\sc HELAS} implementation. The details are presented as follows.

\subsubsection*{TXXXXX}
This subroutine computes the graviton tensor wave function, namely $\epsilon^{\mu\nu}(p,\lambda)$ and $\epsilon^{\mu\nu}(p,\lambda)^*$ in terms of the graviton four-momentum $p$ and its helicity $\lambda$.  The sum of helicity states for the graviton is introduced on the routines to be given by Eq. (\ref{Bmassive}).
%\begin{align} \label{Bmassless}
%B_{\mu\nu,\alpha\beta}(p)= \sum_{\lambda=-2}^{2}
%\epsilon_{\mu\nu}(p,\lambda)
 % \epsilon_{\alpha\beta}(p,\lambda)^*.
%\end{align}
While the massive graviton has five helicity states of polarization, the massless spin-2 particles will only have two physical states. The sum of polarization for the massless graviton is therefore modified on the routines to set $\lambda=\pm 1,\,0$ to zero allowing only $\lambda=\pm 2$ to contribute.

\subsubsection*{UIOXXX, UVVXXX, UIOVXX, UVVVXX}

These subroutines compute off-shell tensor currents {\tt U} by the {\tt FFT}, {\tt VVT}, {\tt FFVT}, and {\tt VVVT} vertices respectively. The main modification here was the inclusion of the zero mass graviton propagator to be the one showed in Eq. (\ref{Prop2}) for the massless case. 
%----------------------------------------------------------------------------------------------------------------

\section{Higgs effective theory new implementation into MadGraph}
\label{heftb}

In Ref.~\cite{Alwall:2007ys}, Higgs effective theory has been
implemented into \madgraph~by reorganizing the Lagrangian with
introducing an extra non-propagating auxiliary tensor, to avoid the
5-point vertex problem. Although it seems more elegant in physics,
it is hard to find a similar way for graviton case (due to more
complicated Lagrangian for graviton interactions). Moreover, the
previous way affects the 4-gluon vertices, reducing them into
subparts corresponding to different color structure, and thus
increases the numbers of Feynman diagrams for process like
$gg\rightarrow gg$. This may slow the running of \madevent, which is
based on single diagram enhanced method and thus sensitive to the
singular diagram numbers.

As mentioned in Sec.~\ref{implem}, we have implemented HEFT into \madgraph~in the same way as for the graviton (see the model directory {\tt heftb} in \madgraph). We add three more  {\sc HELAS}  subroutines as following:

\subsection*{TTSCXX}

This subroutine computes the amplitude of the {\tt $t_A$$t_A$H} vertex,
\begin{align*}
 {\tt CALL\ TTSCXX(T1C,T2C,S,GH, \ VERTEX)}
\end{align*}
The input
{\tt GH} $=\alpha_S/(3\pi v)$ is the Higgs effective coupling constant.
{\tt S(3)} is a complex three-dimensional
array which contain the wavefunctions of the {\tt S}calar bosons,
{\tt S(1)}, and their four-momenta as
\begin{align*}
 p^{\mu} &= (\Re e{\tt S(2)},\Re e{\tt S(3)},
               \Im m{\tt S(3)},\Im m{\tt S(2)}). \\
\end{align*}
The output {\tt VERTEX} is:
\begin{align}
 {\tt VERTEX}=-{\tt GH}\,S(1)\, T_1^{\mu\nu} T_2^{\alpha\beta}(\eta_{\mu\beta}\eta_{\alpha\nu}-\eta_{\mu\alpha}\eta_{\nu\beta})
\label{ttsc}
\end{align}

\subsection*{HTTCXX}

This subroutine computes an off-shell scalar current {\tt H} made from
the interactions of two auxiliary {\tt T}ensor bosons by the
{\tt $t_A$$t_A$H} vertex, and should be called as
\begin{align*}
 {\tt CALL\ HTTCXX(T1C,T2C,GH, SMASS, SWIDTH, \ HTTC)}
\end{align*}
The inputs {\tt SMASS} and {\tt SWIDTH} are the scalar boson mass $m_S$ and width $\Gamma_S$. The output
{\tt HTTC(3)} gives the off-shell scalar current multiplied by the scalar
boson propagator, which is expressed as a complex three-dimensional
array:
\begin{align}
 {\tt HTTC(1)} = -\frac{{\tt GT}}{q^2-m_S^2+im_S\Gamma_S}
\, T_1^{\mu\nu} T_2^{\alpha\beta}(\eta_{\mu\beta}\eta_{\alpha\nu}-\eta_{\mu\alpha}\eta_{\nu\beta})
\label{httc}
\end{align}
and
\begin{align}
 {\tt HTTC(2)} &= {\tt T1C(17)}+{\tt T2C(17)}, \\
 {\tt HTTC(3)} &= {\tt T1C(18)}+{\tt T2C(18)}.
\end{align}
Here the momenta $q$ is
\begin{align*}
 q^{\mu} &= (\Re e{\tt HTTC(2)},\Re e{\tt HTTC(3)},
             \Im m{\tt HTTC(3)},\Im m{\tt HTTC(2)}).
\end{align*}

\subsection*{\tt UTSCXX}

This subroutine computes an off-shell auxiliary tensor current {\tt U} made from
one flowing-out {\tt S}calar bosons and one auxiliary {\tt T}ensor boson by the {\tt $t_A$$t_A$H} vertex, and
should be called as
\begin{align*}
 {\tt CALL\ UTSCXX(T1C,S,GH,XM,XW\ ,\ UTSC)}
\end{align*}
The inputs {\tt XM} and {\tt XW} are two dummy arguments. The output {\tt UTSC(18)}
gives the off-shell tensor current, which is expressed as a complex 18-dimensional array:
\begin{align}
 T^{\alpha\beta} = -{\tt GH}\,S(1)\, T_1^{\mu\nu} (\eta_{\mu\beta}\eta_{\alpha\nu}-\eta_{\mu\alpha}\eta_{\nu\beta})
\label{utsc}
\end{align}
for the first 16 component of {\tt UTSC}, and
\begin{align}
 {\tt UTSC(17)} &= {\tt T1C(17)}+{\tt S(2)},
\label{qUTSC1}\\
 {\tt UTSC(18)} &= {\tt T1C(18)}+{\tt S(3)}.
\label{qUTSC2}
\end{align} \newpage
%
%%%%%%%%%%%%%% Begin References %%%%%%%%%%%%%%%%%%%%%%%%%%%%%%%%%%%%%%%%
%\begin{thebibliography}{00}
%\end{thebibliography}
\bibliographystyle{JHEP}
\bibliography{GJMLM}
\end{document}